\documentclass[11pt,a4paper]{article}
\usepackage{epsfig}
\usepackage[T1]{fontenc}    
\usepackage{graphics}
\usepackage{graphicx}
\usepackage{pstricks,pst-coil,pst-fill,pst-plot}
\usepackage[fleqn]{amsmath}    
\usepackage{amssymb}    
\usepackage{amsthm}    
\usepackage{amsfonts}   
\usepackage{verbatim}   
\usepackage{mathrsfs}   
\usepackage{dsfont}
\usepackage{euscript}
\usepackage{yfonts}
\usepackage{enumerate}     
\usepackage{amsthm}         
\usepackage{txfonts}
\usepackage{marvosym}
\usepackage{stmaryrd}
\usepackage{vmargin}        
\usepackage{wasysym}		

\usepackage{bbm}

\setmarginsrb{1.8cm}{2cm}{1.8cm}{2cm}{1cm}{1cm}{1cm}{1.6cm}
\makeatletter
\@addtoreset{equation}{section}
\makeatother

\providecommand{\bysame}{\leavevmode\hbox to3em{\hrulefill}\thinspace}
\providecommand{\MR}{\relax\ifhmode\unskip\space\fi MR }

\providecommand{\href}[2]{#2}

\let\ua=\uparrow

\let\tend=\rightarrow

\long\def\symbolfootnote[#1]#2{\begingroup%
	\def\thefootnote{\fnsymbol{footnote}}\footnote[#1]{#2}\endgroup}

\newtheorem{theorem}{Theorem}[section]
\newtheorem{prop}[theorem]{Proposition}
\newtheorem*{theorem*}{Theorem}

\newtheorem{defin}[theorem]{Definition}

\newtheorem{conj}[theorem]{Conjecture}

\newtheorem{lemme}[theorem]{Lemma}

\def\Proof{\medskip\noindent {\it Proof --- \ }}

\def\qed{\hfill\rule{2mm}{2mm}}

\newcommand\beq{\begin{equation}}
	\newcommand\enq{\end{equation}}
\newcommand\bem{\begin{multline}}
	\newcommand\enm{\end{multline}}

\def\beqa{\begin{eqnarray}}
	\def\eeqa{\end{eqnarray}}
\def\ba{\begin{array}}
	\def\ea{\end{array}}

\newcommand{\f}[2]{{\ensuremath{%
			\mathchoice%
			{\dfrac{#1}{#2}}
			{\dfrac{#1}{#2}}
			{\frac{#1}{#2}}
			{\frac{#1}{#2}}
}}}

\def\a{\alpha}

\def\be{\beta}
\def\ga{\gamma}
\def\Ga{\Gamma}

\def\de{\delta}

\def\De{\Delta}
\def\eps{\epsilon}
\def\veps{\varepsilon}
\def\la{\lambda}
\def\La{\Lambda}

\def\sg{\sigma}

\def\Ups{\Upsilon}

\def\th{\theta}

\def\vth{\vartheta}

\def\Om{\Omega}

\def\vp{\varphi}

\newcommand{\mc}[1]{\ensuremath{\mathcal{#1}}}
\newcommand{\mf}[1]{\ensuremath{\mathfrak{#1}}}
\newcommand{\msc}[1]{\ensuremath{\mathscr{#1}}}

\newcommand{\bs}[1]{\ensuremath{\boldsymbol{#1}}}

\DeclareFontFamily{OT1}{pzc}{}
\DeclareFontShape{OT1}{pzc}{m}{it}{<-> s * [1.10] pzcmi7t}{}
\DeclareMathAlphabet{\mathpzc}{OT1}{pzc}{m}{it}

\def \i{ \mathrm i}

\newcommand{\ov}[1]{\ensuremath{\overline{#1}}}
\newcommand{\wt}[1]{\ensuremath{\widetilde{#1}}}
\newcommand{\wh}[1]{\ensuremath{\widehat{#1}}}

\newcommand{\Int}[2]{\ensuremath{\int\limits_{#1}^{#2}}}

\newcommand{\sul}[2]{\ensuremath{\sum\limits_{#1}^{#2}}}
\newcommand{\pl}[2]{\ensuremath{\prod\limits_{#1}^{#2}}}

\newcommand{\R}{\ensuremath{\mathbb{R}}}
\newcommand{\Cx}{\ensuremath{\mathbb{C}}}

\newcommand{\Dp}[1]{\ensuremath{\partial_{#1}}}

\newcommand{\limit}[2]{\ensuremath{\underset{#1 \tend #2}{\longrightarrow} }}

\newcommand{\ex}[1]{\ensuremath{\e{e}^{#1}}}

\newcommand{\op}[1]{ \boldsymbol{ \texttt{#1} } }

\newcommand{\norm}[1]{\ensuremath{  || #1 || }}

\newcommand{\dd}{\mathrm{d}}
\newcommand{\e}[1]{\ensuremath{\mathrm{#1}}}

\newcommand{\intff}[2]{\ensuremath{ [  #1 \,; #2 ] }}

\newcommand{\intn}[2]{\ensuremath{[\![ \, #1 \,;\, #2 \,]\!]}}

\begin{document}

\begin{center}
	\begin{LARGE}
			{\bf Wightman axiomatics of the bootstrap construction of the 1+1 dimensional Sinh-Gordon model}
	\end{LARGE}

	\vspace{1cm}
		
	\vspace{4mm}
	{\large Karol K. Kozlowski \footnote{e-mail: karol.kozlowski@ens-lyon.fr}}%
	\\[1ex]
	Univ Lyon, ENS de Lyon, Univ Claude Bernard Lyon 1, CNRS, Laboratoire de Physique, F-69342 Lyon, France \\[2.5ex]

	{\large Alex Simon \footnote{e-mail: alex.simon@ens-lyon.fr}}%
	\\[1ex]
	Univ Lyon, ENS de Lyon, Univ Claude Bernard Lyon 1, CNRS, Laboratoire de Physique, F-69342 Lyon, France \\[2.5ex]

\par 
		
\end{center}
	
\vspace{40pt}

\centerline{\bf Abstract} \vspace{1cm}
\parbox{12cm}{\small}

The integrable bootstrap program allows one to express the tempered distributions associated with the multipoint functions of the integrable  1+1 dimensional Sinh-Gordon quantum field theory
by means of explicit series. The convergence of the latter is an open problem that was only solved for the two-point case.
In this work, by taking for granted the convergence of these series, we show that these expressions satisfy all of the Wightman axioms.
This thus shows that, upon a yet to be proven convergence property, the integrable bootstrap based construction of correlation functions
does lead to a quantum field theory.

\vspace{20pt}
	

\vspace{40pt}
	
\tableofcontents
	
\section{Introduction}

On may formulate the problem of constructing rigorously a \textit{bona fide} quantum field theory as the following requirements:
\begin{itemize}
\item[i)] construct a separable Hilbert space $\mf{h}$
\item[ii)] realise a continuous unitary representation of the Poincaré group on $\mf{h}$
\item[iii)] provide a tempered distribution taking values in unbounded self-adjoint operators on $\mf{h}$, the hermitian field operator $\bs{\Psi}$ (or multiplets thereof).
The latter has to be compatible with the presence of relativistic symmetry in the model.
\end{itemize}
These objects should satisfy several requirements. First of all, there should exists a unique one dimensional subspace $\mf{e}_{\e{vac}} = \big\{ \la \bs{v}_{\e{vac}}, \: \, \la \in \Cx \big\}$
of $\mf{h}$ invariant under the action  of the Poincaré group, called the "vacuum" of the model. Second, the vacuum is cyclic with respect to the action of smeared fields, what means that
\beq
\e{Vect}\Big\{ \bs{\Psi}[f_1]\cdots \bs{\Psi}[f_n] \bs{v}_{\e{vac}} \, : \, n \in \mathbb{N}, f_1,\dots, f_n \in \mc{S}\big( \R^{1 , d} \big) \Big\}
\enq
is dense in $\mf{h}$. Above, $\R^{1 , d}$ refers to $\R^{1 + d}$ endowed with the diagonal metric $(1, -1, \dots, -1)$ and $\mc{S}\big( X \big)$ stands for the
space of Schwartz functions on $X$.

The search for effective ways to provide the above setting for interacting quantum field theories on rigorous mathematical grounds has been an extensive subject since the 60s.
From these investigations, there emerged a mathematical construction of quantum field theories based on objects of prime physical
importance: the correlation functions, see \textit{e.g.}
the monographs \cite{BogoliubiovLogunovTodorovIntroAxiomaticQFT,StreaterWightmanPCTStatisticsAllThat}.
More precisely, if one is given a collection of distributions $\op{W}^{(k)}$ on  $\mc{S}\big( (\R^{1 , d})^k \big)$, $k\in \mathbb{N}$,
satisfying certain axioms, nowadays called the Wightman axioms, then one may automatically realise the aforementioned constructions, points i)-iii) above.
The distributions one starts with then correspond to the constructed theory's correlation functions
\beq
\op{W}^{(k)}[f] \, = \, \Big( \bs{v}_{\e{vac}} , \bs{\Psi}[f_1] \cdots \bs{\Psi}[f_k]  \bs{v}_{\e{vac}}\Big) \qquad \e{with} \qquad
\left\{   \ba{ccc} f(\bs{X}_k) & = & \pl{a=1}{k} f_a(\bs{x}_a)  \vspace{2mm} \\
	 \bs{X}_k  &  =  &  \big( \bs{x}_1,\dots, \bs{x}_k\big) \ea \right. \;,
\enq
and $\bs{x}_a \in \R^{1,d}$. Note that it is convenient to represent these distributions in terms of the associated generalised functions which, formally speaking,
correspond to their integral kernels
\beq
\op{W}^{(k)}[f] \; = \; \Int{ (\R^{1,d})^k   }{} \hspace{-2mm} \dd^k \bs{X} \,  \mc{W}^{(k)}(\bs{X}_k) f(\bs{X}_k)
								\qquad \e{with} \qquad \dd^k \bs{X} \, = \, \pl{a=1}{k}\dd \bs{x}_a \;.
\label{definition noyau generalise distribution}
\enq
The generalised functions $\mc{W}^{(k)}(\bs{X}_k)$ are referred to as Wightman functions.

However, in practice, finding distributions which verify Wightman axioms is an arduous task.
First of all, due to the presence of a probabilistic setting, it is more convenient to construct
Euclidean analogues of the Wightman functions, \textit{aka} Schwinger functions. The $k$-point Schwinger functions are generalised functions on the Euclidean space $(\R^{1+d})^k$
instead of the Minkowski one $(\R^{1,d})^k$.
In such an Euclidean setting, if one is able to construct a collection of distributions on  $\mc{S}\big( (\R^{1+d})^k \big)$ - the mentioned Schwinger functions-
satisfying the so-called Osterwalder-Schrader axioms, then the Osterwalder-Schräder reconstruction theorem
\cite{OsterwalderSchraderAxiomsEQFTIWithError,OsterwalderSchraderAxiomsEQFTIIAdditionalAxiomsAndProof}
ensures that one may use these to construct distributions satisfying the Wightman axioms. This thus leads to the construction of the sought quantum field theory.
The main advantage of working in Euclidean space, is that one
may build on the probabilistic, constructive renormalisation based, rigorous construction of the associated path integral
which then provides one with the Schwinger functions. This program was pioneered by
\cite{GlimmJaffeSpencerP(phi)2ParRenormalisationConstructive} and achieved for many quantum field theories in various dimensions, see \textit{e.g.} the review
 \cite{SummersStateOfTheArtConstructiveQFTAsOf2012}.
It provides one with the existence of the theory but does not lead to closed expressions for the Schwinger, or associated Wightman functions.
The best one can do within the formalism, is to obtain their perturbative expansions in coupling constants around the free theories.
More recently, there emerged various other approaches allowing one to construct Schwinger functions in an alternative way.
Notably, one could cite the rigorous development of the stochastic quantisation, first proposed in \cite{ParisiWuStochasticQuantisationIdea} by the physics litterature, which presents various advantages in respect to the constructive renormalisation based approach  and constituted an important progress in reaching the functional measure allowing one to build the Schwinger functions \cite{GubinelliHofmanovaPhi4inD=2And3EucildianStochasticQuantisation,HairerRegularityStructureOverviewAndPhi4in3D,Jona-LasinioMitterStochasticQuantisationPhi4In2D}.
One should however mention that this alternative construction still provides
a quite indirect access to the Schwinger functions, thus only granting a rather low control on their properties beyond a perturbative regime.
The two mentioned approaches allow one to deal with quantum field theory models in various dimensions, $1+d$, although
going to higher dimensions demands increasing efforts an in various respects, going beyond $1+d=3$ seems quite challenging
for the moment, although some partial results were obtained in $1+d=4$ for instance.

One should definitely mention another progress: the Gaussian multiplicative chaos approach to the construction of Euclidean path integrals
for two-dimensional conformal field theories, which coupled to integrability properties of the latter allowed to access to explicit informations on the Schwinger functions
for the Liouville theory in particular \cite{GuillarmouKupiainenRhodesVargasProofOfDOZZByGaussianMultChaos,KupiainenRhodesVargasProofOfDOZZByGaussianMultChaos}.
Some progress was also achieved  by this approach in respect to the massive Euclidean two-dimensional Sinh-gordon model in finite volume
\cite{GuillarmouGunaratnamVargasGMCConstructionofSinhGMeasure},
although the construction of the infinite volume Schwinger functions through is still open.
It seems fair to say that, as of today and despite important progress on various fronts, there does not exist yet an appropriate unifying framework allowing one to prove the
existence of Schwinger/Wightman functions associated with an arbitrary interacting quantum field theory.

The approaches described above provide one with existence theorems for Schwinger functions associated with various classical Lagrangians.
Still, ideally, one would like to have a better understanding of the Schwinger, resp. the associated Wightman, functions as generalised functions
on $(\R^{1 + d})^k$, resp. $ (\R^{1 , d})^k$, and have closed form expressions for the latter. While this could have been achieved to some extent
for conformal field theories owing to their very specific feature, the matter remains widely open for any massive  infinite volume quantum field theory that would
not be free. Integrable quantum field theories in 1+1 dimensions constitute one instance -the sole know so-far beyond the free theories- where this seems possible.
These theories are constructed within the  $\op{S}$-matrix program \cite{HeisenbergSomeAspectsofSMatrixIdeas,WheelerFirstIntroConceptSmatrix}.
The idea is to take certain \textit{formal} equations which can be argued from the
\textit{formal} Minkowski path-integral construction of these theories, as axioms which allow the theory to be built.
More precisely, starting from a postulated form of a theory's $\op{S}$-matrix, one axiomatises a collection of coupled Riemann--Hilbert problems
for unknown functions in $1, 2,\dots,$ complex variables. These functions arise as building blocks of the Wightman functions for the associated theory.
This is the celebrated bootstrap program  pioneered in \cite{KarowskiWeiszFormFactorsFromSymetryAndSMatrices,KirillovSmirnovFirstCompleteSetbootstrapAxiomsForQIFT}.
We refer to the book \cite{SmirnovFormFactors} for more details.
The main point is that $\op{S}$-matrices of integrable quantum field theories in $1+1$ dimensions may be constructed explicitly. Indeed,
integrability imples elasticity and factorisability of the scattering which, in turn, implies that the $S$-matrix satisfies the celebrated Yang-Baxter equation.
The construction of solutions to the latter is, nowadays, understood quite well, on a deep algebraic level.
In particular, this knowledge allowed to address the Bootstrap program based construction
of the associated theories. Various techniques \cite{BabujianFringKarowskiZapletalExactFFSineGordonBootsstrapI,LukyanovFirstIntroFreeField,SmirnovFormFactors} were devised for that purpose.
This approach led to closed form, explicit, expressions for various two-points arising in the considered integrable quantum field theories.
The two-point functions were expressed in terms of series, whose $n^{\e{th}}$ summand is an $n$-fold integral. The convergence of these
series is however, generically, an open problem that could have been resolved so-far only for the space-like two-point functions of the Sinh-Gordon
quantum field theory \cite{KozConvergenceFFSeriesSinhGordon2ptFcts}.
We remind that the latter  refers to the quantum field theory resulting from the quantisation of the
classical Sinh-Gordon field theory described by the classical Lagrangian
\begin{equation}
\mc{L}[\varphi](\bs{x}) \, = \, \big( \Dp{t}\vp(\bs{x}) \big)^2 \, - \,  \big( \Dp{x}\vp(\bs{x}) \big)^2  \, - \,  \frac{m^2}{g^2}\cosh\big( g\varphi(\bs{x}) \big)
\qquad \e{with} \qquad \bs{x} = (t,x) \in \R^{1,1}\,.
\end{equation}

Thus, apart form the particular example of the space-like two-point functions of the Sinh-Gordon model,
the bootstrap program issued expressions for the two-point functions should be considered as conjectural, in the sense
that one has not established yet their well-definiteness. The construction of multi-point functions, \textit{i.e.} Wightman functions in any number of space-time variables,
was only achieved recently in full detail in \cite{KozSimonMultiPointCorrFctsSinhGordon}, again in the case of the
Sinh-Gordon model, and this up to the convergence conjecture of the obtained expressions.
To be more precise, the multi-point correlation functions are given by series of multiple integrals in the spirit of the results for the two-point functions.
The answer should be considered on the conjectural level in that the convergence of these series is an open problem so far.

The purpose of the present work is as follows. By taking for granted that the series defining the multi-point functions in the Sinh-Gordon quantum field theory
are convergent, we prove that the associated tempered distributions do satisfy all the Wightman axioms. As a consequence, provided convergence holds, we do
establish that the bootstrap approach does lead to a quantum field theory. This is the first time, that the complete Wightman program is achieved
for an integrable quantum field theory. Some aspects of this issue were considered earlier in the literature.
F. Smirnov, \textit{c.f.}  \cite{SmirnovFormFactors}, considered the local commutativity axiom, taken in a non-rigorous way on the level
of operator products. However, while the clustering property of the form factors was largely discussed in the physics litterarture, it turns out that the chain of heuristic ideas
does not adapt sufficienly well to a proof one one has to recourse to much more involved arguemnts.

This paper is organised according to the following. Section \ref{Section Preliminary setup} introduces all the preliminary notions and results that are necessary for obtaining to the main result of the paper.
Subsection \ref{Subsection Wightman Axioms} recalls the Wightman axioms, directly in the form taken for a spinless scalar 1+1 dimensional quantum field theory, such as the Sinh-Gordon model.
Subsection \ref{Subsection bootstrap program results} reviews the results of \cite{KozSimonMultiPointCorrFctsSinhGordon}.
After setting up the necessary notations in Subsection \ref{Subsection building blocks},
closed expressions for the multi-point functions are introduced in Subsection \ref{subsubsection multipointFcts Reminder}.
Section \ref{Section Wightman axioms verification} is then devoted to the proof that the expressions for the multi-point functions obtained through the resolution of the
bootstrap program do satisfy all of the Wightman axioms, this under the convergence conjecture.

\section{Preliminary setup}
\label{Section Preliminary setup}

\subsection{Wightman Axioms}
\label{Subsection Wightman Axioms}

In order to state the Wightman axioms in the setting of interest to us, we need to introduce some preliminary notations. The Minkowski scalar product on $\R^{1,1}$ is defined as
\beq
\bs{x} * \bs{y} \, = \, x_0 y_0 - x_1 y_1   \qquad \e{with} \qquad \bs{x} = (x_0,x_1) \quad \e{and} \quad \bs{y} = (y_0,y_1) \;.
\enq
In particular, the Minkowski norm of $\bs{x}= (x_0,x_1) \in \R^{1,1}$ reads
\beq
	\bs{x}^2 = x_0^2-x_1^2  \;.
\label{ecriture Minkowski norm on R11}
\enq
The Minkowski scalar product is invariant under the action of the orthogonal group $O(1,1)$.
The special orthogonal group
\beq
SO^+(1,1)  \, = \, \Big\{  \La_{\th} \, : \, \th \in \R  \Big\} \qquad \e{where} \qquad
\La_{\th}  \, = \, \left( \ba{cc}  \cosh(\th) & -\sinh(\th) \\ -\sinh(\th) & \cosh(\th) \ea \right)  \;,
\label{definition Lorentz Boost}
\enq
along with space-time translations $\op{T}_{\bs{v}}:\bs{x} \mapsto \bs{x}+\bs{v}$, with $\bs{v} \in \R^{1,1}$, forms the orthochronous
Poincaré group $\mc{P}^+_{\ua}$. It is believed that $\mc{P}^+_{\ua}$ should be a fundamental symmetry of a quantum relativistic theory, \textit{viz}. that
the  action of $\mc{P}^+_{\ua}$ leaves such theories invariant. A group element $\op{U}_{\th, \bs{v}}\in \mc{P}^+_{\ua}$ is parameterised by $(\th,\bs{v}) \in \R \times \R^{1,1}$
and multiplication is defined by $\op{U}_{\th, \bs{v}}\op{U}_{\a, \bs{w}}=\op{U}_{\th+\a, \La_{\th}\bs{w}+\bs{v}}$.
The group element $\op{U}_{\th, \bs{v}}$ acts on $f \in \mc{S}\big( (\R^{1 , 1})^k \big)$ as
\beq
\big( \op{U}_{\th, \bs{v}}\cdot f\big)(\bs{X}_k) \, = \, f\big( \Lambda_{\theta}^{-1}( \bs{x}_1+ \bs{v} ),\dots, \Lambda_{\theta}^{-1} ( \bs{x}_k + \bs{v}) \big) \;.
\label{fonction apres action Groupe Poincaré}
\enq
The permutation group $\mf{S}_{k}$ acts naturally on $\mc{S}\big( (\R^{1 , 1})^k \big)$ by means of coordinate permutations:
\beq
\big(\sg \cdot G\big)(\bs{X}_k) \; = \;  G\big( \sg^{-1} \cdot \bs{X}_k \big) \quad \e{with} \quad
\sg^{-1} \cdot \bs{X}_k \, = \, \big(  \bs{x}_{\sg^{-1}(1)} \, , \cdots \, , \bs{x}_{\sg^{-1}(k)} \big) \;.
\label{ecriture action permutation generale sur fonction}
\enq
Two kinds of permutations will be of use in the following, the inversion $\iota$ and the adjacent transpositions $\tau_{s}$ where it holds:
\beq
\big( \iota \cdot G \big) (\bs{X}_k) \, = \, G(\iota \cdot \bs{X}_k)  \qquad \e{and} \qquad
\big( \tau_s \cdot G \big) (\bs{X}_k) \, = \, G(\tau_s \cdot \bs{X}_k ) \;,
\label{ecriture action inversion et transposition}
\enq
with $s\in \intn{1}{k-1}$ since $\iota^{-1}=\iota$ and $\tau_s^{-1}=\tau_s$. Note that above, it is understood that
\beq
\iota \cdot \bs{X}_k \, = \, \big( \bs{x}_k, \dots, \bs{x}_1 \big) \, = \, \overleftarrow{\bs{X}}_k
\qquad \e{and} \qquad
\tau_s \cdot \bs{X}_k \, = \, (\bs{x}_1,\dots, \bs{x}_{s-1} , \bs{x}_{s+1} , \bs{x}_{s}, \bs{x}_{s+2}, \dots, \bs{x}_k ) \;.
\label{definition action gpe perm sur vect et vect renverse}
\enq
The Fourier transform $\msc{F}$ on $\mc{S}\big( (\R^{1,1})^k\big)$ is defined as
\beq
\msc{F}[f]\big(\bs{q}_1,\cdots, \bs{q}_k\big) \, = \, \Int{ \big( \R^{1,1}\big)^k  }{} \dd^k\bs{X}  \cdot f\big( \bs{X}_k\big)  \cdot  \pl{a=1}{k}  \ex{\i \bs{q}_a * \bs{x}_a}  \;.
\enq
We remind that the Fourier transform of a distribution $\op{D}$ is defined as
\beq
\msc{F}[\op{D}][f] \, = \, \op{D}\big[ \msc{F}[f] \big] \;.
\enq
The spelling out of the Wightman axioms will, among other things, utilise the positive energy subspace of $(\R^{1,1})^k$
\beq
\mc{E}_{k}^+ \, = \, \Big\{(\bs{q}_1,\dots,\bs{q}_{k}) \in (\R^{1,1})^{k} :  \quad \bs{q}_{\ell}^2 \geq 0 \quad \textrm{and} \quad q_{\ell,0} \geq 0 \quad \textrm{for every} \quad \ell \in \intn{1}{k} \Big\} \;.
\label{definition domaine Fourier Energie positive}
\enq

\begin{defin}

A sequence $\big\{ \op{W}^{\, (k)} \big\}_{k \geq 1}$ of tempered distributions,  with $\op{W}^{\, (k)}$ being a distribution on $\mc{S}\big( (\R^{1,1})^k\big)$,
 is said to satisfy the Wightman axioms if the below properties hold:
	\begin{itemize}
\item {\bf Covariance}: For any $(\th,\bs{v}) \in \R \times \R^{1,1}$ and $f \in \mc{S}\big( (\R^{1 ,1})^k \big)$ it holds
\beq
 \op{W}^{\,(k)}\big[ \op{U}_{\th, \bs{v}}\cdot f \big] \; = \;\op{W}^{\,(k)}\big[ f \big] \; .
\enq

\item {\bf Spectral conditions}: there exists a sequence of tempered distributions $\big\{ \op{V}^{\,(k)} \big\} _{k \geq 1}$ such that for every
\beq
f_k(\bs{x}_1,\dots, \bs{x}_k)= g(\bs{x}_k) \cdot G_{k-1}(\bs{x}_1- \bs{x}_2,\dots, \bs{x}_{k-1}-\bs{x}_k)  \quad with \quad
(g,G_{k-1}) \in  \mc{S}\big( \R^{1 ,1} \big) \times \mc{S}\big( (\R^{1 ,1})^{k-1} \big) \, ,
\enq
it holds
\beq
\op{W}^{\,(k)}[f_k] \, = \, \bigg\{ \Int{\R^{1,1} }{} \! \! \dd \bs{x} \, g(x)   \bigg\} \cdot \op{V}^{\,(k-1)}[G_{k-1}] \qquad with \quad  k \geq 2\;.
\enq

Moreover, the Fourier transforms of the distributions $\op{V}^{\,(k-1)}$ satisfy
\beq
\mathrm{supp}\Big\{\msc{F}\big[\op{V}^{\,(k-1)}\big]\Big\} \subseteq \mc{E}^{+}_{k-1}\;,
\enq
with $\mc{E}^{+}_{k-1}$ as introduced in \eqref{definition domaine Fourier Energie positive}, \textit{i.e.} $\msc{F}\big[\op{V}^{\,(k-1)}\big][G]=0$ for any $G\in  \mc{S}\big( (\R^{1 ,1})^{k-1} \big)$
such that $\e{supp}[\msc{F}[G]] \cap \mc{E}^{+}_{k-1} = \emptyset$.
\item {\bf Hermiticity}:
For any $f \in \mc{S}\big( (\R^{1 ,1})^{k} \big)$, it holds
\beq
\ov{ \op{W}^{\,(k)}[ f ]  } \, = \, \op{W}^{\,(k)}\Big[\,  \ov{\iota \cdot f} \,  \Big]\;,
\enq
where $\ov{z}$ refers to the complex conjugate of $z\in \Cx$, $\ov{f}$ is obtained by the pointwise conjugation, and $\iota$ is the action of the inversion \eqref{ecriture action inversion et transposition}.

\item {\bf Local commutativity}: For any $f \in \mc{S}\big( (\R^{1 ,1})^{k} \big)$ such that
\beq
\e{supp}[f] \subset \Om_{s} \, = \, \Big\{  \big( \bs{x}_1,\dots, \bs{x}_k \big) \in (\R^{1 ,1})^{k} \, : \, \big( \bs{x}_{s} - \bs{x}_{s+1}\big)^2 \, < \, 0 \Big\} \;,
\label{definition support purement space like dans Minkowski a la k}
\enq
it holds
\beq
\op{W}^{\,(k)}[f] \, = \,  \op{W}^{\,(k)}[ \tau_{s}\cdot f ]
\enq
in which the  transpositions act as in \eqref{ecriture action inversion et transposition}.

\item {\bf Positive definiteness}: Let $\big\{ f^{(k)} \}_{k \geq 1}$ be any sequence of test functions with $f^{(k)} \in \mc{S}\big( (\R^{1,1})^k \big)$ for each integer $k$,
and such that there is only a finite number of non-zero terms.
Let
\beq
F\!_{p,q}\big(\bs{X}_p , \bs{Y}_q\big) \, = \,  \ov{    f^{(p)}\big(  \overleftarrow{\bs{X}}_p \big) } \cdot f^{(q)} \big( \bs{Y}_q\big) \, ,
\label{definition fct pour positive definiteness}
\enq
then it holds
\beq
\sul{p, q \in \mathbb{N} }{} \op{W}^{\, (p+q)}\big[ F_{p,q} \big] \, \geq \, 0  \;.
\enq
\item {\bf Cluster property}: For any space-like $\bs{v} \in \R^{1,1}$, \textit{i.e.} $\bs{v}^2<0$, and $f \in \mc{S}\big( (\R^{1,1})^p\big)$,  $g\in \mc{S}\big( (\R^{1,1})^q \big)$, it holds
\beq
 \op{W}^{\, (p+q)}\big[ G_{p,q}^{(\la \bs{v})} \big]  \,  \underset{\lambda \rightarrow \pm \infty}{\longrightarrow} \, \op{W}^{\, (p)}\big[ f \big] \,  \op{W}^{\, (q)}\big[ g \big]
\enq
where
\beq
G_{p,q}^{(\la \bs{v})}\big( \bs{X}_p, \bs{Y}_q \big)  \, = \, f\big(\bs{x}_1,\dots,\bs{x}_p) \,  g\big( \bs{y}_1+\la\bs{v},\dots,\bs{y}_q+\la\bs{v}\big) \, .
\label{definition Gpq}
\enq

	\end{itemize}
\end{defin}

We now recall Wightman's reconstruction theorem \cite{StreaterWightmanPCTStatisticsAllThat}, which is the central tool in re-constructing a quantum field theory from the knowledge of its correlation functions. For the needs of this work, we state it here in the case of 1+1 dimensional theories with only one scalar hermitian field.
\begin{theorem}
Let $\big\{ \op{W}^{\, (k)} \big\}_{k \geq 1}$ be a sequence of tempered distributions,    $\op{W}^{\, (k)}$ acting on $\mc{S}\big( (\R^{1,1})^k\big)$,
that satisfy the Wightman axioms. Then, there exist a separable Hilbert space $\mf{h}$, a continuous unitary representation of the orthochronous
Poincaré group $\mc{P}^+_{\ua}$ on $\mf{h}$, a unique one dimensional subspace $\mf{e}_{\e{vac}} = \big\{ \la \bs{v}_{\e{vac}}, \: \, \la \in \Cx \big\}$ of $\mf{h}$ invariant under the action of $\mc{P}^+_{\ua}$, and an unbounded operator valued tempered distribution $\bs{\Psi}$ such that
\begin{equation}
\op{W}^{(k)}[f] \, = \, \Big( \bs{v}_{\e{vac}} , \bs{\Psi}[f_1] \cdots \bs{\Psi}[f_k]  \bs{v}_{\e{vac}}\Big) \qquad \e{with} \qquad 
\left\{   \ba{ccc} f(\bs{X}_k) & = & \pl{a=1}{k} f_a(\bs{x}_a)  \vspace{2mm} \\
	 \bs{X}_k  &  =  &  \big( \bs{x}_1,\dots, \bs{x}_k\big) \in \left(\R^{1,1}\right)^k \ea \right. \;.
\end{equation}
The vacuum correlation functions are exactly the Wightman functions. Moreover, this vacuum is cyclic with respect to the action of the field $\bs{\Psi}$,
and any other quantum field theory with the same Wightman functions is unitarily equivalent to this one.
\end{theorem}

\subsection{The correlation function built from the bootstrap program}
\label{Subsection bootstrap program results}

We now provide the expressions for the multi-point correlation functions of the Sinh-Gordon 1+1 dimensional integrable quantum field theory which were obtained in \cite{KozSimonMultiPointCorrFctsSinhGordon}
within the so-called bootstrap program. In particular, we shall discuss which of their aspects are rigorous and which still remain to be established.
The description goes in several steps. We start by introducing the various building blocks of the
formula for the multi-point functions, and then describe the formula \textit{per se}.

\subsubsection{The $\op{S}$-matrix and form factors}
\label{Subsection building blocks}

The fundamental building block of the Sinh-Gordon model's correlation functions is its $\op{S}$-matrix, first introduced in \cite{GryanikVergelesSMatrixAndOtherStuffForSinhGordon}.
This matrix acts diagonally, with the below scalar factor describing the scattering at rapidity $\be$:
\beq
\op{S}(\beta)\, = \, \f{ \tanh\big[ \tfrac{1}{2}\beta - \i \pi  \mf{b}   \big]  }{ \tanh\big[ \tfrac{1}{2}\beta + \i \pi  \mf{b}   \big]   }
\, = \, \f{ \sinh(\be) - \i \sin[ 2 \pi \mf{b}]   }{ \sinh(\be) +  \i \sin[ 2 \pi \mf{b}]  }
\qquad \e{with} \qquad  \mf{b}\, = \,   \f{1}{2} \f{ g^2  }{ 8\pi + g^{2}  } \, \in \intff{0}{\tfrac{1}{2}}\;.
\label{definition matrice S}
\enq
This $\op{S}$-matrix satisfies the unitarity $\op{S}(\be)\op{S}(-\be)=1$ and crossing  $\op{S}(\be)=\op{S}(\i \pi-\be)$ relations.
Its poles are located at $2\i\pi \mf{b} + 2\i\pi \mathbb{Z}$. In particular, it has no poles in the physical strip
\begin{equation}
\msc{S} = \left\{ 0<\Im(\be)<\pi \right\} \;,
\end{equation}
which translates the absence of bound states in this model. The $\op{S}$-matrix is then used to construct the so-called two-body form factor $\op{F}$ which
can be expressed in closed form in terms of the Barnes $G$- and Euler $\Ga$-functions:
\beq
\op{F}(\be) \, = \, \f{1}{  \Ga\Big( 1+\mf{z}, -\mf{z} \Big)}   G\left( \ba{cccc} 1-\mf{b} - \mf{z} \, ,  &   2 - \mf{b} + \mf{z} \,,  &   1- \hat{\mf{b}} - \mf{z}   \, ,  &   2 - \hat{\mf{b}}  + \mf{z}    \\
  \mf{b} - \mf{z}   \, ,  &   1+  \mf{b} + \mf{z}   \, ,  &   \hat{\mf{b}} - \mf{z}   \, ,  &   1+  \hat{\mf{b}} + \mf{z}    \ea \right)  \quad \e{with} \quad  \mf{z} \, = \, \f{\i \be }{2 \pi } \;, \quad \hat{\mf{b}} = \f{1}{2}-\mf{b} \;.
\label{expression twpo body scattering via Barnes}
\enq
Above, we made use of hypergeometric-like  product conventions
\beq
  \Ga\left(\ba{c} a_1,\dots, a_n \\ b_1,\dots, b_{\ell} \ea \right)  \, = \, \f{ \pl{k=1}{n} \Ga(a_k)  }{  \pl{k=1}{\ell} \Ga(b_k) } \qquad  \e{and}  \qquad
  G\left(\ba{c} a_1,\dots, a_n \\ b_1,\dots, b_{\ell} \ea \right)  \, = \, \f{ \pl{k=1}{n} G(a_k)  }{  \pl{k=1}{\ell} G(b_k) } \;.
\label{ecriture convention produit ratios fcts Gamma}
\enq
$\op{F}$ is the first building block of so-called form factors. The second building block is provided by the so-called $\mc{K}$-transform \cite{BrazhnikovLukyanovFreeFieldRepMassiveFFIntegrable,BabujianKarowskiBreatherFFSineGordon}
which takes as input a function $p_n$ on $\Cx^n\times \{0,1\}^n$ depending on $n$ complex variables
\beq
\bs{\be}_n =\big( \be_1, \dots, \be_n \big) \in \Cx^n
\label{ecriture vecteur n composantes}
\enq
and $n$ discrete variables $\bs{\ell}_n \in \{0,1\}^n$. As an output, the $\mc{K}$-transform generates the below function of $n$ complex variables:
\beq
  \mc{K}_{n}\big[ p_n  \big]\big( \bs{\be}_{n} \big) \, =  \hspace{-2mm} \sul{ \bs{\ell}_n \in \{0,1\}^n }{} (-1)^{\ov{\bs{\ell}}_n}
\pl{k<s}{n} \bigg\{ 1 \, - \, \i \f{ \ell_{ks} \cdot \sin[2\pi \mf{b} ] }{ \sinh(\be_{ks})  }  \bigg\} \cdot p_n\big(\bs{\be}_n\mid \bs{\ell}_n\big)  \;,
\label{definition K transformee fct p}
\enq
in which $\ov{\bs{\ell}}_n \, = \, \sul{a=1}{n} \ell_k$.

Within the bootstrap program setting, the operators, \textit{aka} quantum fields, of the Sinh-Gordon quantum field theory are built from a specific choice of $p_n$
functions, namely those solving the equations
\begin{itemize}

 \item[a)]  $\bs{\be}_n\mapsto p_{n}\big(\bs{\be}_n\mid \bs{\ell}_n\big)$ is a collection of $2\i\pi$ periodic holomorphic functions on $\Cx$,
 $p_{n}\big(\bs{\be}_n + 2\i\pi \bs{e}_1\mid \bs{\ell}_n\big)\, = \,  p_{n}\big(\bs{\be}_n \mid \bs{\ell}_n\big)$ with $\bs{e}_1$ the first basis vector, that are symmetric with respect to simultaneous permutations of the coordinates of the vector $\bs{\be}_n$ and $\bs{\ell}_n$;

  \item[b)] $p_{n}\big(\be_2+\i\pi, \bs{\be}_n^{\prime}\mid \bs{\ell}_n\big)\, = \,g(\ell_1,\ell_2) p_{n-2}\big(\bs{\be}_n^{\prime\prime}\mid \bs{\ell}_n^{\prime\prime}\big)
\, + \, h(\ell_1,\ell_2\mid \bs{\be}_n^{\prime})$ for some function $h$ not depending on the remaining set of variables $\bs{\ell}_n^{\prime\prime} = (\ell_3,\dots,\ell_n)$, with
\beq
g(0,1)\, = \,  g(1,0) \, = \, \f{ -1 }{ \sin (2\pi \mf{b} ) \,  \op{F}(\i\pi)  } \, ,
\enq
and where $\bs{\be}_n^{\prime}= (\be_2,\dots,\be_n)$ while  $\bs{\be}_n^{\prime\prime} = (\be_3,\dots,\be_n)$;

 \item[c)]  $ p_{n}\big( \bs{\be}_n + \theta \, \ov{\bs{e}}_n \mid \bs{\ell}_n \big)  \; = \; \ex{\theta \op{s} } \cdot p_{n}\big(\bs{\be}_n\mid \bs{\ell}_n\big)$ with $\ov{\bs{e}}_n=(1,\dots,1)\in \Cx^n$;

 \item[d)]  $ \big| p_{n}\big(\bs{\be}_n\mid \bs{\ell}_n\big)  \big| \, \leq \, C \cdot  {\displaystyle \pl{a=1}{n} } \big| \cosh\big[ \Re(\be_a) \big]\big|^{ \mf{w} }$.

\end{itemize}
The system of equations is labelled by two numbers characteristic of the class of operators one wants to build:
\begin{itemize}
	\item[$\bullet$] the quantity $\op{s}$ called the spin;
	\item[$\bullet$] the real number  $\mf{w}$ called growth index.
\end{itemize}
We do stress that these two quantities, $\op{s}$ and $\mf{w}$ are $n$-independent.

The system has numerous solutions, although their full classification is still an open problem. We shall henceforth label given solutions with a superscript $\a$
referring to the operator $\op{O}_{\a}$ that they represent,  \textit{viz}.
$p_{n}^{(\a)}\big(\bs{\be}_n\mid \bs{\ell}_n\big)$. Likewise, the spin or growth index of a given operator will be denoted $\op{s}_{\a}$, $\mf{w}_{\a}$.
Below, we only provide the expression for the $p_n$ function associated with the field operator $\bs{\Psi}$
\beq
	\label{fonction p champ}
	p_n^{(\bs{\Psi})}\big(\bs{\be}_n\mid \bs{\ell}_n\big) \, =  \,
\left( \frac{-\i}{\sqrt{2\sin(\pi\mf{b})}} \exp\left\{ \frac{1}{2\pi} \int_0^{2\pi\mf{b}} \hspace{-1mm} \frac{t \dd t}{\sin t}\right\}\right)^n \cdot \frac{2\pi\i\mf{b}}{g} \cdot \sul{p=1}{n} (-1)^{\ell_p} \; .
\enq
We stress that the spin and the growth index in the case of the field $\bs{\Psi}$ $p$-function read
\beq
   \op{s}_{ \bs{\Psi} } \, = \, \mf{w}_{ \bs{\Psi} } \, = \, 0 \;.
\enq
There are many other examples of $p_n$ functions representing other operators of interest to the theory, such as the exponential of the field, its descendents, or the energy-momentum tensor.
see \textit{e.g.} \cite{BabujianKarowskiBreatherFFSineGordon,FeiginLashkevichFreeFieldApproachToDescendents}.

We are finally in position to introduce the fundamental building block of the expressions for the multi-point functions. Given a solution $p_n^{(\a)}\big(\bs{\be}_n\mid \bs{\ell}_n\big)$
to the above mentioned system of equations, the associated form factor takes the form \cite{BrazhnikovLukyanovFreeFieldRepMassiveFFIntegrable,BabujianKarowskiBreatherFFSineGordon}:
\beq
\mc{F}^{(\a)} (\bs{\be}_n) \, = \, \pl{a<b}{n} \op{F}\big( \be_{ab} \big)  \cdot \mc{K}_{n}\big[ p_n^{(\a)}\big]\big( \bs{\be}_{n} \big)  \qquad \e{with} \qquad \be_{ab} = \be_a - \be_b \;,
\label{solution eqns bootstrap via K transformee}
\enq
where $\op{F}$ is as given in \eqref{expression twpo body scattering via Barnes}, $\mc{K}_n$ is as introduced through \eqref{definition K transformee fct p}
while $p_n^{(\a)}$ is the solution to $a)-d)$ associated with the operator indexed by $\a$.

We stress that  $\mc{F}^{(\a)}$ does depend on the dimensionality $n$ of its argument $\bs{\be}_n \in \Cx^n$, however to keep notations simple in what will follow,
we choose to keep this dependence on the dimension implicit. This does not lead to ambiguity since it should simply be read-off from the dimensionality of the evaluated vector.

Given an operator with index $\a$, we shall denote by $\a^{\dagger}$ the index of its adjoint operator (which may or may not coincide with $\a$).
The form factors of the adjoint are built exactly as in \eqref{solution eqns bootstrap via K transformee} with the function
\beq
p_n^{(\a^{\dagger})}( \bs{\be}_n \mid \bs{\ell}_n)  \, = \, \ov{ p_n^{(\a)}\Big( \ov{\bs{\be}_n} + \i\pi \ov{\bs{e}} \mid \overleftarrow{\bs{\ell}}_n \Big) } \;.
\label{definition fct p pour op adjoint}
\enq
Thus, whether the operator is self-adjoint or not is thus a property of its $p$-function. In particular, the field operator, as follows from \eqref{fonction p champ}, is self-adjoint.
\eqref{definition fct p pour op adjoint} implies the relation
\beq
\mc{F}^{(\a^{\dagger})} (\bs{\be}_n) \, = \, \ov{ \mc{F}^{(\a)}\Big( \,  \overleftarrow{\ov{\bs{\be}_n}} + \i\pi \ov{\bs{e}}  \Big) } \; .
\label{ecriture equation correspondance entre FF et celui de adjoint}
\enq

 We now list all of the properties of the form factors playing a role in the proofs to come:
\begin{itemize}
\item  {\bf Exchange property}:
\beq
\mc{F}^{(\a)}(\bs{\be}_n) \; = \; \op{S}(  \be_{a \, a+1}) \cdot \mc{F}^{(\a)}( \tau_{a}\cdot\bs{\be}_n)  \;.
\label{exchange form factor}
\enq
 \item {\bf Pole structure}: For $\{ \be_s \}_{1, \not=a }^{n}$, the maps $\be_a\mapsto \mc{F}^{(\a)}(\bs{\be}_n)$
are meromorphic functions on $\Cx$. For pairwise distinct coordinates, \textit{i.e.} $\be_b\not= \be_c$ for any $b\not=c \in \intn{1}{n}\setminus \{a\}$,
there are only simple poles. These are located in
\beq
\Big\{ \be_{b} + \i \pi   +    2 \i \pi \mathbb{Z} \Big\}_{  \substack{ b=1 \\  \not=a}  }^{n}  \quad \e{and} \quad
\Big\{ \be_{b} - 2 \i \pi\mf{b}  +   2 \i \pi \mathbb{N}^{*} \Big\}_{   b=1   }^{a-1}   \quad \e{and} \quad
\Big\{ \be_{b} - 2 \i \pi \hat{\mf{b}}  +   2 \i \pi \mathbb{N}^{*} \Big\}_{   b=1   }^{a-1}
\enq
as well as in
\beq
\Big\{ \be_{b} + 2 \i \pi \mf{b}  +   2 \i \pi \mathbb{N}^{*} \Big\}_{   b=a+1   }^{n}   \quad \e{and} \quad
\Big\{ \be_{b} +2 \i \pi  \hat{\mf{b}}  +   2 \i \pi \mathbb{N}^{*} \Big\}_{   b= a+1   }^{n}    \;.
\enq
\item {\bf Lorentz boost}: $ \mc{F}^{(\a)}( \bs{\be}_n + \theta\,  \ov{\bs{e}}_n )  \; = \; \ex{\theta \op{s}_{\a} } \cdot \mc{F}^{(\a)}(\bs{\be}_n)$ with
$\ov{\bs{e}}_n=(1,\dots,1)$ and $\op{s}_{\a}$ being the spin of the operator indexed by $\a$.
\item {\bf Growth at infinity}: For every $1 \leq k \leq n$, $\be_k \mapsto \mc{F}^{(\a)}(\bs{\be}_n)$ is bounded at infinity by $ C\cdot \cosh\big( \mf{w}_{\a} \Re(\be_k) \big)$, with
$\mf{w}_{\a}$ being the growth index of the operator labelled by $\a$.
\end{itemize}
We stress that the properties of the form factors are completely transparent to the operator $\a$-label. Some, weak influence of the operator manifests itself on the level
of the Lorentz boost and growth at infinity property.

As we shall demonstrate in the following, the Wightman axioms hold independently on the value taken by the growth index $\mf{w}_{\a}$.

 The elementary exchange property \eqref{exchange form factor} of the form factors allows one to define $\op{S}$-matrices subordinate to the
exchange of two complex valued vectors $\bs{A}=(a_1,\dots,a_n)$ and $\bs{B}=(b_1,\dots,b_m)$ of arbitrary sizes $n$, $m$. For that purpose, we introduce the
concatenation of two vectors
\beq
\bs{A}\cup\bs{B} = (a_1,\dots,a_n,b_1,\dots,b_m) \;.
\enq
Then,
\beq
	\mc{F}(\bs{A} \cup \bs{B}) = \op{S}(\bs{A} \cup \bs{B} \, | \, \bs{B} \cup \bs{A}) \cdot \mc{F}(\bs{B} \cup \bs{A}) \;.
\label{ecriture S propriete FF}
\enq
In fact, these can be computed explicitly and read
\beq
\op{S}(\bs{A} \cup \bs{B} \, | \, \bs{B} \cup \bs{A}) = \pl{j=1 }{n}\pl{\ell=1 }{m} \op{S}(a_j - b_{\ell})   \;.
\label{ecriture expression matrice S permutation}
\enq
 It will also be of use to introduce the operation of reversing a vector's coordinates: given $\bs{A}=(a_1,\dots,a_n)$, we set $\overleftarrow{\bs{A}} = (a_n,\dots,a_1)$.

\subsubsection{The multi-point functions}
\label{subsubsection multipointFcts Reminder}
The work \cite{KozSimonMultiPointCorrFctsSinhGordon} computed, on rigorous grounds, the multi-point correlation functions of appropriately regularised field operators
in the Sinh-Gordon model from the bootstrap axiom formulation of the theory. These were shown to be well-defined tempered distributions.
We first describe this construction and then explain how, conjecturally, these will yield the \textit{per se} expression for the multi-point correlation functions
in the Sinh-Gordon quantum field theory.\\
First of all, given an integer $\bs{n}\; = \, \big( n_{21}, n_{31}, n_{32}, n_{41}, \dots , n_{k k-1} \big) \in  \mathbb{N}^{ \f{k(k-1)}{2} }$, and $G\in \mc{S}\big( (\R^{1,1})^k \big)$,
the momentum transform of $G$ reads
\beq
\mc{R}[ G] \big(   \bs{\ga}    \big)   \; = \hspace{-2mm} \Int{ \big( \R^{1,1} \big)^{k} }{}  \hspace{-2mm} \dd^k \bs{X}
\cdot  G\big(\bs{X}_k \big) \cdot \pl{b>a}{k} \ex{ \i \ov{\bs{p}}(  \bs{\ga}^{(ba)} ) * \bs{x}_{ba}  } \;, \qquad \text{with} \qquad \bs{x}_{ba} = \bs{x}_b - \bs{x}_a \;.
\label{definition TF space-time de la fct test}
\enq
In this expression,
\beq
\bs{\ga} \, = \, \big( \bs{\ga}^{(21)} ,\bs{\ga}^{(31)} ,\bs{\ga}^{(32)} , \dots , \bs{\ga}^{(kk-1)} \big) \in \Cx^{ |\bs{n}| }  \qquad \e{with} \qquad
|\bs{n}| \; = \; \sul{b>a}{k} n_{ba}
\label{definition vecteur gamma tot}
\enq
while $\bs{\ga}^{(ba)} \in  \Cx^{n_{ba}}$. Furthermore, we agree upon
\beq
\ov{\bs{p}}(  \bs{\be}_m ) \, = \, \sul{a=1}{m} \bs{p}(\be_a) \qquad \e{with} \qquad
\bs{p}(\be)\, = \, m \, \big( \cosh(\be), \sinh(\be) \big)
\label{definition 2 impulsion}
\enq
 being the relativistic $2$-momentum of a particle of mass $m$.

Pick solutions $p_n^{(\alpha_1)},\dots, p_n^{(\alpha_k)}$ to the equations $a)-d)$ associated with operator labels $\a_1, \dots, \a_k$ and consider
the associated form factors $\mc{F}^{(\alpha_1)},\dots, \mc{F}^{(\alpha_k)}$ as defined through \eqref{solution eqns bootstrap via K transformee}.
Given $\bs{n}  \in  \mathbb{N}^{ \f{k(k-1)}{2} }$ and agreeing upon $\bs{\a}_{k}=(\a_1,\dots, \a_k)$ this allows one to introduce the integral transform
\beq
\mc{I}_{ \bs{\a}_{k} }^{ (\bs{n} ) } \big[ G \big]  \; = \;
\lim_{ \bs{\veps}_k \tend \bs{0}^+} \Int{  \R^{ |\bs{n}| }   }{}  \f{ \dd^{ |\bs{n}| } \bs{\ga}  }{  \bs{n}! (2\pi)^{|\bs{n}|} } \;
   \Big( \mc{S}\cdot \mc{R}[ G  ] \cdot  \mc{F}_{\bs{\a}_{k};\bs{\veps}_{k} } \Big) \big(   \bs{\ga}     \big)    \; .
\label{definition sommant base dans serie int mult pour fct 2 pts}
\enq
First of all,  the $\bs{\veps}_k\, = \, \big(\veps_1,\dots, \veps_{k}\big) \tend \bs{0}^+$ limits are to be understood as  $\veps_a  \tend 0^+$ for $a=1,\dots,k$.
These can be taken in any order and correspond to computing the $+$ boundary values of multi-dimensional Cauchy transforms.
Its definition involves two additional building blocks. $\mc{S}$ is built from elementary vector exchange products of $\op{S}$-matrices
\beq
\mc{S}\big(   \bs{\ga}     \big) \; = \;
\pl{  \substack{ v >p \\ p \geq 3 } }{ k }  \pl{u>s}{p-1}
\op{S}\big(  \bs{\ga}^{(vu)}  \cup  \bs{\ga}^{(ps)}   \mid   \bs{\ga}^{(ps)}   \cup \bs{\ga}^{(vu)}   \big)  \;.
\label{definition facteur de diffusion complet k pts}
\enq
The second factor is built from a product of form factors:
\beq
\mc{F}_{\bs{\a}_{k};\bs{\veps}_{k} }\big(   \bs{\ga}     \big) \; = \;
\pl{p=1}{k} \mc{F}^{(\alpha_p)}\Big( \, \overleftarrow{ \bs{\ga}^{(pp-1)} } \cup \cdots \cup \overleftarrow{ \bs{\ga}^{(p1)} }  + \i \pi \ov{\bs{e}}_{\veps_{p}} ,
\bs{\ga}^{(kp)}  \cup \cdots \cup \bs{\ga}^{(p+1p)}   \, \Big) \;,
\label{definition F tot eps}
\enq
with  $\bs{\veps}_{k}\, =\, \big(\veps_1,\dots,\veps_{k}\big)$, $\overleftarrow{\bs{X}}_k$ as defined in \eqref{definition action gpe perm sur vect et vect renverse}
and where  $\ov{\bs{e}}_{\veps}$ takes the form
\beq
\ov{\bs{e}}_{\veps} \; = \; \Big( 1 \, - \, \f{\veps}{\pi} \Big) \cdot \big( 1,\dots, 1 \big) \: .
\label{definition ov bs eps avec indice regulateur}
\enq
Note that the dimensionality of the vector $\ov{\bs{e}}_{\veps}$ changes from one form factor to another and is undercurrent  by the dimensionality of the vector to which it is added.

The well-definiteness of \eqref{definition sommant base dans serie int mult pour fct 2 pts} was established in \cite{KozSimonMultiPointCorrFctsSinhGordon}.

%
%
%
%
%
%
%
%
%
%
%
%
%
%
%
%
%
%
%

\begin{prop} \cite{KozSimonMultiPointCorrFctsSinhGordon}
\label{Proposition In sont bien definis}
	For any $\bs{n}  \in  \mathbb{N}^{ \f{k(k-1)}{2} }$, $\mc{I}_{ \bs{\a}_{k} }^{ (\bs{n} ) }$ is a well-defined tempered distribution. In particular, for any $\bs{\veps}_{k}$
small enough and such that $\veps_a>0$, the integrals are strongly convergent.
\end{prop}
We are finally in position to introduce the $\bs{r} \in \mathbb{N}^{k-1}$-truncated correlation functions.
Given a $k-1$ dimensional integer valued vector $\bs{r}\, =\, \big( r_1,\dots, r_{k-1} \big) \in \mathbb{N}^{k-1}$, one associates with it a subset of $ \mathbb{N}^{ \f{k(k-1)}{2} } $:
\beq
\mc{N}_{\bs{r}}  \; = \; \bigg\{ \bs{n} \; = \, \big( n_{21}, n_{31}, n_{32}, n_{41}, \dots , n_{k k-1} \big) \; : \; \sul{u=p+1}{k} \sul{s=1}{p} n_{us} \, = \, r_p \quad p=1,\dots, k-1 \bigg\} \;.
\label{definition ensemble sommation restreint pour troncation fct a k pts}
\enq
\begin{defin}
\label{Definition  rep serie int mult tronquee pour fct multipts tronquee}
Given $G \in  \mc{S}\big( (\R^{1,1})^k \big) $, the $\bs{r}$-truncated correlation functions corresponds to the below sum
\beq
\label{def trunc corr}
\mc{W}^{ \, (\bs{r}) }_{ \bs{\a}_k}[G]  \; = \;
\sul{ \bs{n} \in \mc{N}_{\bs{r}} }{}    \mc{I}_{ \bs{\a}_{k} }^{ (\bs{n} ) }\big[ G\big]     \; .
\enq
Above one sums over integer valued vectors $\bs{n}$  belonging  to $\mc{N}_{\bs{r}} \subset  \mathbb{N}^{ \f{k(k-1)}{2} } $ defined in \eqref{definition ensemble sommation restreint pour troncation fct a k pts}.

\end{defin}

By virtue of Proposition \ref{Proposition In sont bien definis}, being defined as finite linear combinations of the $\mc{I}_{ \bs{\a}_{k} }^{ (\bs{n} )}$, the $\mc{W}^{ (\bs{r}) }_{ \bs{\a}_k }$ are
well-defined tempered distributions. On this level, the role played by multi-index truncation parameter $\bs{r}$ becomes apparent in that it imposes dealing only
with \textit{finite} sums. Thus well-definiteness of the expression follows from the the one of the individual building blocks $ \mc{I}_{ \bs{\a}_{k} }^{ (\bs{n} ) }$.
In the bootstrap formalism, $\mc{W}^{ \, (\bs{r}) }_{ \bs{\a}_k}$ corresponds to a correlation function of certain "regularised" field operators
which only connect a finite portion of the Fock Hilbert space. We refer to \cite{KozSimonMultiPointCorrFctsSinhGordon} for more details.
The \textit{per se} correlation functions of the Sinh-Gordon quantum field theory are only obtained by summing up the contributions of each individual $\bs{r}$-truncated
terms $\mc{W}^{ \, (\bs{r}) }_{ \bs{\a}_k}$, \textit{viz}.
\beq
\mc{W}_{ \bs{\a}_k}[G] \, = \, \sul{ \bs{r} \in \mathbb{N}^{k-1} }{} \mc{W}^{ \, (\bs{r}) }_{ \bs{\a}_k}[G] \;.
\enq
The convergence problem of such series is, in general, a hard open problem that will \textit{not} be addressed here.
So far, convergence was only proven in the case of two-point functions \cite{KozConvergenceFFSeriesSinhGordon2ptFcts},
\textit{viz}. for $k=2$, and for test functions $G$ having mutually purely space-like support (\textit{i.e.} $G(\bs{x},\bs{y}) \not=0$ only if $(\bs{x}-\bs{y})^2<0$)

This work established that the collection of tempered distributions $\big\{ \mc{W}_{ \bs{\a}_k } \big\}_{\bs{\a}_k}$ defined above
fulfills the system of Wightman axioms, under their well-definiteness conjecture which we state below:
\begin{conj}
\label{Conjecture convergence}
For any  $k \in \mathbb{N}$, any $G \in  \mc{S}\big( (\R^{1,1})^k \big) $ and any labels $\a_1, \dots, \a_k$ of solutions to the system a)-d),
the $k$-point correlation function
\beq
\label{full corr func}
\mc{W}_{ \bs{\a}_k}\big[ G \big] = \sul{ \bs{n} \in \mathbb{N}^{\frac{k(k-1)}{2} }}{}   \mc{I}_{ \bs{\a}_{k} }^{ (\bs{n} )}\big[ G\big] \;,
\enq
is given by an absolutely converging series and defines a tempered distribution.
\end{conj}
Below, given $g_1,\dots,g_k\in \mc{S}(\R^{1,1})$,  we shall often use the short-hand notation
\beq
G_k\big(\bs{X}_k \big) \, = \,  \pl{s=1}{k} g_s\big( \bs{x}_s \big) \;.
\label{definition fonction Gk}
\enq

\section{Proof of the Wightman Axioms}
\label{Section Wightman axioms verification}

The present section and those that will follow will be dedicated to proving that the $k$-point correlation functions $\mc{W}_{\bs{\a}_k}$ \eqref{full corr func}
do satisfy the Wightman axioms
under the well-definiteness Conjecture \ref{Conjecture convergence}. For that, one considers a choice of $k$ initial $p^{(\a_a)}$ functions, $a=1,\dots, k$,
leading to form factors $\mc{F}^{(\alpha_1)},\dots,\mc{F}^{(\alpha_k)}$
according to the construction \eqref{solution eqns bootstrap via K transformee} and thus satisfying the corresponding properties (exchange, pole structure, Lorentz boost, growth at infinity).
The spins $\op{s}_{\alpha_1},\dots,\op{s}_{\alpha_k}$ and the growth indices $\mf{w}_{\alpha_1},\dots,\mf{w}_{\alpha_k}$ will thus be taken generic, what allows us to
show that the Wightman axioms are satisfied by a larger class of correlation functions than those of the fields alone.

\subsection{The covariance axiom}
We start with the first axiom.

\begin{prop}
For any operator indices $\bs{\a}_k \, =\, \big( \a_1,\dots, \a_k \big)$, any $G \in \mc{S}\big( (\R^{1,1})^k \big)$ and  $(\theta , \bs{v}) \in  \R\times \R^{1,1}$, it holds
\beq
\mc{I}_{\bs{\a}_k}^{(\bs{n})} \big[ \op{U}_{\th, \bs{v}} \cdot G \big] \, = \, \pl{a=1}{k} \Big\{ \ex{ -\th \op{s}_{\a_k} } \Big\} \cdot \mc{I}_{\bs{\a}_k}^{(\bs{n})} \big[  G \big]
\qquad with \qquad
\bs{n} \in \mathbb{N}^{k\f{k-1}{2}} \;.
\enq
$\mc{I}_{\bs{\a}_k}^{(\bs{n})}$ has been introduced in \eqref{definition sommant base dans serie int mult pour fct 2 pts}, while the action of the orthochronous Poincaré
group on $\mc{S}\big( (\R^{1,1})^k \big)$ is defined in \eqref{fonction apres action Groupe Poincaré}.

Moreover, provided Conjecture \ref{Conjecture convergence} holds, the covariance property is inherited by the per se correlation function
\beq
\mc{W}_{ \bs{\a}_k}\big[ \op{U}_{\th, \bs{v}} \cdot G \big]  \,  =  \,  \pl{a=1}{k} \Big\{ \ex{ \th \op{s}_{\a_k} } \Big\} \cdot  \mc{W}_{ \bs{\a}_k}\big[ G \big] \;.
\enq
\end{prop}
\Proof
Starting from \eqref{definition TF space-time de la fct test}, one gets
\beq
\mc{R}\big[ \op{U}_{\th, \bs{v}} \cdot G \big]  \big(   \bs{\ga}    \big) \,  =  \hspace{-2mm} \Int{ \big( \R^{1,1} \big)^{k} }{}  \hspace{-2mm} \dd^k \bs{X}
\cdot  G \big(\La_{\th}^{-1}\!\cdot(\bs{x}_1+\bs{v}),\dots, \La_{\th}^{-1}\!\cdot(\bs{x}_k + \bs{v}) \big) \cdot \pl{b>a}{k} \ex{ \i \ov{\bs{p}}(  \bs{\ga}^{(ba)} ) * \bs{x}_{ba}  } \;.
\enq
At this stage, one proceeds with the change of variable whose Jacobian is 1:
\beq
\bs{x}_s \, = \, \La_{\th} \cdot \bs{y}_s - \bs{v}  \qquad \e{for} \qquad  s \in \intn{1}{k}
\enq
leading to
\beq
\mc{R}\big[ \op{U}_{\th, \bs{v}} \cdot G \big]  \big(   \bs{\ga}    \big) \, =  \hspace{-2mm} \Int{ \big( \R^{1,1} \big)^{k} }{}  \hspace{-2mm} \dd^k \bs{Y}
	\cdot  G \big(\bs{Y}_k \big) \cdot \pl{b>a}{k} \ex{ \i \ov{\bs{p}}(  \bs{\ga}^{(ba)} ) *  (\La_{\th} \cdot \bs{y}_{ba})    } \;.
\enq
Now using the invariance of the Minkowski scalar product under the action of the $O(1,1)$ group $(\La_{\th}\cdot \bs{x} )* (\La_{\th}\cdot \bs{y} )= \bs{x}* \bs{y}$ and the fact that
$\La_{\th}^{-1} = \La_{-\th}$, one gets
\beq
\ov{\bs{p}}(  \, \bs{\ga}^{(ba)} ) *  \big(\La_{\th}  \cdot \bs{y}_{ba}\big)  \, = \,  \big( \La_{\th}^{-1}\cdot \ov{\bs{p}}( \, \bs{\ga}^{(ba)} ) \big) *   \cdot \bs{y}_{ba}
\, = \, \ov{\bs{p}}( \, \bs{\ga}^{(ba)} +  \th \, \ov{\bs{e}}  \,  ) *     \bs{y}_{ba} \;.
\enq
It then remains to translate the integration variables  $\bs{\ga} \mapsto \bs{\ga} -  \th \ov{\bs{e}}$ and observe that
\beq
\mc{S}\big(   \bs{\ga} -  \th \, \ov{\bs{e}}  \,  \big) \; = \;\mc{S}\big(   \bs{\ga}     \big)
\enq
since its explicit expression only depends on the difference of $\bs{\ga}$ entries \textit{c.f.}
\eqref{ecriture expression matrice S permutation} and \eqref{definition facteur de diffusion complet k pts}, while
\bem
\mc{F}_{\bs{\a}_{k};\bs{\veps}_{k} }\big(   \bs{\ga}  -  \th \ov{\bs{e}}   \big) \; = \;
\pl{p=1}{k} \mc{F}^{(\alpha_p)}\Big( \, \overleftarrow{ \bs{\ga}^{(pp-1)} } \cup \cdots \cup \overleftarrow{ \bs{\ga}^{(p1)} }  + \i \pi \ov{\bs{e}}_{\veps_{p}} -  \th \ov{\bs{e}}  ,
\bs{\ga}^{(kp)}  \cup \cdots \cup \bs{\ga}^{(p+1p)}   -\th \ov{\bs{e}}  \, \Big) \\
\; = \;  \pl{a=1}{k} \Big\{ \ex{ -\th \op{s}_{\a_k} } \Big\} \cdot \mc{F}_{\bs{\a}_{k};\bs{\veps}_{k} }\big(   \bs{\ga}   \big)  \;,
\end{multline}
due to the Lorentz  boost property of the form factors (axiom ${\bf IV)}$). This entails the claim. \qed

\subsection{Spectral condition}

We now establish the spectral condition axiom.
For that purpose, we introduce the linear map
\beq
\mc{L}: \mc{S}\big( (\R^{1,1})^k \big) \tend \mc{S}\big( (\R^{1,1})^{k-1} \big)
\enq
with
\beq
\mc{L}[G](\bs{Y}_{k-1})\,= \, \Int{ \R^{1,1} }{} \hspace{-2mm} \dd \bs{y}_k \; G\Big( \sul{s=1}{k} \bs{y}_s , \dots, \bs{y}_k \Big) \:.
\label{definition linear map L}
\enq
Further, we set
\beq
\op{P}_{\ell}\big( \bs{\ga}  \big) \; = \; \sul{ b = \ell + 1 }{k}  \sul{ a = 1 }{ \ell } \ov{\bs{p}}(  \bs{\ga}^{(ba)} ) \; = \;
\Big(  \op{P}_{\ell}^{(0)}\big( \bs{\ga}  \big) ,  \op{P}_{\ell}^{(1)}\big( \bs{\ga}  \big)  \Big)   \qquad \e{for} \qquad \ell \in \intn{1}{ k-1} \;.
\label{definition impulsion reduite}
\enq
This allows us to introduce a reduced momentum transform on $\mc{S}\big( (\R^{1,1})^{k-1} \big)$
\beq
\widehat{\mc{R}}[H] \big(   \bs{\ga}    \big)   \; = \hspace{-2mm} \Int{ \big( \R^{1,1} \big)^{k-1} }{}  \hspace{-4mm} \dd^{k-1} \bs{Y}
\;H  \big( \bs{Y}_{k-1} \big)  \pl{ \ell = 1 }{ k - 1 } \Big\{ \ex{ -  \i  \bs{y}_{\ell}  * \op{P}_{\ell}( \bs{\ga}  ) } \Big\} \;.
\label{definition TF space-time de la fct test differentielle}
\enq
We then introduce the tempered distribution on $\mc{S}\big( (\R^{1,1})^{k-1} \big)$
\beq
\mc{J}^{ (\bs{n}) }_{\bs{\a}_k } [H] \, = \,  \lim_{\bs{\veps}_k \tend 0^+} \Int{  \R^{ |\bs{n}| }   }{} \hspace{-1mm} \f{ \dd^{ |\bs{n}| } \bs{\ga}  }{ \bs{n}! (2\pi)^{|\bs{n}|}} \,
\Big( \mc{S} \cdot \widehat{\mc{R}}[H] \cdot \mc{F}_{\bs{\a}_k;\bs{\veps}_{k} } \Big) \big(   \bs{\ga}     \big) \; .
\label{definition distrib diff variables}
\enq
The fact that $\mc{J}^{ (\bs{n}) }_{\bs{\a}_k }$ is well-defined will be established below.

\begin{prop}
\label{prop spectral conditions}
For every $\bs{r} \in \mathbb{N}^{ k-1 }$, the distribution
\beq
\mc{V}_{\bs{\a}_k}^{(\bs{r})}\big[ H \big] \, = \,  \sul{ \bs{n} \in \mc{N}_{\bs{r}} }{}    \mc{J}^{ (\bs{n}) }_{\bs{\a}_k}\big[ H\big]
\label{new distrib}
\enq
on $\mc{S}\big( (\R^{1,1})^{k-1} \big)$ is well-defined and such that it holds:
\beq
\mc{W}_{\bs{\a}_k}^{(\bs{r})}\big[ G \big] \, = \,  \mc{V}_{\bs{\a}_k}^{(\bs{r})}\big[ \mc{L}[G] \big]  \qquad with \qquad  G \in \mc{S}\big( (\R^{1,1})^{k} \big) \;.
\enq
Moreover, the Fourier transform of $\mc{V}_{\bs{\a}_k}^{(\bs{r})}$ satisfies the support condition
\beq
\mathrm{supp}\Big(\msc{F}\big[ \mc{V}_{\bs{\a}_k}^{(\bs{r})} \big] \Big) \subseteq \mc{E}^{+}_{k-1} \;,
\enq
with $\mc{E}^{+}_{k-1}$ as given in \eqref{definition domaine Fourier Energie positive}.

In particular, provided Conjecture \ref{Conjecture convergence} holds, the spectral condition property is inherited by the per se correlation function
\beq
\mc{W}_{ \bs{\a}_k}\big[   G \big]  \,  =  \,   \mc{V}_{ \bs{\a}_k}\big[ \mc{L}[G] \big]
\qquad with \qquad   \mc{V}_{ \bs{\a}_k} \, = \hspace{-3mm}  \sul{ \bs{n} \in \mathbb{N}^{  \f{1}{2} k(k-1)  } }{} \hspace{-3mm}   \mc{J}^{ (\bs{n}) }_{\bs{\a}_k}
  \;.
\enq
\end{prop}

\Proof

We first focus on the momentum transform \eqref{definition TF space-time de la fct test}  $\mc{R}[G]$ in which we implement the change of variables :
\beq
\label{cdv position}
\bs{x}_s  \, = \, \sul{p=s}{k} \bs{y}_p \quad \e{for} \quad s=1,\dots, k  \;, \quad viz. \quad \left\{ \ba{ccc} \bs{y}_k & = & \bs{x}_k  \\
																									\bs{y}_s & = & \bs{x}_s \, -\,  \bs{x}_{s+1} \ea \right.  \;.
\enq
In particular, one has $\bs{x}_{ba} \; = \; - \sul{p=a}{b-1} \bs{y}_p$ for $b>a$. This yields
\beq
\sul{b>a}{k} \ov{\bs{p}}(  \bs{\ga}^{(ba)} ) * \bs{x}_{ba} \; = \; - \sul{b=2}{k} \sul{a=1}{b-1}\sul{\ell = a}{ b-1 } \ov{\bs{p}}(  \bs{\ga}^{(ba)} ) * \bs{y}_{\ell}
\; = \; - \sul{b=2}{k}\sul{\ell = 1}{ b-1 }  \bs{y}_{\ell}  *  \sul{a=1}{\ell} \ov{\bs{p}}(  \bs{\ga}^{(ba)} )
\; = \; - \sul{\ell = 1}{ k-1 }  \bs{y}_{\ell}  * \op{P}_{\ell}\big( \bs{\ga} \big) \;,
\enq
where $\op{P}_{\ell}$ is as introduced in \eqref{definition impulsion reduite}.
Since the change of variables has unit Jacobian, one gets the equality
\beq
\mc{R}[ G] \big(   \bs{\ga}    \big)   \; = \; \wh{\mc{R}}\big[  \mc{L}[G] \big] \big(   \bs{\ga}    \big)   \;.
\label{definition TF space-time de la fct test diff variables}
\enq
The latter immediately yields
\beq
\mc{I}^{ (\bs{n}) }_{\bs{\a}_k } [G] \, = \, \mc{J}^{ (\bs{n}) }_{\bs{\a}_k }\big[ \mc{L}[G] \big] \;.
\label{ecriture distribution I en terme de distr J reduite}
\enq
Upon applying this identity to functions
\beq
G(\bs{X}_k) \, = \, g(\bs{x}_k) \cdot H\big(\bs{x}_1-\bs{x}_2,\dots, \bs{x}_{k-1} - \bs{x}_k \big) \qquad \e{where} \qquad
(g, H) \in \mc{S}( \R^{1,1}  ) \times \mc{S}\big( ( \R^{1,1} )^{k-1} \big) \;,
\enq
and observing that in such a case $\mc{L}[G]$ takes the particularly simple form
\beq
\mc{L}[G](\bs{Y}_{k-1}) \, = \,  H(\bs{Y}_{k-1}) \Int{ \R^{1,1} }{ } \hspace{-1mm} \dd \bs{y} \,  g(\bs{y})     \;,
\enq
one gets
\beq
\mc{I}^{ (\bs{n}) }_{\bs{\a}_k } [G] \, = \, \mc{J}^{ (\bs{n}) }_{\bs{\a}_k }\big[ H  \big]   \cdot \Int{ \R^{1,1} }{ } \dd \bs{y} g(\bs{y})  \;.
\enq
This entails that $ \mc{J}^{ (\bs{n}) }_{\bs{\a}_k }$ is a well-defined tempered distribution, and so is $\mc{V}_{\bs{\a}_k}^{(\bs{r})}$.\\
Turning on to the properties of the Fourier transform, one first observes that
\beq
\widehat{\mc{R}}\big[  \msc{F}[H] \big] \big(   \bs{\ga}    \big)   \; = \hspace{-2mm} \Int{ \big( \R^{1,1} \big)^{k-1} }{}  \hspace{-4mm} \dd^{k-1} \bs{Y}
\; \msc{F}[H]  \big( \bs{Y}_{k-1} \big)  \pl{ \ell = 1 }{ k - 1 } \Big\{ \ex{ -  \i  \bs{y}_{\ell}  * \op{P}_{\ell}( \bs{\ga}  ) } \Big\}
 \, = \, (2\pi)^{2(k-1)}  H\big(  \op{P}_{1}( \bs{\ga}), \dots,  \op{P}_{k-1}( \bs{\ga} )  \big)  \;.
\label{reecriture R hat transform Fourier}
\enq
This implies that
\beq
\Big(\msc{F}\big[\mc{J}^{ (\bs{n}) }_{\bs{\a}_k } \big] \Big) \big[H\big] \, =  \,
\mc{J}^{ (\bs{n}) }_{\bs{\a}_k }  \Big[\msc{F}\big[H\big] \Big]  \, =  \,
\lim_{\bs{\veps}_k \tend 0^+} \Int{  \R^{\bs{n}}   }{} \f{ \dd^{ \bs{n} } \bs{\ga}  }{ \bs{n}! (2\pi)^{|\bs{n}|}} \Big( \mc{S} \cdot  \mc{F}_{\bs{\a}_k;\bs{\veps}_{k} } \Big) \big(   \bs{\ga}     \big) \cdot
\;(2\pi)^{2(k-1)} H  \big( \op{P}_{1}( \bs{\ga}  ),\dots,\op{P}_{k-1}( \bs{\ga}  ) \big)  \;.
\label{expression TF distrib}
\enq
It is obvious from \eqref{definition 2 impulsion} and \eqref{definition impulsion reduite} that $\op{P}_{\ell}^{(0)}\big( \bs{\ga}  \big) > 0$. Further, the obvious lower bound
$\cosh(\ga_{k}^{(ba)}) \, > \, |\sinh(\ga_{k}^{(ba)})|$ leads to
\beq
\sul{b=\ell+1}{k} \sul{a=1}{\ell} \sul{ s=1 }{ n_{ba} }\cosh(\ga_{s}^{(ba)}) \, > \, \Big|\sul{b=\ell+1}{k} \sul{a=1}{\ell} \sul{ s=1 }{ n_{ba} } \sinh(\ga_{s}^{(ba)}) \Big|
\enq
what ensures  that, for any $\ell \in \intn{1}{k-1}$,  $\op{P}_{\ell}^2 \big( \bs{\ga}  \big) = \big(\op{P}_{\ell}^{(0)}\big( \bs{\ga}  \big)\big)^2 - \big(\op{P}_{\ell}^{(1)}\big( \bs{\ga}  \big)\big)^2 >0$.
This means that
\beq
\Big(\msc{F}\big[\mc{J}^{ (\bs{n}) }_{\bs{\a}_k } \big] \Big) \big[H\big] \, =  \, 0 \qquad \e{if} \qquad \e{supp}\big[ H \big] \cap \mc{E}_{k-1}^{+} \, = \, \emptyset \;,
\enq
\textit{i.e.}  that $\e{supp}\Big[\msc{F}\big[\mc{J}^{ (\bs{n}) }_{\bs{\a}_k } \big]  \Big] \subset  \mc{E}_{k-1}^{+}$.  \qed

\subsection{Hermiticity}

This subsection establishes that the truncated correlation functions satisfy the hermiticity axiom.

\begin{prop}
Let $G \in \mc{S}\big( (\R^{1,1})^k \big)$, $\bs{r} \in \mathbb{N}^{k-1}$ and pick operator indices $\a_1,\dots, \a_k$ associated with solutions to the $a)-d)$ constrains.
Let $\a^{\dagger}_1, \dots, \a^{\dagger}_k$ be the associated indices of the adjoint operators.

Then, the truncated correlation functions enjoy the below form of hermiticity:
\beq
\mc{W}^{(\bs{r})}_{ \bs{\a}_k }\big[ G \big] \, = \,  \ov{  \mc{W}^{(\overleftarrow{\bs{r}})}_{ \overleftarrow{\bs{\a}}_k^{\,\dagger} }\Big[\, \ov{ \iota \cdot G} \Big] } \;.
\enq
Above, $\overleftarrow{\bs{\a}}_k^{\, \dagger} = \big( \a_k^{\dagger}, \dots, \a_1^{\dagger} \big)$.
\end{prop}
\Proof

The truncated correlator of interest admits the representation
\beq
\ov{  \mc{W}^{(\overleftarrow{\bs{r}})}_{ \overleftarrow{\bs{\a}}_k^{\,\dagger} }\Big[\, \ov{ \iota \cdot G} \Big] } \,  = \,
\sul{\bs{n}\in \mc{N}_{\overleftarrow{\bs{r}}}}{}   \ov{ \mc{I}_{ \overleftarrow{\bs{\a}}_k^{\,\dagger} }^{ (\bs{n} ) }\Big[\, \ov{ \iota \cdot G} \Big]   }
\enq
where the summand can be represented as
\beq
 \ov{ \mc{I}_{ \overleftarrow{\bs{\a}}_k^{\,\dagger} }^{ (\bs{n} ) }\Big[\, \ov{ \iota \cdot G} \Big]   }  \; = \;
\lim_{ \bs{\veps}_k \tend \bs{0}^+} \Int{  \R^{ |\bs{n}| }   }{}  \f{ \dd^{ |\bs{n}| } \bs{\ga}  }{  \bs{n}! (2\pi)^{|\bs{n}|} } \;
   \ov{ \Big( \mc{S}\cdot \mc{R}\Big[\, \ov{ \iota \cdot G} \Big]  \cdot  \mc{F}_{ \overleftarrow{\bs{\a}}_k^{\,\dagger} ; \overleftarrow{ \bs{\veps} }_{k} } \Big) \big(   \bs{\ga}     \big) }   \; .
\enq
Note that the presence of $\overleftarrow{ \bs{\veps} }_{k}$ follows from the relabelling $\veps_{p} \hookrightarrow \veps_{k+1-p}$ of the regularisation
parameters in the limit.
We now study the building blocks one-by-one. First of all, one has
\beq
\ov{\mc{R}\Big[\, \ov{ \iota \cdot G} \Big](\bs{\ga}) }  \, =   \Int{ \big( \R^{1,1} \big)^{k} }{}  \hspace{-3mm} \dd^k \bs{X} \, G \big( \overleftarrow{\bs{X}}_k \big)
 \cdot \pl{b>a}{k} \ex{-  \i \ov{\bs{p}}(  \bs{\ga}^{(ba)} )* \bs{x}_{ba}  }  \;.
\label{fourier transform adjoint}
\enq
There, we introduce new variables $\bs{y}_{a}=\bs{x}_{k+1-a}$ and denote
\beq
\bs{\xi}^{(ba)}=\bs{\ga}^{(k+1-a \hspace{0.5mm} k+1-b)} \quad  \e{for} \; b>a \quad \e{as}\,  \e{well}\, \e{as} \quad  m_{ba} \, = \, n_{k+1-a \hspace{0.5mm} k+1-b} \; .
\label{definition variables xi reordonnees}
\enq
This yields
\beqa
\ov{\mc{R}\Big[\, \ov{ \iota \cdot G} \Big](\bs{\ga}) }  & =  & \hspace{-3mm} \Int{ \big( \R^{1,1} \big)^{k} }{}  \hspace{-3mm} \dd^k \bs{Y} \, G \big( \bs{Y}_k \big)
 \cdot \pl{b>a}{k} \ex{-  \i \ov{\bs{p}}(  \bs{\xi}^{(k+1-a  \hspace{0.5mm}k+1-b)} ) * \bs{y}_{k+1-b \hspace{0.5mm} k+1-a}  }   \\
& =  & \hspace{-3mm} \Int{ \big( \R^{1,1} \big)^{k} }{}  \hspace{-3mm} \dd^k \bs{Y} \, G \big( \bs{Y}_k \big)
 \cdot \pl{b>a}{k} \ex{  \i \ov{\bs{p}}(  \bs{\xi}^{(ba)} ) * \bs{y}_{b a}  } \;  = \; \mc{R}[G](\bs{\xi}) \;.
\eeqa

We now turn to the $\mc{S}$-factor. By using that $\op{S}(\be)=\ov{ \op{S}(-\be)}$ and introducing the variables $\bs{\xi}^{(ba)}$ as in \eqref{definition variables xi reordonnees}, one gets
\beqa
\ov{ \mc{S} \big( \bs{\ga} \big) } &  =  & \pl{    v > p > u > s    }{ k }
\op{S}\Big( \bs{\xi}^{(k+1-s\hspace{0.5mm}k+1-p)}  \cup \bs{\xi}^{(k+1-u\hspace{0.5mm}k+1-v)}  \mid   \bs{\xi}^{(k+1-u\hspace{0.5mm}k+1-v)}  \cup  \bs{\xi}^{(k+1-s \hspace{0.5mm}k+1-p)}   \Big)   \\
&= & \pl{    t > r > c > \ell    }{ k }  \op{S}\Big( \bs{\xi}^{(tc)}  \cup \bs{\xi}^{(r\ell)}  \mid   \bs{\xi}^{(r \ell)}  \cup  \bs{\xi}^{(t c)}   \Big)
 \, = \,  \mc{S} \big( \bs{\xi} \big) \;.
\eeqa

Finally, we turn towards the product of form factors which we re-express in terms of the variables $\bs{\xi}^{(ba)}$:
\beqa
 \mc{F}_{ \overleftarrow{\bs{\a}}_k^{\,\dagger} ; \overleftarrow{ \bs{\veps} }_{k} } \big( \bs{\ga} \big)    & = &
\pl{p=1}{k}   \mc{F}^{(\alpha^{\dagger}_{k+1-p})}\Big( \, \overleftarrow{ \bs{\ga}^{(pp-1)} }
																	\cup \cdots \cup \overleftarrow{ \bs{\ga}^{(p1)} }  + \i \pi \ov{\bs{e}}_{\veps_{k+1-p}} ,
																	\bs{\ga}^{(kp)}  \cup \cdots \cup \bs{\ga}^{(p+1p)}   \, \Big)   \\
& = &  \pl{p=1}{k}   \mc{F}^{(\alpha^{\dagger}_{p})}\Big( \, \overleftarrow{ \bs{\ga}^{(k+1-p\hspace{0.5mm}k-p)} }
																	\cup \cdots \cup \overleftarrow{ \bs{\ga}^{(k+1-p\hspace{0.5mm}1)} }  + \i \pi \ov{\bs{e}}_{\veps_{p}} ,
																	\bs{\ga}^{( k\hspace{0.5mm}k+1-p )}  \cup \cdots \cup \bs{\ga}^{(k+2-p\hspace{0.5mm}k+1-p)}   \, \Big)   \\
& =&   \pl{p=1}{k}   \mc{F}^{(\alpha^{\dagger}_{p})}\Big( \, \overleftarrow{ \bs{\xi}^{(p+1p)} } \cup \cdots \cup \overleftarrow{ \bs{\xi}^{(kp)} }  + \i \pi \ov{\bs{e}}_{\veps_{p}} ,
	\bs{\xi}^{(p1)}  \cup \cdots \cup \bs{\xi}^{(pp-1)}   \, \Big)     \;.
\eeqa
One then invokes the relation between the form factors of an operator with label $\a$ and its adjoint \eqref{ecriture equation correspondance entre FF et celui de adjoint}, so as to get
\beqa
\ov{ \mc{F}_{ \overleftarrow{\bs{\a}}_k^{\,\dagger} ; \overleftarrow{ \bs{\veps} }_{k} } \big( \bs{\ga} \big)  }  & = &
\pl{p=1}{k}  \mc{F}^{(\alpha_{p})}\Big( \, \overleftarrow{ \bs{\xi}^{(pp-1)} } \cup \cdots \cup \overleftarrow{ \bs{\xi}^{(p1)} }  + \i \pi \ov{\bs{e}}  ,
	\bs{\xi}^{(kp)}  \cup \cdots \cup \bs{\xi}^{(p+1p)}   - \i \pi \ov{\bs{e}}_{\veps_{p}}    + \i \pi \ov{\bs{e}}   \, \Big) \\
 & = & \pl{p=1}{k} \Big\{  \ex{\i \veps_p \op{s}_{\a_p} } \Big\} \cdot
 \pl{p=1}{k}  \mc{F}^{(\alpha_{p})}\Big( \, \overleftarrow{ \bs{\xi}^{(pp-1)} } \cup \cdots \cup \overleftarrow{ \bs{\xi}^{(p1)} }  + \i \pi \ov{\bs{e}}_{\veps_{p}}   ,
	\bs{\xi}^{(kp)}  \cup \cdots \cup \bs{\xi}^{(p+1p)}     \, \Big)   \\
& = & \pl{p=1}{k} \Big\{  \ex{\i \veps_p \op{s}_{\a_p} } \Big\}  \cdot \mc{F}_{ \bs{\a}_k ; \bs{\veps}_{k} } \big( \bs{\xi} \big) \;.
\eeqa
The last line follows by invoking the Lorentz symmetry property of the form factors.

Thus, all-in-all, we have established that with $\bs{m}$ as defined through \eqref{definition variables xi reordonnees}
\beq
 \ov{ \mc{I}_{ \overleftarrow{\bs{\a}}_k^{\,\dagger} }^{ (\bs{n} ) }\Big[\, \ov{ \iota \cdot G} \Big]   }  \; = \;
\lim_{ \bs{\veps}_k \tend \bs{0}^+}  \bigg\{ \pl{p=1}{k} \Big\{  \ex{\i \veps_p \op{s}_{\a_p} } \Big\}  \cdot   \Int{  \R^{ |\bs{m}| }   }{}  \f{ \dd^{ |\bs{m}| } \bs{\xi}  }{  \bs{m}! (2\pi)^{|\bs{m}|} } \;
   \Big( \mc{S}\cdot \mc{R} [    G   ]  \cdot  \mc{F}_{\bs{\a}_k ;  \bs{\veps} _{k} } \Big) \big(   \bs{\xi}     \big)
\; = \; \mc{I}_{  \bs{\a}_k }^{ (\bs{m} ) } [ G  ]  \; ,
\enq
since both $\bs{\veps}_k$ dependent factors admit $\bs{\veps}_k \tend \bs{0}^+$ limits.

Finally, one observes that if
\beq
r_{k-p} \, = \, \sul{u=p+1}{k} \sul{s=1}{p} n_{us} \quad \e{for} \, \quad  p\in \intn{1}{k-1}  \qquad \e{then} \qquad
r_{t} \, = \, \sul{u=t+1}{k} \sul{s=1}{t} m_{us} \quad \e{for} \, \quad  t\in \intn{1}{k-1}  \;,
\enq
with $m_{ba}$ as defined in \eqref{definition variables xi reordonnees}. Thus the sets $\mc{N}_{\bs{r}}$ and $\mc{N}_{ \overleftarrow{ \bs{r} } }$
are in one-to-one correspondence under the map $\bs{n} \hookrightarrow \bs{m}$. The above entails that
\beq
\ov{  \mc{W}^{(\overleftarrow{\bs{r}})}_{ \overleftarrow{\bs{\a}}_k^{\,\dagger} }\Big[\, \ov{ \iota \cdot G} \Big] } \,  = \,
\sul{\bs{m}\in \mc{N}_{\overleftarrow{\bs{r}}}}{}  \mc{I}_{  \bs{\a}_k }^{ (\bs{m} ) } [ G  ]  \, = \, \mc{W}^{(\bs{r})}_{ \bs{\a}_k  }[ G  ] \;,
\enq
hence leading to the claim. \qed

\subsection{Local Commutativity}

We first start by recalling an alternative representation for the truncated $k$-point function obtained in \cite{KozSimonMultiPointCorrFctsSinhGordon}.

\begin{prop} \cite{KozSimonMultiPointCorrFctsSinhGordon}
	\label{Proposition alternate representation for the truncated}
Let $G \in \mc{S}\big( (\R^{1,1})^k \big)$ and pick $t \in \intn{1}{k}$ and $\bs{r}\in \mathbb{N}^{k-1}$. Then, one has the below, well-defined, representation for the
$\bs{r}$-truncated $k$-point function. This alternative representation does coincide with the one given in \eqref{def trunc corr}.
\beq
\mc{W}^{ (\bs{r}) }_{ \bs{\a}_k }[G]  \; = \;
\sul{ \bs{n} \in \mc{N}_{\bs{r}} }{}    \mc{I}^{(\bs{n})}_{\bs{\a}_k; t}[ G ]
\enq
The building blocks are expressed as
\beq
\mc{I}^{(\bs{n})}_{\bs{\a}_k; t}[G]  \, = \,
\lim_{\bs{\veps}_k \tend \bs{0}^+} \hspace{-1mm} \Int{  \R^{ |\bs{n}|}    }{} \hspace{-1mm}  \f{ \dd^{|\bs{n}| } \bs{\ga} }{ \bs{n}! (2\pi)^{ |\bs{n}| }  }
\Big( \mc{S}^{(t)}\cdot \mc{R}[ G ] \cdot  \mc{F}_{\bs{\a}_k;\bs{\veps}_{k} }^{(t)} \Big) \big(   \bs{\ga}     \big) \;,
\label{ecriture distribution type t pour fct k pts}
\enq
with
\beq
\mc{S}^{(t)}(\bs{\ga}) \, = \,
\mc{S}(\bs{\ga}) \cdot \pl{v=t+1}{k} \pl{u=1}{t-1} \Bigg\{ \pl{s=1}{t-1} S\big(\bs{\ga}^{(ts)} \cup \bs{\ga}^{(vu)}   \mid \bs{\ga}^{(vu)} \cup \bs{\ga}^{(ts)} \big)
\cdot \pl{s=t+1}{k} S\big(\bs{\ga}^{(vu)} \cup \bs{\ga}^{(st)} \mid \bs{\ga}^{(st)} \cup \bs{\ga}^{(vu)} \big) \Bigg\} \; ,
\label{S product representation for alternating signs}
\enq
$\mc{S}$  as defined in \eqref{definition facteur de diffusion complet k pts} and
\bem
\mc{F}_{\bs{\a}_k;\bs{\veps}_{k} }^{(t)} \big(   \bs{\ga}     \big) \; = \;
\pl{\substack{p=1 \\ p \neq t }}{k} \mc{F}^{(\alpha_p)}\Big( \, \overleftarrow{ \bs{\ga}^{(pp-1)} } \cup \cdots \cup \overleftarrow{ \bs{\ga}^{(p1)} }  + \i \pi \ov{\bs{e}}_{\veps_{p}} ,
\bs{\ga}^{(kp)}  \cup \cdots \cup \bs{\ga}^{(p+1p)} \, \Big) \\
\times \mc{F}^{(\alpha_t)}\Big( \, \bs{\ga}^{(kt)}  \cup \cdots \cup \bs{\ga}^{(t+1t)} ,
\overleftarrow{ \bs{\ga}^{(tt-1)} } \cup \cdots \cup \overleftarrow{ \bs{\ga}^{(t1)} }  - \i \pi \ov{\bs{e}}_{\veps_{t}}  \, \Big) \; .
\label{alternate form factor product}
\end{multline}
Finally,  $\mc{R}[G]$ is as defined through \eqref{definition TF space-time de la fct test}.

\end{prop}
We will also make use of another representation for the general term to have more freedom with the deformation of contours:
\begin{prop} \cite{KozSimonMultiPointCorrFctsSinhGordon}
\label{proposition generale shifts pour causalite}
Let $G \in \mc{S}\big( (\R^{1,1})^k \big)$ and pick $t \in \intn{1}{k}$ and $\bs{n}\in \mathbb{N}^{\f{k(k-1)}{2}}$. Then,
 $\mc{I}^{(\bs{n})}_{\bs{\a}_k; t}[G]$ as defined in \eqref{ecriture distribution type t pour fct k pts}
admit the below integral representation:
	\beq
	\mc{I}^{(\bs{n})}_{\bs{\a}_k; t}[G]  \, = \,
	\lim_{\bs{\eta}_1, \bs{\eta}_2 \tend \bs{0}^+} \lim_{\bs{\delta}_{\mc{S}}, \bs{\delta}_{\mc{R}} \tend \bs{0}^+} \hspace{-0.5mm} \Int{  \R^{ |\bs{n}|}    }{} \hspace{-1mm}  \f{ \dd^{|\bs{n}| } \bs{\ga} }{ \bs{n}! (2\pi)^{ |\bs{n}| }  }
	\mc{S}^{(t)} \big(   \bs{\ga} + \i \bs{\delta}_\mc{S}    \big)\cdot \mc{R}[ G ] \big(   \bs{\ga}  + \i \bs{\delta}_\mc{R}   \big) \cdot  \mc{F}_{\bs{\a}_k }^{(t)}  \big(   \bs{\ga}  ; \bs{\eta}_1, \bs{\eta}_2   \big) \;,
	\label{ecriture plus generale integrande type t avec deformations}
	\enq
	where all limits can be freely interchanged and we introduced the following deformation parameters:
\begin{itemize}
\item $\bs{\delta}_{\mc{S}} \in \mathbb{R}^{|\bs{n}|}$ is small enough so that $\bs{\ga} \mapsto \mc{S}^{(t)} \big(   \bs{\ga} + \i \bs{\delta}_\mc{S}    \big)$ remains bounded on $\mathbb{R}^{|\bs{n}|}$;
\item $\bs{\delta}_{\mc{R}} \in \mathbb{R}^{|\bs{n}|}$ is chosen such that $\bs{\ga}^{(ba)} \mapsto \Re\left[\i \ov{\bs{p}}(\bs{\ga}^{(ba)}+\i\bs{\delta}^{(ba)}_{\mc{R}}) * \bs{x}_{ba} \right]$ is bounded from above for every $b>a$ uniformly in $(\bs{x}_1,\dots,\bs{x}_k) \in \e{supp}[G]$, ensuring that $\mc{R}[ G ] \big(   \bs{\ga}  + \i \bs{\delta}_\mc{R}   \big)$ is well-defined;
\item $\bs{\eta}_1 = (\eta^{(21)}_1,\dots,\eta^{(kk-1)}_1) \in \mathbb{R}^{\f{k(k-1)}{2}}_+$ and
$\bs{\eta}_2 = (\eta^{(21)}_2,\dots,\eta^{(kk-1)}_2) \in \mathbb{R}^{\f{k(k-1)}{2}}_+$ are small and under the constraints that
$\eta^{(cb)}_1 + \eta^{(dc)}_2 > 0$ for all $1 \leq b < c < d \leq k$.
\end{itemize}
Finally:
\bem
\mc{F}_{\bs{\a}_k }^{(t)}  \big(   \bs{\ga}  ; \bs{\eta}_1, \bs{\eta}_2   \big) \; = \;
\pl{\substack{p=1 \\ p \neq t }}{k} \bigg\{ \mc{F}^{(\alpha_p)}\Big( \, \overleftarrow{ \bs{\ga}^{(pp-1)} } + \i (\pi - \eta^{(pp-1)}_1)\ov{\bs{e}} ,
 \, \dots \, , \, \overleftarrow{ \bs{\ga}^{(p1)} }  + \i (\pi - \eta^{(p1)}_1)\ov{\bs{e}} , \\
\hspace{8cm} 	\bs{\ga}^{(kp)}  + \i \eta^{(kp)}_2 \ov{\bs{e}} \, ,  \dots \, , \, \bs{\ga}^{(p+1p)} + \i \eta^{(p+1p)}_2 \ov{\bs{e}} \, \Big) \bigg\} \\
\times \mc{F}^{(\alpha_t)}\Big( \, \bs{\ga}^{(kt)}  - \i \eta^{(kt)}_2 \ov{\bs{e}} ,\, \dots \, , \,  \bs{\ga}^{(t+1t)}  - \i \eta^{(t+1t)}_2 \ov{\bs{e}} \, ,
	\overleftarrow{ \bs{\ga}^{(tt-1)} } - \i (\pi - \eta^{(tt-1)}_1)\ov{\bs{e}} \, , \,  \dots \, , \,  \overleftarrow{ \bs{\ga}^{(t1)} }  - \i (\pi - \eta^{(t1)}_1)\ov{\bs{e}}  \, \Big) \; .
\end{multline}
\end{prop}
The representation for the $\bs{r}$-truncated $k$-point function provided by
propositions \ref{Proposition alternate representation for the truncated}-\ref{proposition generale shifts pour causalite}  is well-tailored for proving
the local commutativity axiom of the $k$-point function. The proof we shall give below follows the line of ideas developed in the context of
studying the weak local commutativity of two local operators  in the Sinh-Gordon model: first in the context of the quantum Gelfand-Levitan-Marchenko
equation construction of the model \cite{KhamitovGLMeqnsForSinhGordonAndProofOfLocalCommutativity}, second through certain identities
which translate weak local commutativity into one of the bootstrap axioms \cite{KirillovCombinatorialIdenititesForLocalCommutativityinSinhGordonQFTByKhamitov}
and third for the Thirring model in \cite{KirillovSmirnovUseOfbootstrapAxiomsForQIFTToGetMassiveThirringFF}. The main idea is to relate
the correlation functions with operator indices $\a_t$ and $\a_{t+1}$ swapped and with the test function $\tau_{t}\cdot G$ to the original
one by performing a contour integral deformation in \eqref{ecriture distribution type t pour fct k pts}. However, the technical details between considering the weak commutativity of the local operator and
the one within a correlation function are different. Moreover, the mentioned work did not consider various rigorous aspects of the handlings.

\begin{prop}
\label{local commut prop}
Let $t \in \intn{1}{k-1}$ and pick $G\in \mc{S}\big((\R^{1,1})^k \big)$ such that $\e{supp}[G] \subset \Om_{t}$
with $\Om_{t}$ as given in \eqref{definition support purement space like dans Minkowski a la k}

Then, provided that Conjecture \ref{Conjecture convergence} holds, the correlation functions satisfy
\beq
\mc{W}_{\bs{\a}_k}[G]  \; = \;
\mc{W}_{\tau_t \cdot \bs{\a}_k }[ \tau_t \cdot G] \;,
\enq
where $\tau_t$ is the transposition $(tt+1)$, \textit{c.f.}
\eqref{ecriture action inversion et transposition}-\eqref{definition action gpe perm sur vect et vect renverse}.

\end{prop}

\Proof

To start, observe that
\beq
\Om_{t} \, = \,  \Om_t^{+} \cup \Om_t^{-} \qquad \e{with} \qquad
\Om_t^{\pm} \, = \,  \Big\{   \bs{X}_k \in \Om_{t}  \, : \, \pm \big( x_{t;1} - x_{t+1;1} \big) >0 \Big\} \;.
\enq
$ \Om_t^{\pm}$ are open and connected. It is thus enough to focus on the case when $\e{supp}[G] \subset \Om_{t}^{\pm}$. We shall carry out the proof in the case when
 $\e{supp}[G] \subset \Om_{t}^{+}$, the other case can be dealt with quite similarly.

We start from the representation for the $\bs{r}$-truncated $k$ point function provided by Proposition \ref{Proposition alternate representation for the truncated}
with the $t+1^{\e{st}}$ operator singled out, which involves

\beq
\label{S matrix factors}
\mc{S}^{(t+1)}(\bs{\ga})
\, = \, \mc{S}(\bs{\ga}) \cdot \pl{v=t+2}{k} \pl{u=1}{t} \bigg\{
\pl{s=1}{t} \op{S}\big(\bs{\ga}^{(t+1s)} \cup \bs{\ga}^{(vu)}  \mid  \bs{\ga}^{(vu)}\cup \bs{\ga}^{(t+1s)} \big)
\cdot \pl{s=t+2}{k} \op{S}\big(\bs{\ga}^{(vu)} \cup \bs{\ga}^{(st+1)}  \mid  \bs{\ga}^{(st+1)}\cup \bs{\ga}^{(vu)} \big)   \bigg\} \;,
\enq
as well as 
\bem
\mc{F}_{\bs{\a}_k;\bs{\veps}_{k} }^{(t+1)} \big(   \bs{\ga}     \big) \, = \, \pl{ \substack{ p=1  \\  \not= t, t+1} }{k}\mf{f}^{(p)}(\bs{\ga})
\cdot \mc{F}^{(\alpha_t)}\Big( \, \overleftarrow{ \bs{\ga}^{(tt-1)} } \cup \cdots \cup \overleftarrow{ \bs{\ga}^{(t1)} }  + \i \pi \ov{\bs{e}}_{\veps_{t}} ,
\bs{\ga}^{(kt)}  \cup \cdots \cup \bs{\ga}^{(t+1t)}  \, \Big) \\
\times
\mc{F}^{(\alpha_{t+1})}\Big( \, \bs{\ga}^{(kt+1)}  \cup \cdots \cup \bs{\ga}^{(t+2t+1)} ,
\overleftarrow{ \bs{\ga}^{(t+1t)} } \cup \cdots \cup \overleftarrow{ \bs{\ga}^{(t+11)} }  - \i \pi \ov{\bs{e}}_{\veps_{t+1}}  \, \Big) \;,
\label{definition FF t dependant}
\end{multline}
where we agreed upon
\beq
\mf{f}^{(p)}(\bs{\ga}) \, = \, \mc{F}^{(\alpha_p)}\Big( \, \overleftarrow{ \bs{\ga}^{(pp-1)} } \cup \cdots \cup \overleftarrow{ \bs{\ga}^{(p1)} }  + \i \pi \ov{\bs{e}}_{\veps_{p}} ,
\bs{\ga}^{(kp)}  \cup \cdots \cup \bs{\ga}^{(p+1p)} \, \Big) \;.
\enq
One now applies the contour deformation for the variables $\bs{\ga}^{(t+1t)}$ from
\beq
\R^{n_{t+1t}} \qquad  \e{to} \qquad  \Big\{ \R+ \i(\pi-\veps_{t}-\veps_{t+1}) \Big\}^{ n_{t+1t} } .
\enq
We check that no poles of the integrand  are crossed in the process.
The pole structure of the form factors given in Subsection \ref{Subsection building blocks} indicates that the contour deformation crosses
the simple poles of \eqref{definition FF t dependant}
located at
\begin{equation}
	\ga^{(vt)}_p - \ga^{(t+1t)}_q \in \left\{-2\i\pi\mf{b},-2\i\pi\hat{\mf{b}}\right\} \;, \quad v=t+2,\dots,k \;, \quad p \in \intn{1}{n_{vt}} \quad \text{and} \quad q \in \intn{1}{n_{t+1t}} \;.
\end{equation}
These are compensated by zeroes of the $\op{S}$-matrix factors \eqref{S matrix factors} at the corresponding points. Thus, the choice of the representation
one starts with ensures that no poles are crossed in the process of the contour deformation. Further, one observes that the exponential
factor present in the momentum transform part $ \mc{R}[ G ]$ with $\mc{R}$ as introduced in  \eqref{definition TF space-time de la fct test}
remains bounded due to the condition $x_{t;1}-x_{t+1;1} >0$ being always fulfilled throughout the support of $G$.
The contour deformation modifies the expression of the integrand. The $\bs{\ga}^{(t+1t)}$ contribution to the momentum
modifies as $\mc{R}[G](\bs{\ga})\, \hookrightarrow \, \wt{\mc{R}}[G](\bs{\ga} \mid \veps_{t}, \veps_{t+1})$  with
\beq
\wt{\mc{R}}[G](\bs{\ga} \mid \veps_{t}, \veps_{t+1}) \, = \,
\Int{ \big( \R^{1,1} \big)^{k} }{}  \hspace{-2mm} \dd^k \bs{X}
\cdot  G\big(\bs{X}_k \big)  \cdot
\ex{\i \ov{\bs{p}}(\bs{\ga}^{(t+1t)}-\i (\veps_t+\veps_{t+1})\, \ov{\bs{e}} ) * \bs{x}_{tt+1} }
\cdot \hspace{-3mm} \prod_{\substack{ d>c \\ (d,c) \neq (t+1,t)}}\hspace{-3mm} \ex{\i \ov{\bs{p}}(\bs{\ga}^{(dc)}) * \bs{x}_{dc} } \:.
\enq
one has that
\bem
\mc{F}_{\bs{\a}_k;\bs{\veps}_{k} }^{(t+1)} \big(   \bs{\ga}     \big) \, \hookrightarrow \, \pl{ \substack{ p=1  \\  \not= t, t+1} }{k}\mf{f}^{(p)}(\bs{\ga})
\cdot \mc{F}^{(\alpha_t)}\Big( \, \overleftarrow{ \bs{\ga}^{(tt-1)} } \cup \cdots \cup \overleftarrow{ \bs{\ga}^{(t1)} }  + \i \pi \ov{\bs{e}}_{\veps_{t}} ,
\bs{\ga}^{(kt)}  \cup \cdots \cup \bs{\ga}^{(t+2t)}, \bs{\ga}^{(t+1t)} + \i \pi \ov{\bs{e}}_{\veps_{t}+\veps_{t+1}}   \, \Big) \\
\times
\mc{F}^{(\alpha_{t+1})}\Big( \, \bs{\ga}^{(kt+1)}  \cup \cdots \cup \bs{\ga}^{(t+2t+1)} ,
\overleftarrow{ \bs{\ga}^{(t+1t)} } - \i \veps_{t} \ov{\bs{e}} , \overleftarrow{ \bs{\ga}^{(t+1t-1)} } \cup \cdots \cup \overleftarrow{ \bs{\ga}^{(t+11)} }  - \i \pi \ov{\bs{e}}_{\veps_{t+1}}  \, \Big) \;.
\end{multline}
Indeed, only the form factors labelled by $\a_t$ and $\a_{t+1}$ depend on $\bs{\ga}^{(t+1t)}$.
Finally, focusing on the transformation incurred by $\mc{S}^{(t+1)}(\bs{\ga})$, one observes that the
expression \eqref{definition facteur de diffusion complet k pts} for $\mc{S}(\bs{\ga})$ involves pairs of variables $\bs{\ga}^{(vu)}$ and $\bs{\ga}^{(ps)}$
with indices ordered as $v>p>u>s$. Hence, the $v,u$  or  $p,s$ integers are at least distant by 2: the expression for $\mc{S}(\bs{\ga})$  does not involve
the variables $\bs{\ga}^{(t+1t)}$. It thus remains unaltered in the contour deformation process.
Therefore, by unitarity of $\op{S}$, one gets that
\bem
\mc{S}^{(t+1)}(\bs{\ga})
\, \hookrightarrow \, \mc{S}(\bs{\ga}) \cdot \pl{v=t+2}{k} \pl{u=1}{t} \bigg\{
\op{S}\big(\bs{\ga}^{(vu)} \cup \big[\bs{\ga}^{(t+1t)} -\i \ov{\veps}_{t t+1} \ov{\bs{e}}  \big]  \mid  \big[\bs{\ga}^{(t+1t)} -\i \ov{\veps}_{t t+1} \ov{\bs{e}} \big] \cup \bs{\ga}^{(vu)} \big) \\
\times \pl{s=1}{t-1} \op{S}\big(\bs{\ga}^{(t+1s)} \cup \bs{\ga}^{(vu)}  \mid  \bs{\ga}^{(vu)}\cup \bs{\ga}^{(t+1s)} \big)
\cdot \pl{s=t+2}{k} \op{S}\big(\bs{\ga}^{(vu)} \cup \bs{\ga}^{(st+1)}  \mid  \bs{\ga}^{(st+1)}\cup \bs{\ga}^{(vu)} \big)   \bigg\} \;.
\end{multline}
Above, we have set $\ov{\veps}_{t t+1}=\veps_{t}+\veps_{t+1}$.

Next,  one swaps the positions of $\bs{\ga}^{(t+1p)}$ and  $\bs{\ga}^{(tp)}$ in the form factor labelled by $\a_{p}$ with $p<t$,
by using the exchange property, \textit{c.f.} \eqref{ecriture S propriete FF}:
\bem
\mf{f}^{(p)}(\bs{\ga})  \, = \,  \op{S}\big(\bs{\ga}^{(t+1p)} \cup \bs{\ga}^{(tp)}  \mid  \bs{\ga}^{(tp)}\cup \bs{\ga}^{(t+1p)} \big) \\
\times \mc{F}^{(\alpha_p)}\Big( \, \overleftarrow{ \bs{\ga}^{(pp-1)} } \cup \cdots \cup \overleftarrow{ \bs{\ga}^{(p1)} }  + \i \pi \ov{\bs{e}}_{\veps_{p}} ,
\bs{\ga}^{(kp)}  \cup \cdots \cup \bs{\ga}^{(tp)} \cup \bs{\ga}^{(t+1p)} \cup \cdots \cup \bs{\ga}^{(p+1p)} \, \Big) \;.
\end{multline}
Further, one swaps $\bs{\ga}^{(t+1p)}$ and  $\bs{\ga}^{(tp)}$ in each form factor labelled by $\a_p$ with $p>t+1$,
\bem
\mf{f}^{(p)}(\bs{\ga})  \, = \, \op{S}\big(\bs{\ga}^{(pt+1)} \cup \bs{\ga}^{(pt)}  \mid  \bs{\ga}^{(pt)}\cup \bs{\ga}^{(pt+1)} \big)  \\
\times \mc{F}^{(\alpha_p)}\Big( \, \overleftarrow{ \bs{\ga}^{(pp-1)} } \cup \cdots \cup \overleftarrow{ \bs{\ga}^{(pt)} }
\cup \overleftarrow{ \bs{\ga}^{(pt+1)} } \cup \cdots \cup \overleftarrow{ \bs{\ga}^{(p1)} }  + \i \pi \ov{\bs{e}}_{\veps_{p}} ,
\bs{\ga}^{(kp)} \cup \cdots \cup \bs{\ga}^{(p+1p)} \, \Big) \, .
\end{multline}
Finally, we move $\bs{\ga}^{(t+1t)}$ through all other variables in the form factor labelled by $\a_t$:
\bem
\mc{F}^{(\alpha_t)}\Big( \, \overleftarrow{ \bs{\ga}^{(tt-1)} } \cup \cdots \cup \overleftarrow{ \bs{\ga}^{(t1)} }  + \i \pi \ov{\bs{e}}_{\veps_{t}} ,
\bs{\ga}^{(kt)}  \cup \cdots \cup \bs{\ga}^{(t+2t)}, \bs{\ga}^{(t+1t)} + \i \pi \ov{\bs{e}}_{ \ov{\veps}_{t t+1} }   \, \Big) \\
= \pl{a=1}{t-1} \op{S}\Big(\bs{\ga}^{(ta)} \cup \big[\bs{\ga}^{(t+1t)} -\i \veps_{t+1} \ov{\bs{e}} \big]  \mid  \big[\bs{\ga}^{(t+1t)} -\i \veps_{t+1} \ov{\bs{e}} \big] \cup \bs{\ga}^{(ta)} \Big)
\pl{a=t+2}{k} \op{S}\Big(\big[\bs{\ga}^{(t+1t)} -\i \ov{\veps}_{t t+1} \ov{\bs{e}}  \big] \cup \bs{\ga}^{(at)} \mid \bs{\ga}^{(at)} \cup \big[\bs{\ga}^{(t+1t)} -\i \ov{\veps}_{t t+1} \ov{\bs{e}}  \big]     \Big)  \\
\times
S\Big( \bs{\ga}^{(t+1t)} \mid \overleftarrow{\bs{\ga}^{(t+1t)}} \Big) \cdot
\mc{F}^{(\alpha_t)}\Big( \, \overleftarrow{\bs{\ga}^{(t+1t)}} + \i \pi \ov{\bs{e}}_{ \ov{\veps}_{t t+1} },
\overleftarrow{ \bs{\ga}^{(tt-1)} } \cup \cdots \cup \overleftarrow{ \bs{\ga}^{(t1)} }  + \i \pi \ov{\bs{e}}_{\veps_{t}} ,
\bs{\ga}^{(kt)}  \cup \cdots \cup \bs{\ga}^{(t+2t)}   \, \Big)
\end{multline}
while we reorder $\bs{\ga}^{(tt-1)}$ in the form factor labelled by $\a_{t+1}$
\bem
\mc{F}^{(\alpha_{t+1})}\Big( \, \bs{\ga}^{(kt+1)}  \cup \cdots \cup \bs{\ga}^{(t+2t+1)} ,
\overleftarrow{ \bs{\ga}^{(t+1t)} } - \i \veps_{t} \ov{\bs{e}}  , \overleftarrow{ \bs{\ga}^{(t+1t-1)} } \cup \cdots \cup \overleftarrow{ \bs{\ga}^{(t+11)} }  - \i \pi \ov{\bs{e}}_{\veps_{t+1}}  \, \Big)
= \op{S}\Big( \,  \overleftarrow{\bs{\ga}^{(t+1t)}} \mid \bs{\ga}^{(t+1t)} \Big)  \\
\times \mc{F}^{(\alpha_{t+1})}\Big( \, \bs{\ga}^{(kt+1)}  \cup \cdots \cup \bs{\ga}^{(t+2t+1)} ,
\bs{\ga}^{(t+1t)} - \i \veps_{t} \ov{\bs{e}} , \overleftarrow{ \bs{\ga}^{(t+1t-1)} } \cup \cdots \cup \overleftarrow{ \bs{\ga}^{(t+11)} }  - \i \pi \ov{\bs{e}}_{\veps_{t+1}}  \, \Big) \, .
\end{multline}
Gathering together all of the $\op{S}$-matrix terms leads us to introduce
\bem
\mc{S}_{\e{mod}}\big( \bs{\ga} \mid \veps_t,\veps_{t+1}\big) \, = \, \mc{S}(\bs{\ga}) \cdot \pl{v=t+2}{k} \pl{u=1}{t} \bigg\{
\op{S}\big(\bs{\ga}^{(vu)} \cup \big[\bs{\ga}^{(t+1t)} -\i \ov{\veps}_{t t+1} \ov{\bs{e}}  \big]  \mid  \big[\bs{\ga}^{(t+1t)} -\i \ov{\veps}_{t t+1} \ov{\bs{e}} \big] \cup \bs{\ga}^{(vu)} \big) \\
\times \pl{s=1}{t-1} \op{S}\big(\bs{\ga}^{(t+1s)} \cup \bs{\ga}^{(vu)}  \mid  \bs{\ga}^{(vu)}\cup \bs{\ga}^{(t+1s)} \big)
\cdot \pl{s=t+2}{k} \op{S}\big(\bs{\ga}^{(vu)} \cup \bs{\ga}^{(st+1)}  \mid  \bs{\ga}^{(st+1)}\cup \bs{\ga}^{(vu)} \big)   \bigg\}  \\
\times \pl{a=1}{t-1}  \bigg\{ \op{S}\big(\bs{\ga}^{(t+1a)} \cup \bs{\ga}^{(ta)}  \mid  \bs{\ga}^{(ta)}\cup \bs{\ga}^{(t+1a)} \big)
\cdot
\op{S}\big(\bs{\ga}^{(ta)} \cup \big[\bs{\ga}^{(t+1t)} -\i \veps_{t+1} \ov{\bs{e}} \big]  \mid  \big[\bs{\ga}^{(t+1t)} -\i \veps_{t+1} \ov{\bs{e}} \big] \cup \bs{\ga}^{(ta)} \big) \bigg\}  \\
\times \pl{a=t+2}{k} \bigg\{  \op{S}\big(\bs{\ga}^{(at+1)} \cup \bs{\ga}^{(at)}  \mid  \bs{\ga}^{(at)}\cup \bs{\ga}^{(at+1)} \big)
\cdot \op{S}\big(\big[\bs{\ga}^{(t+1t)} -\i \ov{\veps}_{t t+1} \ov{\bs{e}}  \big] \cup \bs{\ga}^{(at)} \mid \bs{\ga}^{(at)} \cup \big[\bs{\ga}^{(t+1t)} -\i \ov{\veps}_{t t+1} \ov{\bs{e}}  \big]     \big) \bigg\} \;.
\end{multline}
It is also handy to set
\bem
\wt{\mc{F}}_{\bs{\a}_k;\bs{\veps}_{k} }^{(t+1)} \big(   \bs{\ga}   \mid \nu_t, \nu_{t+1}  \big) \, = \,
\pl{  p=1  }{t-1}\mc{F}^{(\alpha_p)}\Big( \, \overleftarrow{ \bs{\ga}^{(pp-1)} } \cup \cdots \cup \overleftarrow{ \bs{\ga}^{(p1)} }  + \i \pi \ov{\bs{e}}_{\veps_{p}} ,
\bs{\ga}^{(kp)}  \cup \cdots \cup \bs{\ga}^{(tp)} \cup \bs{\ga}^{(t+1p)} \cup \cdots \cup \bs{\ga}^{(p+1p)} \, \Big)  \\
\times \pl{  p= t+2  }{ k }\mc{F}^{(\alpha_p)}\Big( \, \overleftarrow{ \bs{\ga}^{(pp-1)} } \cup \cdots \cup \overleftarrow{ \bs{\ga}^{(pt)} }
\cup \overleftarrow{ \bs{\ga}^{(pt+1)} } \cup \cdots \cup \overleftarrow{ \bs{\ga}^{(p1)} }  + \i \pi \ov{\bs{e}}_{\veps_{p}} ,
\bs{\ga}^{(kp)} \cup \cdots \cup \bs{\ga}^{(p+1p)} \, \Big)   \\
\times \mc{F}^{(\alpha_t)}\Big( \, \overleftarrow{\bs{\ga}^{(t+1t)}} + \i \pi \ov{\bs{e}}_{ \veps_{t} +\nu_{t+1}},
\overleftarrow{ \bs{\ga}^{(tt-1)} } \cup \cdots \cup \overleftarrow{ \bs{\ga}^{(t1)} }  + \i \pi \ov{\bs{e}}_{\veps_{t}} ,
\bs{\ga}^{(kt)}  \cup \cdots \cup \bs{\ga}^{(t+2t)}   \, \Big) \\
\times \mc{F}^{(\alpha_{t+1})}\Big( \, \bs{\ga}^{(kt+1)}  \cup \cdots \cup \bs{\ga}^{(t+2t+1)} ,
\bs{\ga}^{(t+1t)} - \i \nu_{t} \ov{\bs{e}} , \overleftarrow{ \bs{\ga}^{(t+1t-1)} } \cup \cdots \cup \overleftarrow{ \bs{\ga}^{(t+11)} }  - \i \pi \ov{\bs{e}}_{\veps_{t+1}}  \, \Big) \;.
\end{multline}
Thus, we have just shown that
\beq
\mc{W}^{(\bs{r})}_{\bs{\a}_k}[G] \; = \;
\sul{ \bs{n} \in \mc{N}_{\bs{r}} }{}
\lim_{ \bs{\veps}_{k} \tend \bs{0}^+}  \Int{  \R^{ |\bs{n}| }   }{}  \hspace{-2mm}\f{ \dd^{ |\bs{n}| }  \bs{\ga} }{ \bs{n}! (2\pi)^{|\bs{n}|}  }
\;  \Big( \, \wt{\mc{R}}[G] \cdot \mc{S}_{\e{mod}}  \cdot \wt{\mc{F}}_{\bs{\alpha}_k;\bs{\veps}_{k} }^{(t+1)} \Big) \big( \bs{\ga} \mid \veps_t,\veps_{t+1}\big) \;.
\label{intermediate commut}
\enq
At this stage, we change variables in the integral $\bs{\ga} \hookrightarrow \bs{\ga}^{\sg_t}$ so that
\beq
\left( \ba{c}  \bs{\ga}^{(pt)}  \\ \bs{\ga}^{(pt+1)} \ea \right)  \hookrightarrow \left( \ba{c}  \bs{\ga}^{(pt+1)}  \\ \bs{\ga}^{(pt)} \ea \right)  \quad \e{for} \quad  p \geq t+2
\enq
and
\beq
\left( \ba{c}  \bs{\ga}^{(tp)} \\ \bs{\ga}^{(t+1p)} \ea \right)  \hookrightarrow \left( \ba{c}  \bs{\ga}^{(t+1p)} \\ \bs{\ga}^{(tp)} \ea \right) \quad \e{for} \quad  p \leq t-1 \;,
\enq
all other variables being unchanged. We then swap the space-time coordinates in the integral corresponding to  $\wt{\mc{R}}[G]\big( \bs{\ga}^{\sg_t} \mid \veps_t , \veps_{t+1}\big)$:
\beq
\bs{x}_t \leftrightarrow \bs{x}_{t+1}
\enq
leading to
\beq
\wt{\mc{R}}[G]\big( \bs{\ga}^{\sg_t} \hspace{-0.5mm} \mid \veps_t , \veps_{t+1}\big) \, = \,  \mc{R}\big[ \tau_t\cdot G \big]\big( \bs{\ga} +\i \bs{\delta}_{\mc{R}}\big) \;.
\enq
Here $\bs{\delta}_{\mc{R}}$ is chosen with $\delta^{(ba)}_{\mc{R},\ell}= -(\veps_t + \veps_{t+1})\delta_{b,t+1}\delta_{a,t}$. Further, one gets that
\beq
\wt{\mc{F}}_{\bs{\a}_k;\bs{\veps}_{k} }^{(t+1)}  \big( \bs{\ga}^{\sg_t} \mid \veps_t, \veps_{t+1}\big) \, = \, \mc{F}_{ \tau_t\cdot \bs{\a}_k }^{(t)}  \big(   \bs{\ga}  ; \bs{\eta}_1, \bs{\eta}_2   \big)
\enq
with $\bs{\eta}_1$ and $\bs{\eta}_2$ given by $\eta^{(ba)}_1 = \veps_b + (\veps_{t+1}-\veps_t)(\delta_{b,t}-\delta_{b,t+1}) + \veps_{t+1}\delta_{b,t+1}\delta_{a,t}$ and $\eta^{(ba)}_2 = \veps_t \delta_{b,t+1}\delta_{a,t}$.\\
%
%
%
Dealing with the $\op{S}$-matrix contributions demands more care. One has that
\bem
\mc{S}_{\e{mod}}\big( \bs{\ga}^{\sg_t} \mid \veps_t,\veps_{t+1}\big) = \mc{S}(\bs{\ga}^{\sg_t}) \cdot \pl{v=t+2}{k} \Bigg\{ \pl{u=1}{t-1}
\bigg[ \op{S}\big(\bs{\ga}^{(vu)} \cup  \big[\bs{\ga}^{(t+1t)} -\i \ov{\veps}_{t t+1} \ov{\bs{e}}  \big]      \mid  \big[\bs{\ga}^{(t+1t)} -\i \ov{\veps}_{t t+1} \ov{\bs{e}}  \big]      \cup \bs{\ga}^{(vu)} \big)   \\
\times \op{S}\big(\bs{\ga}^{(tu)} \cup \bs{\ga}^{(vt+1)}  \mid  \bs{\ga}^{(vt+1)}\cup \bs{\ga}^{(tu)} \big) \cdot\pl{s=1}{t-1}   \op{S}\big(\bs{\ga}^{(ts)} \cup  \bs{\ga}^{(vu)}   \mid  \bs{\ga}^{(vu)}  \cup \bs{\ga}^{(ts)} \big)
\cdot  \pl{s=t+2}{k}    \op{S}\big(\bs{\ga}^{(vu)} \cup  \bs{\ga}^{(st)}   \mid  \bs{\ga}^{(st)}  \cup \bs{\ga}^{(vu)} \big)  \bigg]  \\
\times   \pl{s=t+2}{k} \op{S}\big(\bs{\ga}^{(vt+1)} \cup  \bs{\ga}^{(st)}   \mid  \bs{\ga}^{(st)}  \cup \bs{\ga}^{(vt+1)} \big)\Bigg\}
\cdot  \pl{s=t+2}{k} \op{S}\big(\bs{\ga}^{(st)} \cup \bs{\ga}^{(st+1)}  \mid  \bs{\ga}^{(st+1)}\cup \bs{\ga}^{( s t)} \big)  \\
\times \pl{u=1}{t-1} \bigg\{  \op{S}\big(\bs{\ga}^{(tu)} \cup \bs{\ga}^{(t+1u)}  \mid  \bs{\ga}^{(t+1u)}\cup \bs{\ga}^{(t u)} \big)
\cdot \op{S}\big(\bs{\ga}^{(t+1u)} \cup \big[\bs{\ga}^{(t+1t)} -\i \ov{\veps}_{t t+1} \ov{\bs{e}}  \big]      \mid  \big[\bs{\ga}^{(t+1t)} -\i \ov{\veps}_{t t+1} \ov{\bs{e}}  \big]    \cup \bs{\ga}^{(t+1u)} \big)  \bigg\}  \;.
\end{multline}
Then, one observes that
\bem
\mc{S}(\bs{\ga}^{\sg_t}) \, = \,\mc{S}(\bs{\ga}) \cdot  \pl{ \substack{ p, v =t+2 \\ p \not= v}  }{k}  \op{S}\big(\bs{\ga}^{(vt)} \cup \bs{\ga}^{(pt+1)}  \mid  \bs{\ga}^{(pt+1)}\cup \bs{\ga}^{(vt)} \big)
\pl{ \substack{ u, s = 1\\ u \not= s}  }{ t-1 }  \op{S}\big(\bs{\ga}^{(tu)} \cup \bs{\ga}^{(t+1s)}  \mid  \bs{\ga}^{(t+1s)}\cup \bs{\ga}^{(tu)} \big)  \\
\pl{  v =t+2   }{k}  \pl{  s = 1  }{t-1} \bigg\{  \op{S}\big(\bs{\ga}^{(vt+1)} \cup \bs{\ga}^{(ts)}  \mid  \bs{\ga}^{(ts)}\cup \bs{\ga}^{(vt+1)} \big)
\cdot \op{S}\big(\bs{\ga}^{(t+1s)} \cup \bs{\ga}^{(vt)}  \mid  \bs{\ga}^{(vt)}\cup \bs{\ga}^{(t+1s)} \big)  \bigg\} \;.
\end{multline}
These intermediate decompositions lead to $\mc{S}_{\e{mod}}\big( \bs{\ga}^{\sg} \mid \veps_t,\veps_{t+1}\big) \, = \,  \mc{S}^{(t)}(\bs{\ga}+\i\bs{\delta}_{\mc{S}}) \cdot \mc{S}_{\e{add}}(\bs{\ga})$ with the introduced shift $\delta^{(ba)}_{\mc{S},\ell} = -(\veps_t + \veps_{t+1})\delta_{b,t+1}\delta_{a,t}$ and
\bem
\mc{S}_{\e{add}}(\bs{\ga}) \; = \;   \f{ \mc{S}(\bs{\ga}^{\sg_t}) }{ \mc{S}(\bs{\ga}) }  \pl{v=t+2}{k} \bigg\{ \pl{u=1}{t-1}
\op{S}\big(\bs{\ga}^{(tu)} \cup \bs{\ga}^{(vt+1)}  \mid  \bs{\ga}^{(vt+1)}\cup \bs{\ga}^{(tu)} \big)
\cdot  \pl{u=t+2}{k}  \op{S}\big(\bs{\ga}^{(vt+1)} \cup  \bs{\ga}^{(ut)}   \mid  \bs{\ga}^{(ut)}  \cup \bs{\ga}^{(vt+1)} \big)  \bigg\}  \\
\times \pl{u,v=1}{t-1}  \op{S}\big(\bs{\ga}^{(t+1u)} \cup \bs{\ga}^{(tv)}  \mid  \bs{\ga}^{(tv)}\cup \bs{\ga}^{(t+1u)} \big)
\pl{v=t+1}{k}  \pl{u=1}{t-1}  \op{S}\big( \bs{\ga}^{(vt)}\cup \bs{\ga}^{(t+1u)} \mid \bs{\ga}^{(t+1u)} \cup \bs{\ga}^{(vt)} \big) \\
\times  \pl{u=1}{t-1} \bigg\{ \op{S}\big(\bs{\ga}^{(t+1u)} \cup  \bs{\ga}^{(t+1t)}   \mid  \bs{\ga}^{(t+1t)}  \cup \bs{\ga}^{(t+1u)} \big)
\op{S}\big(\bs{\ga}^{(tu)} \cup  \bs{\ga}^{(t+1u)}   \mid  \bs{\ga}^{(t+1u)}  \cup \bs{\ga}^{(tu)} \big)   \bigg\}  \\
\times \pl{u=t+2}{k}  \op{S}\big(\bs{\ga}^{(ut)} \cup  \bs{\ga}^{(ut+1)}   \mid  \bs{\ga}^{(ut+1)}  \cup \bs{\ga}^{(ut)} \big)  \;.
\end{multline}
After simplifications, one gets that $ \mc{S}_{\e{add}}(\bs{\ga}) \; = \;  1$.

Finally, one should accomodate for the impact of the change of variables on the dimensionality of the various integrations.
This leads to introduce a new collection of integers:
\beqa
&& m_{bt} = n_{bt+1} \, , \quad  m_{b t+1} = n_{bt} \, ,  \quad b \geq t+2 \\
&& m_{ta} = n_{t+1a} \, , \quad  m_{t+1a} = n_{ta} \, , \quad a \leq t-1 \\
&& \qquad m_{dc} = n_{dc} \, , \quad \e{if} \quad d, c \not\in \{t, t+1\} \;.
\label{new indices}
\eeqa
We denote the inverse map by $\bs{n}(\bs{m})$. The above recasts the truncated correlation function as:
\beq
\mc{W}^{(\bs{r})}_{\bs{\a}_k}[G] \;  = \;
\sul{\bs{m} : \bs{n}(\bs{m})  \in \mc{N}_{\bs{r}} }{} 
\lim_{\bs{\eta}_1, \bs{\eta}_2 \tend \bs{0}^+} \lim_{\bs{\delta}_{\mc{S}}, \bs{\delta}_{\mc{R}} \tend \bs{0}^+} \hspace{-0.5mm} \Int{  \R^{ |\bs{m}|}    }{} \hspace{-1mm}
\f{ \dd^{|\bs{m}| } \bs{\ga} }{ \bs{m}! (2\pi)^{ |\bs{m}| }  }
	\mc{S}^{(t)} \big(   \bs{\ga} + \i \bs{\delta}_\mc{S}    \big)\cdot \mc{R}[ \tau_t \cdot G ] \big(   \bs{\ga}  + \i \bs{\delta}_\mc{R}   \big) \cdot  \mc{F}_{\tau_t \cdot \bs{\a}_k }^{(t)}  \big(   \bs{\ga}  ; \bs{\eta}_1, \bs{\eta}_2   \big) \;.
\enq

Therefore, under the assumption that the series defining the correlation function converges, one can sum over all integers $r_1,\dots,r_{k-1}$.
This relaxes the $\bs{n}(\bs{m}) \in \mc{N}_{\bs{r}}$ constraint to summing up over $\bs{m} \in \mathbb{N}^{ \f{k(k-1)}{2} }$,
namely that
\beq
\mc{W}_{\bs{\a}_k}[G]
= \sul{\bs{m}   \in\mathbb{N}^{ \f{k(k-1)}{2} } }{}
\lim_{\bs{\eta}_1, \bs{\eta}_2 \tend \bs{0}^+} \lim_{\bs{\delta}_{\mc{S}}, \bs{\delta}_{\mc{R}} \tend \bs{0}} \hspace{-0.5mm} \Int{  \R^{ |\bs{m}|}    }{} \hspace{-1mm}  \f{ \dd^{|\bs{m}| } \bs{\ga} }{ \bs{m}! (2\pi)^{ |\bs{m}| }  }
	\mc{S}^{(t)} \big(   \bs{\ga} + \i \bs{\delta}_\mc{S}    \big)\cdot \mc{R}[ \tau_t \cdot G ] \big(   \bs{\ga}  + \i \bs{\delta}_\mc{R}   \big) \cdot  \mc{F}_{\tau_t \cdot \bs{\a}_k }^{(t)}  \big(   \bs{\ga}  ; \bs{\eta}_1, \bs{\eta}_2   \big)  \;.
\enq
The above is exactly the representation for $\mc{W}_{\tau_t \cdot \bs{\a}_k}\big[  \tau_t\cdot G \big]$
that follows by summing up over $\bs{r}\in \mathbb{N}^{k-1}$ the result of Proposition \ref{proposition generale shifts pour causalite} with
the $t^{\e{th}}$ operator being singled out and our adequate choices of shifts $\bs{\delta}_{\mc{S}}$, $\bs{\delta}_{\mc{R}}$, $\bs{\eta}_1$ and $\bs{\eta}_2$, one gets the claim. \qed

\subsection{Positivity}

We now pass on to proving the positivity property:

\begin{prop}

Fix $N\in \mathbb{N}$ and pick $f^{(p)} \in \mc{S}\big( (\R^{1,1})^p \big)$, $p=0,\dots, N$. Let  $F\!_{p,q}\in \mc{S}\big( (\R^{1,1})^{p+q} \big)$ be as introduced
in \eqref{definition fct pour positive definiteness}. Further, pick labels $\big\{ \a_{ab} \big\}_{a\leq b}^{N}$ corresponding to solutions to \textit{a)-d)}.

If Conjecture \ref{Conjecture convergence} holds, the below linear combination of correlators is non-negative:
\begin{equation}
\sul{p,q=0}{N} \mc{W}_{\bs{\tau}_{p+q}}  [F_{p,q}] \, \geq\,  0 \qquad with \qquad  \bs{\tau}_{p+q} \, = \, \big( \alpha^{\dagger}_{pp},\dots,\alpha^{\dagger}_{1p}, \alpha_{1q},\dots,\alpha_{qq}  \big) \;.
\end{equation}

\end{prop}

\proof 

We start by writing a convenient representation for the quantity of interest. First of all, because of absolute convergence, it holds
\beq
 \mc{W}_{\bs{\tau}_{p+q}}  [F_{p,q}] \, = \, \lim_{R \tend + \infty} \lim_{M \tend + \infty} \sul{  \bs{n}_{p-1} \in \mc{I}_M^{p-1} }{  }   \sul{   \bs{m}_{q-1}\in \mc{I}_M^{q-1} }{  }  \sul{r=0}{R}
 \mc{W}_{\bs{\tau}_{p+q}}^{(\bs{r})}  [F_{p,q}] \qquad \e{where} \qquad \mc{I}_M \, = \,  \intn{0}{M} \;.
\label{ecriture correlateur comme limite se sommes finies}
\enq
Moreover, $\bs{r}$ arising above is to be understood as $\bs{r} \,  = \, \big( \overleftarrow{\bs{n}}_{p-1}, r , \bs{m}_{q-1} \big) \in \mathbb{N}^{p+q-1}$.
It is shown in Propositions 3.1, 3.2 and 4.7 of \cite{KozSimonMultiPointCorrFctsSinhGordon}, that the $\bs{r}$-truncated correlator $\mc{W}_{\bs{\tau}_{p+q}}^{(\bs{r})}  [F_{p,q}] $
admits the below, well-defined distributional integral representation -see equation (3.7) of that paper-
\bem
\mc{W}_{\bs{\tau}_{p+q}}^{(\bs{r})}  [F_{p,q}] = \lim_{\bs{\eta}_{p+q} \tend \bs{0}^+}  \pl{a=1}{p+q-1} \bigg\{  \Int{ \R^{r_a} }{} \hspace{-2mm} \f{ \dd^{r_a} \a_a }{ r_a! (2\pi)^{r_a} }  \bigg\}
\, \mc{M}^{(\tau_1)}_{0;r_1}\big(  \emptyset ; \bs{\a}_{1} \big)_{\eta_1}  \cdot \mc{M}^{(\tau_2)}_{r_1;r_2}\big(  \bs{\a}_1 ; \bs{\a}_{2} \big)_{\eta_2}  \\
\cdots \mc{M}^{(\tau_{p+q-1})}_{r_{p+q-2};r_{p+q-1}}\big(  \bs{\a}_{p+q-2} ; \bs{\a}_{p+q-1} \big)_{\eta_{p+q-1}} \cdot \mc{M}^{(\tau_{p+q})}_{r_{p+q-1};0}\big(  \bs{\a}_{p+q-1};  \emptyset  \big)_{ \eta_{p+q} }
  \cdot \msc{R}[F_{\! p,q} ]\Big( \{ \bs{\a}_a \}_{ a = 1 }^{ p + q - 1 }  \Big)\; .
\end{multline}
The order in which the $\bs{\a}_a$ integrals are taken is irrelevant and the $\eta_{k}\tend 0^+$ limit may be moved in front of the $\bs{\a}_k$ integration.
The first building block of this representation corresponds to the integral transform
\beq
 \msc{R}[F_{\! p,q} ]\Big( \{ \bs{\a}_a \}_{ a = 1 }^{ p + q - 1 }  \Big) \, = \hspace{-5mm} \Int{ \big( \R^{1,1} \big)^{p+q-1} }{} \hspace{-4mm} \dd^{p+q-1} \bs{Z}  \; F_{p,q}\big( \bs{Z}_{p+q-1} \big)
 \pl{s=1}{p+q-1} \Big\{  \ex{\i \ov{\bs{p}}( \bs{\a}_{s} ) * \bs{z}_{s+1s} }\Big\} \;.
\label{ecriture R momentum transform forme originale}
\enq
The remaining building blocks are given by the below distributions
\beq
\mc{M}^{(\alpha)}_{n;m}\big(  \bs{\lambda}_n ; \bs{\be}_{m} \big)_{\veps} \; = \hspace{-2mm}   \sul{ A= A_1 \cup A_2  }{}  \sul{ B = B_1 \underset{1}{\cup} B_2  }{}
\hspace{-2mm}  \De\big( \bs{A}_1 \mid \bs{B}_1  \big) \cdot \op{S}\Big( \, \overleftarrow{ \bs{A} }  \mid \overleftarrow{ \bs{A}_2 }\cup  \overleftarrow{ \bs{A}_1 } \Big)
\cdot \op{S}\big(  \bs{B} \mid \bs{B}_1\cup  \bs{B}_2 \big)
\cdot \mc{F}^{(\alpha)} \Big( \overleftarrow{ \bs{A}_2 } + \i \pi \ov{\bs{e}}_{\veps}, \bs{B}_2  \Big) \;.
\label{ecriture representation combinatoire pour MO via reduction de type 1}
\end{equation}
The expression is to be understood as follows. First of all, one associates with the vectors $\bs{\lambda}_n \in \R^{n}$, $\bs{\be}_{m} \in \R^m$
the sets $A \, = \, \{\la_a\}_1^n$ and $B\, = \, \{ \be_a\}_1^m$. One then sums over all partitions $A_1 \cup A_2$ of $A$ and all partitions $B_1 \underset{1}{\cup} B_2$  of $B$.
Moreover, the superscript $1$ in the partitions of $B$ indicates that one should, in addition, sum-up over all permutations of the elements of $B_1$.
These partitions are constrained  by the  condition $|A_1| \; = \; |B_1|$.  The formula uses for an ordered set
$C \, = \, \big\{ c_{k_a} \big\}_{a=1}^{|C|}$, $\bs{C}\, =\, \big( c_{k_1}, \dots, c_{k_{|C|}} \big)$.

Finally, given $\de_{x,y}$ the Dirac mass centred at $x=y$, we have set
\beq
\De\big( \bs{A}_1 \mid \bs{B}_1  \big) \; = \; \pl{a=1}{|A_1|} \Big\{ 2\pi \de_{ \la_{k_a},\,  \be_{i_a}} \Big\} \qquad \e{where} \qquad
\left\{\ba{ccccc}  A_1 & =& \big\{ \la_{k_a} \big\}_{a=1}^{|A_1|} &, &  k_1<\dots< k_{|A_1|} \vspace{2mm}  \\
B_1 &  = & \big\{ \be_{i_a} \big\}_{a=1}^{|B_1|}  & , &  i_1\not= \dots \not= i_{|B_1|}   \ea \right.  \; .
\label{definition facteur masse dirac globale}
\enq
Upon implementing the relabelling of the integration variables
\beq
\bs{\a}_a \, = \, \bs{\la}_{p-a} \qquad \e{for} \quad a\in \intn{1}{p-1}\,, \quad \bs{\a}_p  \, = \,  \bs{\vth} \, , \quad  \e{and} \quad
\bs{\a}_{p+a} \, = \, \bs{\be}_{a} \qquad \e{for} \quad a \in \intn{1}{q-1} \;,
\enq
changing the space-time integration variables in \eqref{ecriture R momentum transform forme originale} to $\bs{Z}_{p+q-1}=\big( \overleftarrow{\bs{X}}_p, \bs{Y}_q \big)$ and
reparameterising the regularisation parameters $\bs{\eta}_{p+q} \, = \, \big( \overleftarrow{\bs{\veps}}_{p}, \bs{\de}_q \big)$, the $\bs{r}$-truncated correlator
is recast as
\bem
 \mc{W}_{\bs{\tau}_{p+q}}^{(\bs{r})}  [F_{p,q}]  \, =  \Int{  \R^{r}   }{} \hspace{-1mm} \frac{\dd^{ r}  \bs{\vth}}{(2\pi)^{r}r!}
 \lim_{\bs{\veps_p} \tend 0^+} \prod_{s=1}^{p-1}\Bigg\{ \Int{  \hspace{1mm} \R^{n_{s}}   }{} \hspace{-1mm}  \frac{\dd^{ n_s}  \bs{\lambda}_{s}}{(2\pi)^{n_s}n_s!}  \Bigg\} \cdot
 \lim_{\bs{\delta_q} \tend 0^+}  \prod_{j=1}^{q-1} \Bigg\{  \Int{\hspace{1mm} \R^{m_{j}}   }{} \hspace{-1mm}  \frac{\dd^{ m_j}  \bs{\beta}_{j}}{(2\pi)^{m_j}m_j!}  \Bigg\}
 \cdot \hspace{-3mm}  \Int{ \big( \R^{1,1} \big)^{p} }{}  \hspace{-3mm} \dd^{p} \bs{X} \cdot \hspace{-3mm}  \Int{ \big( \R^{1,1} \big)^{q} }{}  \hspace{-3mm} \dd^q \bs{Y}  \\
\mc{M}^{(\alpha^{\dagger}_{pp})}_{0;n_{p-1}} \big( \emptyset ; \bs{\lambda}_{p-1} \big)_{\veps_p} \cdot \mc{M}^{(\alpha^{\dagger}_{p-1p})}_{n_{p-1};n_{p-2}} \big( \bs{\lambda}_{p-1} ; \bs{\lambda}_{p-2} \big)_{\veps_{p-1}}
\cdots \mc{M}^{(\alpha^{\dagger}_{2p})}_{n_{2};n_{1}} \big( \bs{\lambda}_{2} ; \bs{\lambda}_{1} \big)_{\veps_{2}} \cdot \mc{M}^{(\alpha^{\dagger}_{1p})}_{n_{1};r} \big( \bs{\lambda}_{1} ; \bs{\vth} \big)_{\veps_1}
\cdot \ov{ f^{(p)}\big(\bs{X}_p\big) }\\
\times \mc{M}^{(\alpha_{1q})}_{r;m_1} \big( \bs{\vth} ; \bs{\beta}_{1} \big)_{\delta_1} \cdot \mc{M}^{(\alpha_{2q})}_{m_1;m_2} \big( \bs{\beta}_{1} ; \bs{\beta}_{2} \big)_{\delta_2} \cdots
\mc{M}^{(\alpha_{q-1q})}_{m_{q-2};m_{q-1}} \big( \bs{\beta}_{q-2} ; \bs{\beta}_{q-1} \big)_{\delta_{q-1}} \cdot \mc{M}^{(\alpha_{qq})}_{m_{q-1};0} \big( \bs{\beta}_{q-1} ; \emptyset \big)_{\delta_{q}} \cdot
f^{(q)}\big( \bs{Y}_q \big)\\
\times \ex{ \i  ( \ov{\bs{p}}(\bs{\lambda}_1)-\ov{\bs{p}}(\bs{\gamma}) ) *  \bs{x}_{1}  } \cdot \prod_{s=2}^{p-1} \ex{ \i  ( \ov{\bs{p}}(\bs{\lambda}_s)-\ov{\bs{p}}(\bs{\lambda}_{s-1}) ) * \bs{x}_{s}  }
\cdot \ex{ -\i \ov{\bs{p}}(\bs{\lambda}_{p-1}) * \bs{x}_{p}  } \cdot \ex{ -\i ( \ov{\bs{p}}(\bs{\beta}_1)-\ov{\bs{p}}(\bs{\gamma})) * \bs{y}_{1}  } \cdot
\prod_{s=2}^{q-1} \ex{ -\i ( \ov{\bs{p}} (\bs{\beta}_s)-\ov{\bs{p}}(\bs{\beta}_{s-1}) ) * \bs{y}_{s}  } \cdot \ex{ \i \ov{\bs{p}}(\bs{\be}_{q-1}) * \bs{y}_{q}  } \; .
\label{equation positivity}
\end{multline}

By using the expression for the form factors of the adjoint operators \eqref{ecriture equation correspondance entre FF et celui de adjoint}, the elementary behaviour of
the $\op{S}$-factors under complex conjugation
\beq
\ov{ \op{S}\Big( \, \overleftarrow{ \bs{A} }  \mid \overleftarrow{ \bs{A}_2 }\cup  \overleftarrow{ \bs{A}_1 } \Big) } \; = \;  \op{S}\big(  \bs{A} \mid \bs{A}_1\cup  \bs{A}_2 \big)
\enq
the Lorentz boost property of the form factors and elementary properties of sums over partitions one infers from \eqref{ecriture representation combinatoire pour MO via reduction de type 1} that
\beq
\mc{M}^{(\a^{\dagger})}_{n;m} \big( \bs{\lambda}_n ; \bs{\beta}_m \big)_{\veps}  \,  = \,
  \overline{\mc{M}^{(\a)}_{m;n} \big(  \bs{\beta}_m ; \bs{\lambda}_n \big)_{\veps}} \cdot \ex{ - \i \veps \op{s}_{\a} } \;.
\label{epsilon matrix element adjoint}
\enq
By applying this relation and introducing
\bem
\msc{G}^{(\bs{n}_{p-1})}_{\bs{\a}_p}\big[ f^{(p)} \big]\big( \bs{\vth} \big)\, = \, \lim_{\bs{\veps_p} \tend 0^+}
\prod_{s=1}^{p-1}\Bigg\{ \Int{  \hspace{1mm} \R^{n_{s}}   }{} \hspace{-1mm}  \frac{\dd^{ n_s}  \bs{\lambda}_{s}}{(2\pi)^{n_s}n_s!}  \Bigg\}
 \Int{ \big( \R^{1,1} \big)^{p} }{}  \hspace{-3mm} \dd^{p} \bs{X}  \; f^{(p)}\big(\bs{X}_p\big)
\,  \ex{ \i  ( \ov{\bs{p}}(\bs{\lambda}_1)-\ov{\bs{p}}(\bs{\gamma}) ) *  \bs{x}_{1}  }\\
\times \prod_{s=2}^{p-1} \ex{ \i  ( \ov{\bs{p}}(\bs{\lambda}_s)-\ov{\bs{p}}(\bs{\lambda}_{s-1}) ) * \bs{x}_{s}  }
\cdot \ex{ -\i \ov{\bs{p}}(\bs{\lambda}_{p-1}) * \bs{x}_{p}  } \cdot
\mc{M}^{(\a_{1p})}_{r;n_1} \big( \bs{\vth} ; \bs{\la}_{1} \big)_{\veps_1} \cdots \mc{M}^{(\a_{pp})}_{n_{p-1};0} \big( \bs{\la}_{p-1} ; \emptyset \big)_{\veps_{q}}
\end{multline}
one gets that
\beq
 \mc{W}_{\bs{\tau}_{p+q}}^{(\bs{r})}  [F_{\! p,q}]  \, =  \Int{  \R^{r}   }{} \hspace{-1mm} \frac{\dd^{ r}  \bs{\vth}}{(2\pi)^{r}r!}
\ov{ \msc{G}^{(\bs{n}_{p-1})}_{\bs{\a}_p}\big[ f^{(p)} \big]\big( \bs{\vth} \big) } \cdot
\msc{G}^{(\bs{m}_{q-1})}_{\bs{\a}_p}\big[ f^{(q)} \big]\big( \bs{\vth} \big) \;.
\enq
Note that the factors $\ov{ \msc{G}^{(\bs{n}_{p-1})}_{\bs{\a}_p}\big[ f^{(p)} \big]\big( \bs{\vth} \big) } $ decay as $\bs{\vth}\tend \infty$
faster than any polynomial in $\cosh(\vth_1), \dots, \cosh(\vth_r)$ as can be inferred from direct estimates of the momentum transform $\mc{R}$
and the calculation of the Dirac mass distributions as well as the taking of the $\bs{\veps}_p\tend \bs{0}^+$ limits.
We refer to \cite{KozSimonMultiPointCorrFctsSinhGordon} for further details.

We are now in position to insert this result into \eqref{ecriture correlateur comme limite se sommes finies}, what gives
\beq
 \sul{p,q=0}{N} \mc{W}_{\bs{\tau}_{p+q}}  [F_{ \! p,q}] \, = \, \lim_{R \tend + \infty} \lim_{M \tend + \infty}
  \Int{  \R^{r}   }{} \hspace{-1mm} \frac{\dd^{ r}  \bs{\vth}}{(2\pi)^{r}r!}
\big| \msc{I}^{(M)}_{r}[f](\bs{\vth})\big|^2
\enq
where
\beq
\msc{I}^{(M)}_{r}[f](\bs{\vth}) \, = \, \sul{p=0}{N} \sul{  \bs{n}_{p-1} \in \mc{I}_M^{p-1} }{  } \msc{G}^{(\bs{n}_{p-1})}_{\bs{\a}_p}\big[ f^{(p)} \big]\big( \bs{\vth} \big)
\enq
and $\mc{I}_M$ is as defined in \eqref{ecriture correlateur comme limite se sommes finies}.

On the left hand side, the result is well-defined by assumption. On the right hand side, one takes the limit of a strictly increasing sequence. This limit is thus well-defined,
and one gets that the sum in the left hand side is non-negative. \qed

\subsection{Cluster Decomposition}

We finally check the cluster decomposition axiom which translates the loss of correlations at sufficiently large space-like interval separations.
The proof relies on a few auxiliary lemmas.

\begin{lemme}
\label{split distrib lemma}
Pick $p,q\in \mathbb{N}^*$. Let $(f,g) \in \mc{S}\big( (\R^{1,1})^p \big) \times \mc{S}\big( (\R^{1,1})^q \big)$, $\bs{v} \in \R^{1,1}$ and $G_{p,q}^{(\la \bs{v})}$ as introduced in \eqref{definition Gpq}.
Let $\bs{\a}_p$, resp. $\bs{\be}_q$, be $p$, resp. $q$, dimensional vectors of operators labels corresponding to solutions to $a)-d)$ and
\beq
\bs{\ell} =\big( \ell_{21}, \ell_{31},\dots, \ell_{p+q\, p+q -1} \big)\in \mathbb{N}^{ (p+q) \f{p+q-1}{2} } \qquad be\; such\; that \quad
\ell_{dc} \, =\,  0 \quad for \; all \quad  d \, > \,  p \, \geq\,  c \, .
\label{definition vecteur ell dans split lemma}
\enq
Then with $\bs{\tau}_{p+q} \, = \,  \big( \bs{\a}_p, \bs{\be}_q \big)$, one has the factorisation
\begin{equation}
\mc{I}^{ (\bs{\ell}) }_{ \bs{\tau}_{p+q} }\big[ G_{p,q}^{(\la \bs{v})}\big] \, =  \, \mc{I}^{ (\bs{n}) }_{ \bs{\a}_{p} }\big[ f \big] \cdot   \mc{I}^{ (\bs{m}) }_{ \bs{\be}_{q} }\big[ g \big] \;.
\label{split distribution}
\end{equation}
The factorising distributions are labelled by the vector integers
	\begin{equation}
		\bs{n} \,  =  \, \big( \ell_{21}, \ell_{31}, \ell_{32}, \dots, \ell_{pp-1} \big)
		\qquad and \qquad 	\bs{m} \,  =  \, \big( \ell_{p+2\hspace{0.5mm}p+1}, \ell_{p+3\hspace{0.5mm}p+1}, \ell_{p+3\hspace{0.5mm}p+2}, \dots, \ell_{p+q\hspace{0.5mm} p+q-1} \big)  \;.
	\end{equation}
\end{lemme}

\Proof

Taken that $\bs{\ga}^{(ba)} \in \R^{\ell_{ba}}$, there are no integration variables $\bs{\ga}^{(ba)}$ such that $b \geq p+1$ and $a \leq p$
in the integral representation \eqref{definition sommant base dans serie int mult pour fct 2 pts} for
$\mc{I}^{ (\bs{\ell}) }_{ \bs{\tau}_{p+q} }\big[ G_{p,q}^{(\la \bs{v})}\big] $ as soon as $\bs{\ell}$ is as in \eqref{definition vecteur ell dans split lemma}.
Thus, one has that
\beq
\sul{b>a}{p+q} \ov{\bs{p}}\big( \bs{\ga}^{(ba)} \big) * \bs{z}_{ba} \, = \, \sul{b>a}{p} \ov{\bs{p}}\big( \bs{\ga}^{(ba)} \big) * \bs{z}_{ba}
\, + \, \sul{  \substack{ b> a  \\ p+1} }{p+q} \ov{\bs{p}}\big( \bs{\ga}^{(ba)} \big) * \bs{z}_{ba}  \;.
\enq
Thus, by setting
\beq
\bs{\la}\, = \, \big( \bs{\la}^{(21)}, \bs{\la}^{(31)}, \dots, \bs{\la}^{(pp-1)}   \big) \quad \e{with} \quad \bs{\la}^{(ba)} \, = \, \bs{\ga}^{(ba)}  \quad 1 \leq a < b \leq p
\enq
and
\beq
\bs{\vth}\, = \, \big( \bs{\vth}^{(21)}, \bs{\vth}^{(31)}, \dots, \bs{\vth}^{(qq-1)}   \big) \quad \e{with} \quad \bs{\vth}^{(ba)} \, = \, \bs{\ga}^{(b+p \, a+p)}  \quad 1 \leq a < b \leq q
\enq
and reparameterising $\bs{Z}_{p+q} \, = \, \big( \bs{X}_p, \bs{Y}_q + \la \bs{v}_q \big)$ with $\bs{v}_q=\big(\bs{v}, \dots, \bs{v} \big)\in \big( \R^{1,1} \big)^q$, one thus gets
\beq
\mc{R}\big[ G_{p,q}^{(\la \bs{v})}\big] \big(   \bs{\ga}    \big) \,  = \hspace{-2mm} \Int{ \big( \R^{1,1} \big)^{p} }{}  \hspace{-3mm} \dd^p \bs{X}
\hspace{-3mm} \Int{ \big( \R^{1,1} \big)^{q} }{}  \hspace{-3mm} \dd^q \bs{Y}
f\big(\bs{X}_p\big)  \cdot g\big(\bs{Y}_q \big)  \;  \pl{b>a}{p}  \Big\{ \ex{ \i \ov{\bs{p}}(  \bs{\la}^{(ba)} ) * \bs{x}_{ba}  } \Big\}
\cdot \pl{b>a}{q} \Big\{ \ex{ \i \ov{\bs{p}}(  \bs{\vth}^{(ba)} ) * \bs{y}_{ba}  } \Big\}
\, = \, \mc{R}\big[f \big] \big(   \bs{\la}    \big) \cdot  \mc{R}\big[g \big] \big(   \bs{\vth}    \big)    \;.
\enq
Further, the product of $\op{S}$-matrices given in \eqref{definition facteur de diffusion complet k pts} may be presented as
\beq
\mc{S}\big(\bs{\ga}\big) \; =  \hspace{-2mm}
\pl{  \substack{ v >r >u>s} }{ p+q } \hspace{-2mm}
\op{S}\big(  \bs{\ga}^{(vu)}  \cup  \bs{\ga}^{(rs)}   \mid   \bs{\ga}^{(rs)}   \cup \bs{\ga}^{(vu)}   \big)   \;.
\enq
A given $\op{S}$-matrix factor will be present if both $\bs{\ga}^{(vu)}$ and $\bs{\ga}^{(rs)} $ are non trivial.
$\bs{\ga}^{(vu)}$ is non-trivial if, either, $p \geq v$ or $u \geq p+1$. In the first case one necessarily has
$p \geq  r > s $. In the second case scenario, one necessarily has $r > p+1$ so that  $\bs{\ga}^{(rs)} $ will be non-trivial
provided that $s \geq p+1$. This leads to the factorisation
\bem
\mc{S}\big(\bs{\ga}\big)  =  \hspace{-2mm} \pl{  \substack{ v >r >u>s} }{ p  }  \hspace{-2mm}
\op{S}\big(  \bs{\ga}^{(vu)}  \cup  \bs{\ga}^{(rs)}   \mid   \bs{\ga}^{(rs)}   \cup \bs{\ga}^{(vu)}   \big)
\, \cdot   \hspace{-3mm}
\pl{  \substack{ v >r >u>s  \\ \geq p+1} }{ p+q } \hspace{-2mm}
\op{S}\big(  \bs{\ga}^{(vu)}  \cup  \bs{\ga}^{(rs)}   \mid   \bs{\ga}^{(rs)}   \cup \bs{\ga}^{(vu)}   \big)
\, = \,\mc{S}\big(\bs{\la}\big) \cdot \mc{S}\big(\bs{\vth}\big) \;.
\end{multline}
Finally, we focus on the building blocks of $\mc{F}_{\bs{\tau}_{p+q};\bs{\eta}_{p+q}}(\bs{\ga})$ while reparameterising $\bs{\eta}_{p+q} \, = \, \big(  \bs{\veps}_{p} , \bs{\de}_q \big)$.
Then, for $s \leq p$, one has:
\beqa
\msc{F}_s( \bs{\ga} ) & \equiv & \mc{F}^{(\tau_s)}\Big( \, \overleftarrow{ \bs{\ga}^{(ss-1)} } \cup \cdots \cup \overleftarrow{ \bs{\ga}^{(s1)} }  + \i \pi \ov{\bs{e}}_{\eta_{s}} ,
\bs{\ga}^{(p+qs)}  \cup \cdots \cup \bs{\ga}^{(s+1s)}   \, \Big) \\
\hspace{1.5cm} & = &  \mc{F}^{(\alpha_s)}\Big( \, \overleftarrow{ \bs{\ga}^{(ss-1)} } \cup \cdots \cup \overleftarrow{ \bs{\ga}^{(s1)} }  + \i \pi \ov{\bs{e}}_{\veps_{s}} ,
\bs{\ga}^{(ps)}  \cup \cdots \cup \bs{\ga}^{(s+1s)}   \, \Big)  \\
&= &  \mc{F}^{(\alpha_s)}\Big( \, \overleftarrow{ \bs{\la}^{(ss-1)} } \cup \cdots \cup \overleftarrow{ \bs{\la}^{(s1)} }  + \i \pi \ov{\bs{e}}_{\veps_{s}} ,
\bs{\la}^{(ps)}  \cup \cdots \cup \bs{\la}^{(s+1s)}   \, \Big)  \;.
\eeqa
Further, for $s=u+p$ with  $u \in \intn{1}{q}$, one rather has
\beqa
\msc{F}_s( \bs{\ga} ) & = &  \mc{F}^{(\be_u)}\Big( \, \overleftarrow{ \bs{\ga}^{(u+p \, u+p-1)} } \cup \cdots \cup \overleftarrow{ \bs{\ga}^{(u+p \, p+1)} }  + \i \pi \ov{\bs{e}}_{\de_{u}} ,
\bs{\ga}^{(p+q \, u+p)}  \cup \cdots \cup \bs{\ga}^{(p+u+1 \, p+u)}   \, \Big) \\
& = &  \mc{F}^{(\be_u)}\Big( \, \overleftarrow{ \bs{\vth}^{(u u-1)} } \cup \cdots \cup \overleftarrow{ \bs{\vth}^{(u1)} }  + \i \pi \ov{\bs{e}}_{\de_{u}} ,
\bs{\vth}^{(q u)}  \cup \cdots \cup \bs{\vth}^{(u+1 u)}   \, \Big) \;.
\eeqa
This thus leads to the factorisation of the form factor contribution
\beq
\mc{F}_{\bs{\tau}_{p+q},\bs{\eta}_{p+q}}(\bs{\ga}) \, = \, \mc{F}_{\bs{\a}_{p},\bs{\veps}_{p}}(\bs{\la}) \cdot \mc{F}_{\bs{\be}_{q},\bs{\de}_{q}}(\bs{\vth}) \;.
\enq
The factorisation of the integrand that we have just demonstrated ensures the claimed factorisation of the integral transform  we started with.  \qed

\vspace{4mm}

\begin{lemme}
	\label{Lemme support distributions}
Let $p,q\in \mathbb{N}^*$, $(f,g) \in \mc{S}\big( (\R^{1,1})^p \big) \times \mc{S}\big( (\R^{1,1})^q \big) $, $\bs{v} \in \R^{1,1}$ and set $G_{p,q}^{(\la \bs{v})}$ as introduced in \eqref{definition Gpq}.
Let $\bs{\a}_p$, resp. $\bs{\be}_q$, be $p$, resp. $q$, dimensional vectors of operators labels corresponding to solutions to $a)-d)$ and $\bs{\tau}_{p+q} \, = \,  \big( \bs{\a}_p, \bs{\be}_q \big)$
be their concatenation. Let $\sg \in \mf{S}_{p+q}$ refer to the below permutation of the coordinates
\beq
\sg \cdot (\bs{X}_p,\bs{Y}_q) = (\bs{Y}_q,\bs{X}_p) \;.
\label{permutation des coordonnee Xp Yq sigma}
\enq
Assume Conjecture \ref{Conjecture convergence} holds. Then, the distributions
\begin{equation}
\mf{M}^{(\la\bs{v})}[f,g]  \, = \, \mc{W}_{\bs{\tau}_{p+q}}[G_{p,q}^{(\la \bs{v})}]  \, -  \, \mc{W}_{\bs{\alpha}_p}[f] \cdot \mc{W}_{\bs{\be}_q}[g]
\label{definition difference distrib}
\end{equation}
and
\begin{equation}
\mf{M}^{(\la\bs{v})}_{\sg}[f,g] \,  =  \, \mc{W}_{\sg \cdot \bs{\tau}_{p+q}}[\sg \cdot G_{p,q}^{(\la \bs{v})}] \, -  \, \mc{W}_{\bs{\be}_q}[g] \cdot \mc{W}_{\bs{\alpha}_p}[f]  \;,
\label{definition difference distrib sigma permutee}
\end{equation}
where the action of the symmetric group on $\mc{S}\big( (\R^{1,1})^{p+q} \big)$ has been introduced in \eqref{ecriture action permutation generale sur fonction},
can be recast as
\beq
\mf{M}^{(\la\bs{v})}[f,g]  \, = \, \mf{N}^{(\la\bs{v})}\big[ \mc{L}[ G_{p,q}^{ (\bs{0})} ] \big]  \qquad and  \qquad
\mf{M}^{(\la\bs{v})}_{\sg}[f,g]  \, = \, \mf{N}^{(\la\bs{v})}_{\sg}\big[ \mc{L}[ G_{p,q}^{ (\bs{0})} ] \big] \;.
\label{reecriture distribution M et Msg via dist N et N sg}
\enq
The operator $\mc{L}$ is as introduced in \eqref{definition linear map L}, while the auxiliary distributions $\mf{N}^{(\la\bs{v})}$, $\mf{N}^{(\la\bs{v})}_{\sg}$ are
defined by means of a summation over the subset of $\mathbb{N}^{\f{ (p+q)(p+q-1)}{2}}$:
\beq
\mc{T}^{(p)} \,= \, \bigg\{ \bs{n} \in \mathbb{N}^{\f{ (p+q)(p+q-1)}{2}} \; : \; \sul{b=p+1}{p+q}  \sul{a=1}{p} n_{ba} \, \geq \, 1 \bigg\} \;,
\enq
as follows
\beq
\mf{N}^{(\la \bs{v})}[H] \, = \, \sul{ \bs{n} \in \mc{T}^{(p)} }{} \mc{J}^{(\bs{n};p)}_{\bs{\tau}_{p+q},  \la \bs{v} }\big[H]
\qquad  and \qquad
\mf{N}^{(\la \bs{v})}_{\sg} [H] \, = \, \sul{ \bs{n} \in \mc{T}^{(q)} }{} \mc{J}^{(\bs{n};q)}_{\sg \cdot \bs{\tau}_{p+q},  -\la \bs{v} }\big[H].
\enq
There, we agree upon
\beq
 \mc{J}^{(\bs{n};p)}_{\bs{\tau}_{p+q},  \la \bs{v} }\big[H] \, = \,  \lim_{\bs{\veps}_{p+q} \tend 0^+} \Int{  \R^{ |\bs{n}| }   }{} \hspace{-1mm} \f{ \dd^{ |\bs{n}| } \bs{\ga}  }{ \bs{n}! (2\pi)^{|\bs{n}|}} \,
\ex{-\i\la \bs{v} * \op{P}_{p}(\bs{\ga}) }
\Big( \mc{S} \cdot \widehat{\mc{R}}[H] \cdot \mc{F}_{\bs{\tau}_{p+q};\bs{\veps}_{p+q} } \Big) \big(   \bs{\ga}     \big) \; .
\enq
Above, $\op{P}_{p}$ is as introduced in \eqref{definition impulsion reduite}, the $\wh{\mc{R}}$ transform has been defined in \eqref{definition TF space-time de la fct test differentielle}
while $\mc{S}$ and $\mc{F}_{\bs{\tau}_{p+q};\bs{\veps}_{p+q} }$ have been, respectively, introduced in \eqref{definition facteur de diffusion complet k pts} and \eqref{definition F tot eps}.

$\mf{N}^{(\la\bs{v})}$, $\mf{N}^{(\la\bs{v})}_{\sg}$ are well-defined tempered distributions on $\mc{S}\big( (\R^{1,1})^{p+q-1}  \big)$. Their Fourier transforms are supported
on  forward or backward lightcones:
\begin{equation}
\mf{N}^{(\la\bs{v})}\big[ \msc{F}[H] \big]   \, =  \, 0 \qquad if \qquad \e{supp}\big[ H \big] \cap \mc{E}_{p+q-1, p}^{+, +} \, = \, \emptyset \;,
\label{equation vanishing support N lambda v}
\end{equation}
where given $\bs{Q}_{k-1}\, = \, (\bs{q}_1,\dots,\bs{q}_{k-1}) \in (\R^{1,1})^{k-1}$, one has
\begin{equation}
\mc{E}_{p-1, p}^{+, +} \, = \, \left\{ \bs{Q}_{k-1}\in (\R^{1,1})^{k-1} :  \ba{cc} q_{p,0} \geq m \, , & \bs{q}_{p}^2 \geq m^2  \vspace{1mm}  \\
												 q_{\ell,0} \geq 0\;,  &  \bs{q}_{\ell}^2 \geq 0 \ea  \quad \textrm{for} \quad \ell \in \intn{1}{k-1} \setminus \{p \}   \right\} \;.
\label{definition support masse}
\end{equation}
Similarly
\begin{equation}
\mf{N}^{(\la\bs{v})}_{\sg}\big[ \msc{F}[H] \big]   \, =  \, 0 \qquad if  \qquad \e{supp}\big[ H \big] \cap \mc{E}_{p+q-1, p}^{+, -} \, = \, \emptyset \;,
\label{equation vanishing support N sigma lambda v}
\end{equation}
where we set
\begin{equation}
	\mc{E}_{k-1, p}^{+, -} = \, \Big\{ \, \bs{Q}_{k-1} \in (\R^{1,1})^{k-1} :  \quad q_{p,0} \leq -m \, , \quad \bs{q}_{p}^2 \geq m^2  \Big\} \;.
\label{definition support masse 2}
\end{equation}
\end{lemme}

\Proof 

The fact that $\mc{J}^{(\bs{n};p)}_{\bs{\tau}_{p+q},  \la \bs{v} }$, $\mf{N}^{(\la\bs{v})}$ and $\mf{N}^{(\la\bs{v})}_{\sg}$ are all
well-defined follows from similar handlings as those outlined in Proposition \ref{prop spectral conditions}.

\vspace{2mm}

We first focus on the distribution $\mf{M}$. Starting from the expansion \eqref{full corr func} applied to $\mc{W}_{\bs{\tau}_{p+q}}[G_{p,q}^{(\la \bs{v})}]$, $\mc{W}_{\bs{\alpha}_p}[f]$, $\mc{W}_{\bs{\be}_q}[g]$
taking the product of the series and recalling Lemma \ref{split distrib lemma} which ensures the factorisation \eqref{split distribution} provided that condition \eqref{definition vecteur ell dans split lemma} holds,
one gets that when computing $\mf{M}^{(\la\bs{v})}[f,g]$ through series expansions, only the complement of the condition \eqref{definition vecteur ell dans split lemma} contributes, thus leading to
\beq
\mf{M}^{(\la\bs{v})}[f,g]  \, = \,  \sul{ \bs{n} \in \mc{T}^{(p)}}{}   \mc{I}_{ \bs{\tau}_{p+q} }^{ (\bs{n} )}\big[ G^{(\la\bs{v})}_{p,q}\big]  \, =\,
\sul{ \bs{n} \in \mc{T}^{(p)}}{} \mc{J}^{ (\bs{n}) }_{\bs{\tau}_{p+q}}\big[ \mc{L} [G^{(\la\bs{v})}_{p,q}]\big]
	\label{representation somme moins origine}
\enq
where the second equality follows from \eqref{ecriture distribution I en terme de distr J reduite}, and $\mc{J}^{ (\bs{n}) }_{\bs{\tau}_{p+q}}$ is as defined in \eqref{definition distrib diff variables}.
Further, by implementing the change of variables $\bs{y}_{s} \hookrightarrow \bs{y}_{s} + \la \bs{v} \big(\de_{s,p} - \de_{s,p+q} \big)$ for $s\in \intn{1}{p+q}$, one infers that
\begin{equation}
	\wh{\mc{R}}\big[ \mc{L} [G^{(\la\bs{v})}_{p,q}]\big] \big(   \bs{\ga}    \big) \, =  \,
	\ex{ -\i\la \bs{v} * \op{P}_{p}( \bs{\ga})  } \cdot \wh{\mc{R}}\big[ \mc{L} [G^{(\bs{0})}_{p,q}] \big] \big(   \bs{\ga}    \big) \;.
\label{ecriture factorisation la v dependence dans R hat}
\end{equation}
This leads to the identity $\mc{J}^{ (\bs{n}) }_{\bs{\tau}_{p+q}}\big[ \mc{L} [G^{(\la\bs{v})}_{p,q}]\big] \, = \, \mc{J}^{ (\bs{n};p) }_{ \bs{\tau}_{p+q}, \la \bs{v} }\big[ \mc{L} [G^{(\bs{0})}_{p,q}]\big]$
which, along with \eqref{representation somme moins origine}, yields the first equality given in \eqref{reecriture distribution M et Msg via dist N et N sg}.

Further, owing to \eqref{reecriture R hat transform Fourier}, one infers that  $\mc{J}^{ (\bs{n};p) }_{ \bs{\tau}_{p+q}, \la \bs{v} }\big[  \msc{F}[H] \big] $
will be non zero only if $\e{supp}[H]$ intersects the range of $\bs{\ga} \mapsto \big(  \op{P}_{1}( \bs{\ga}), \dots,  \op{P}_{p+q-1}( \bs{\ga} )  \big)$.
Now, recalling the expression for $\op{P}_{s}(\bs{\ga})$ \eqref{definition impulsion reduite}, one readily infers that
\beq
\op{P}_{p}^{(0)}(\bs{\ga}) \, = \, m \sul{ b = p + 1 }{ p + q } \sul{ a=1 }{ p } \sul{ s=1 }{ n_{ba} } \cosh\big( \ga_{s}^{(ba)} \big) \, \geq \, m \sul{ b = p + 1 }{ p + q } \sul{ a=1 }{ p }   n_{ba}  \;,
\label{ecriture valeurs de P ell 0}
\enq
as well as
\beq
\op{P}^2_p\big(\bs{\ga}\big)  \, = \, m^2 \sul{ b = p + 1 }{ p + q } \sul{ a=1 }{ p }   n_{ba}
\, + \, m^2 \sul{  \substack{ b, b^{\prime} \\  = p+1 } }{p+q} \sul{a,a^{\prime}=1}{p} \sul{ s=1 }{ n_{ba} } \sul{ s^{\prime}=1 }{ n_{b^{\prime}a^{\prime} } }
\cosh\big(  \ga_{s}^{(ba)} \, - \, \ga_{s^{\prime}}^{(b^{\prime}a^{\prime})}  \big)  \geq m^2\;,
\label{ecriture valeurs de P ell square}
\enq
for any $\bs{n} \in \mc{T}^{(p)}$. This entails the claim relatively to $\mf{N}^{(\la \bs{v})}$, \textit{viz}. \eqref{equation vanishing support N lambda v}.

\vspace{2mm}

We now move on to proving the analogous properties of the distribution $\mf{M}^{(\la\bs{v})}_{\sg}$. Similar reasoning ensure that
\beq
\mf{M}^{(\la\bs{v})}_{\sg}[f,g] = \sul{ \bs{n} \in \mc{T}^{(p)}}{}   \mc{I}_{\sg \cdot \bs{\tau}_{p+q}}^{ (\bs{n} )}\big[ \sg \cdot G^{(\la\bs{v})}_{p,q}\big]
\, =  \, \sul{ \bs{n} \in \mc{T}^{(p)}}{} \mc{J}^{ (\bs{n}) }_{\sg \cdot \bs{\tau}_{p+q}}\big[ \mc{L}  [ \sg \cdot  G_{p,q}^{(\la \bs{v})} ] \big] \;.
\enq
At this stage, one observes that
\bem
\mc{R}\big[   \sg \cdot G_{p,q}^{(\la \bs{v})} \big] \big(   \bs{\ga}    \big) \, = \hspace{-3mm}  \Int{ \big( \R^{1,1} \big)^{p+q} }{}  \hspace{-4mm} \dd^p \bs{X} \cdot \dd^q \bs{Y}
 \;  G_{p,q}^{(\la \bs{v})}\big( \bs{X}_p, \bs{Y}_q \big)
\, \pl{b>a}{q} \Big\{ \ex{ \i \ov{\bs{p}}(  \bs{\ga}^{(b  a)} ) * \bs{y}_{ba}  } \Big\} \, \pl{b>a}{p}  \Big\{ \ex{ \i \ov{\bs{p}}(  \bs{\ga}^{(b+q\, a+q)} ) * \bs{x}_{ba}  } \Big\} \\
 \times  \pl{b=1}{p} \pl{a=1}{q} \Big\{ \ex{ \i \ov{\bs{p}}(  \bs{\ga}^{(b+q \, a)} ) * (\bs{x}_b - \bs{y}_a)  } \Big\} \;.
\end{multline}
We now proceed with the relabelling $\bs{Z}_{p+q} \, = \, \big( \bs{X}_p, \bs{Y}_q \big)$, which yields
\bem
	\mc{R}\big[   \sg \cdot G_{p,q}^{(\la \bs{v})} \big] \big(   \bs{\ga}    \big) \, =\hspace{-3mm}  \Int{ \big( \R^{1,1} \big)^{p+q} }{}  \hspace{-4mm}\dd^{p+q} \bs{Z}
\, G_{p,q}^{(\la \bs{v})}\big( \bs{Z}_{p+q} \big) \, \pl{  \substack{ b>a   \\ \geq p+1 } }{p+q}  \Big\{ \ex{ \i \ov{\bs{p}}(  \bs{\ga}^{(b-p \, a-p)} ) * \bs{z}_{ba}  }  \Big\} \,
\pl{b>a}{p}  \Big\{ \ex{ \i \ov{\bs{p}}(  \bs{\ga}^{(b+q \, a+q)} ) * \bs{z}_{ba}  } \Big\} \\
\times \pl{b=1}{p} \pl{a=1+p}{q+p} \Big\{   \ex{ \i \ov{\bs{p}}(  \bs{\ga}^{(b+q \, a-p)}  + \i \pi \ov{\bs{e}}) * \bs{z}_{ab}  } \Big\} \;.
\end{multline}
Thus, by introducing the change of variables $\bs{\ga} \mapsto \bs{\Ga}(\bs{\ga})$ defined as:
\beq
\ba{ccccc}
 \bs{\Ga}^{(ba)}(\bs{\ga}) & = &  \bs{\ga}^{(b+q \, a+q)}  & ,    &  1 \leq a < b \leq p  \vspace{1mm} \\
 \bs{\Ga}^{(ba)}(\bs{\ga}) & = & \bs{\ga}^{(a+q \, b-p)} + \i\pi \ov{\bs{e}}  & ,  & 1\leq  a  < 1+p  \leq  b \leq p+q   \vspace{1mm} \\
 \bs{\Ga}^{(ba)}(\bs{\ga}) & = & \bs{\ga}^{(b-p \, a-p)}  & ,  & p+1 \leq a  < b \leq p+q  \ea
\enq
one gets the chain of equalities
\beq
\mc{R}\big[   \sg \cdot G_{p,q}^{(\la \bs{v})} \big] \big(   \bs{\ga}    \big) \,  =  \, \mc{R}\big[ G_{p,q}^{(\la \bs{v})} \big] \big(   \bs{\Ga}(\bs{\ga})   \big)
\, = \, \wh{\mc{R}}\big[ \mc{L}  [ \sg \cdot  G_{p,q}^{(\la \bs{v})} ]  \big] \big(   \bs{\Ga}(\bs{\ga})     \big) \, =  \,
\ex{ -\i\la \bs{v} * \op{P}_{p}(  \bs{\Ga}(\bs{\ga}) )  } \cdot \wh{\mc{R}}\big[  \mc{L} \cdot G_{p,q}^{(\bs{0})} \big] \big(   \bs{\Ga}(\bs{\ga})     \big) \;,
\enq
where we invoked \eqref{ecriture factorisation la v dependence dans R hat}. Then, observing through direct calculations that
$ \op{P}_{p}(  \bs{\Ga}(\bs{\ga}) ) \, = \, - \op{P}_{q}(  \bs{\ga} )$, one eventually gets that
\beq
 \mc{J}^{ (\bs{n}) }_{\sg \cdot \bs{\tau}_{p+q}}\big[ \mc{L}  [ \sg \cdot  G_{p,q}^{(\la \bs{v})} ] \big]  \; = \;
 \mc{J}^{ (\bs{n};q) }_{\sg \cdot \bs{\tau}_{p+q}, -\la \bs{v} }\big[ \mc{L}  [ \sg \cdot  G_{p,q}^{(\bs{0})} ] \big]  \;.
\enq
This entails  the second equality given in \eqref{reecriture distribution M et Msg via dist N et N sg}.

Again, the representation \eqref{reecriture R hat transform Fourier} implies that  $\mc{J}^{ (\bs{n};q) }_{ \bs{\tau}_{p+q}, -\la \bs{v} }\big[  \msc{F}[H] \big] $
will be non zero only if $\e{supp}[H]$ intersects the range of $\bs{\ga} \mapsto \big(  \op{P}_{1} \big(   \bs{\Ga}(\bs{\ga})     \big), \dots,  \op{P}_{p+q-1} \big(   \bs{\Ga}(\bs{\ga})     \big)  \big)$.
The identity $ \op{P}_{p}(  \bs{\Ga}(\bs{\ga}) ) \, = \, - \op{P}_{q}(  \bs{\ga} )$ and equations \eqref{ecriture valeurs de P ell 0}-\eqref{ecriture valeurs de P ell square}
thus ensure \eqref{equation vanishing support N sigma lambda v}. \qed

\begin{prop}
Let $p,q\in \mathbb{N}^*$, $(f,g) \in \mc{S}\big( (\R^{1,1})^p \big) \times \mc{S}\big( (\R^{1,1})^q \big) $, $\bs{v} \in \R^{1,1}$ and let $G_{p,q}^{(\la \bs{v})}$ be as introduced in \eqref{definition Gpq}.
Let $\bs{\a}_p$, resp. $\bs{\be}_q$, be $p$, resp. $q$, dimensional vectors of operators labels corresponding to solutions to $a)-d)$.
Assume that Conjecture \ref{Conjecture convergence} is valid. Given $\bs{\tau}_{p+q} \, = \,  \big( \bs{\a}_p, \bs{\be}_q \big)$, it holds
\beq
\mc{W}_{ \bs{\tau}_{p+q} }[ G_{p,q}^{(\la \bs{v})} ]  \underset{\lambda \rightarrow \pm \infty}{\longrightarrow} \mc{W}_{\bs{\alpha}_p}[f] \cdot \mc{W}_{\bs{\be}_q}[g]  \;.
\enq
\end{prop}

\Proof The proof follows closely the chain of reasoning outlined in \cite{StreaterWightmanPCTStatisticsAllThat}.
By Conjecture \ref{Conjecture convergence}, $\mc{W}_{ \bs{\tau}_{p+q} }$ is a tempered distribution on $\mc{S}\big( (\R^{1,1})^{p+q} \big)$.
By the structure theorem for tempered distributions (see \textit{e.g.} Theorem 8.3.1 in \cite{FriedlanderIntroDistributions}), there exists $\bs{\xi}_{p+q}\in \mathbb{N}^{p+q}$,
a polynomially bounded continuous function $\Phi$ on $(\R^{1,1})^{p+q}$, \textit{i.e.} for some $C, M>0$
\beq
\big| \Phi\big(\bs{Z}_{p+q}\big) \big| \, \leq \, C \Big( 1+\norm{ \bs{Z}_{p+q} } \Big)^M \qquad \e{with} \qquad
\norm{ \bs{Z}_{p+q} }^2 \, = \, \sul{s=1}{p+q} \big( z_{s;0}^2\, + \, z_{s;1}^2 \big)
\label{ecriture controle croissance facteur distribution temperee}
\enq
referring to the  Euclidean norm of the vector, such that
\beq
\mc{W}_{ \bs{\tau}_{p+q} }[F] \, =  \hspace{-4mm}  \Int{ (\R^{1,1})^{p+q} }{} \hspace{-4mm} \dd^{p+q} \bs{Z} \,   \Phi\big(\bs{Z}_{p+q}\big) \, \Dp{ \bs{Z}_{p+q} }^{\bs{\xi}_{p+q}} F(\bs{Z}_{p+q})   \;.
\enq
Here, it is understood that $\bs{\xi}_{p+q}\, =\, \big(\xi_{1;0},\xi_{1;1}, \xi_{2;0}, \dots , \xi_{p+q;1} \big) $ and
\beq
\Dp{ \bs{Z}_{p+q} }^{\bs{\xi}_{p+q}} \, = \, \pl{s=1}{p+q} \pl{a=0}{1} \Dp{ Z_{s;a} }^{\xi_{s;a}} \;.
\enq
The vectors $\bs{\xi}_{p+q}$ and the function $\Phi$ along with its control \eqref{ecriture controle croissance facteur distribution temperee} do depend, in principle, on the given choice of the
operator indices $\bs{\tau}_{p+q}$. It thus follows readily from  changes of variables and integrations by parts that
$\de \mc{W}_{ \sg }[F] \, = \, \mc{W}_{ \bs{\tau}_{p+q} }[F]  \, - \, \mc{W}_{ \sg \cdot \bs{\tau}_{p+q} }[ \sg \cdot F]$
admits the representation
\beq
\de \mc{W}_{ \sg } [F]  \, =   \hspace{-2mm} \Int{ (\R^{1,1})^{p } }{} \hspace{-2mm}  \dd^{p} \bs{X} \hspace{-2mm}  \Int{ (\R^{1,1})^{q} }{} \hspace{-2mm}  \dd^{q} \bs{Y}
\, \Psi \big( \bs{X}_{p}, \bs{Y}_q \big) \, \Dp{ \bs{X}_{p} }^{\bs{\mu}_{p}} \Dp{ \bs{Y}_{q} }^{\bs{\rho}_{q}} F (\bs{X}_{p}, \bs{Y}_q) \;.
\label{ecriture rep canonique distribution delta W sg}
\enq
There $(\bs{\mu}_{p},\bs{\rho}_{q})\in \mathbb{N}^{2p} \times \mathbb{N}^{2q}$, $\Psi$ is continuous on $(\R^{1,1})^{p+q}$ and of polynomial growth in the sense of
\eqref{ecriture controle croissance facteur distribution temperee}, with possibly different values for $C, M$.

Let
\beq
\Om_{p,q} \, = \, \Big\{  \big( \bs{X}_{p}, \bs{Y}_q \big) \in (\R^{1,1})^{p+q}  \; : \; \big( \bs{x}_s - \bs{y}_r)^2 \, < \, 0 \quad \forall  (s,r) \in \intn{1}{p} \times \intn{1}{q} \Big\} \:.
\enq
It follows from Proposition \ref{local commut prop} that $\de \mc{W}_{ \sg } [F] = 0$ for any $F$ such that  $\e{supp}[F] \subset \Om_{p,q}$.
By proposition \ref{Proposition support fct dans thm structurel distributions temperees}, one may pick $\Psi$ arising in the representation
\eqref{ecriture rep canonique distribution delta W sg} such that it satisfies $ \e{supp}[\Psi]\subset \Big( \Om_{p,q}^{\e{c}} \big)_{\eps}$
with $\eps>0$ and small enough and
\beq
\Big( \Om_{p,q}^{\e{c}} \big)_{\eps} =  \Big\{ \bs{Z}_{p,q} \in  \big( \R^{1,1}\big)^{p+q}  \; : \; \e{d}\big( \Om_{p,q}^{\e{c}} , \bs{Z}_{p,q}  \big) \leq \eps \, \Big\} \;.
\enq
Further, it is direct to check that
\beq
\de\mf{M}^{(\la \bs{v})}[f,g] \, = \,  \mf{M}^{(\la \bs{v})}[f,g] \, - \, \mf{M}^{(\la \bs{v})}_{\sg}[f,g]
\enq
with $  \mf{M}^{(\la \bs{v})}, \mf{M}^{(\la \bs{v})}_{\sg}$ as introduced in \eqref{definition difference distrib}, \eqref{definition difference distrib sigma permutee},
can be recast as $\de\mf{M}^{(\la \bs{v})}[f,g] \, = \, \de \mc{W}_{ \sg } [G_{p,q}^{(\la \bs{v})}] $ with $G_{p,q}^{(\la \bs{v})}$ as introduced in \eqref{definition Gpq}.
Thus, after shifting the $\bs{Y}_q$ variables, one gets
\beq
\de\mf{M}^{(\la \bs{v})}[f,g] \, = \,\hspace{-2mm} \Int{ (\R^{1,1})^{p } }{} \hspace{-2mm}  \dd^{p} \bs{X} \hspace{-2mm}  \Int{ (\R^{1,1})^{q} }{} \hspace{-2mm}  \dd^{q} \bs{Y}
\, \Psi \big( \bs{X}_{p}, \bs{Y}_q -\la \bs{v} \big) \, \Dp{ \bs{X}_{p} }^{\bs{\mu}_{p}}f(\bs{X}_p) \,  \Dp{ \bs{Y}_{q} }^{\bs{\rho}_{q}} g (\bs{Y}_q) \;.
\enq
It is direct to infer from the polynomial bound on $\Psi$ that
\beq
\big| \Psi\big( \bs{X}_{p}, \bs{Y}_q -\la \bs{v} \big) \big| \, \leq \, C \,  \big( 1+\norm{ \bs{Z}_{p+q} } \big)^M  \big( 1 + |\la| \big)^M \qquad \e{with} \qquad
 \bs{Z}_{p+q} \, = \,  \big( \bs{X}_{p}, \bs{Y}_q  \big)  \;.
\label{ecriture borne sup fct Psi shiftee par lambda}
\enq
Likewise, the Schwartz class of $f$ and $g$ ensures that for any $Q>0$ there exist $C_Q>0$ such that
\beq
\big| \Dp{ \bs{X}_{p} }^{\bs{\mu}_{p}}f(\bs{X}_p) \,  \Dp{ \bs{Y}_{q} }^{\bs{\rho}_{q}} g (\bs{Y}_q) \big| \, \leq \,
\f{  C_Q  }{   (1 \,+ \, \norm{\bs{X}_p })^Q \cdot (1 \,+ \, \norm{\bs{Y}_q })^Q   } \, \leq \, \f{  C_Q  }{   (1 \,+ \, \norm{\bs{Z}_{p+q} })^Q  }
\label{ecriture borne sup produit fonctions f et g}
\enq
with $ \bs{Z}_{p+q} $ as introduced above. Further, observe that
\beq
\big( \bs{x}_r - \bs{y}_s+\la \bs{v} \big)^2 \, = \,   x_{s;0}^2 \, + \, y_{r;0}^2 \, - \, 2 x_{s;0}\cdot y_{r;0} \, -  \,  \big( x_{s;1} - y_{r;1} \big)^2 + \bs{v}^2 \la^2
\, - \, 2 \la\, \big[ v_0  (x_{s;0}-y_{r;0})  \, - \,   v_1 (x_{s;1}-y_{r;1}) \big] \;.
\enq
Thus, for any $ \bs{Z}_{p+q} \, = \,  \big( \bs{X}_{p}, \bs{Y}_q  \big) $ such that  $\norm{\bs{Z}_{p+q}} \, < \, \alpha |\la|$ with $\a$ small enough, one has
\begin{equation}
\big( \bs{x}_r - \bs{y}_{s} - \la \bs{v} \big)^2 \, < \, 4 \la^2 \Big[\a^2-  |\bs{v}^2| + 2 \a (\lvert v_0 \rvert + \lvert v_1 \rvert) \Big] \,  <  \,  - \f{\la^2 }{2}|\bs{v}^2| \;.
\end{equation}
Thus,  $\Psi\big( \bs{X}_{p}, \bs{Y}_q -\la \bs{v} \big)$ is identically zero whenever $\norm{\bs{Z}_{p+q}} \, < \, \alpha |\la|$. As a consequence, it holds
\beq
\de\mf{M}^{(\la \bs{v})}[f,g] \, = \,\hspace{-3mm} \Int{ (\R^{1,1})^{p+q } }{} \hspace{-3mm}  \dd^{p+q} \bs{Z} \mathbbm{1}_{\Ups}( \bs{Z}_{p+q} )
\, \Psi \big( \bs{X}_{p}, \bs{Y}_q -\la \bs{v} \big) \, \Dp{ \bs{X}_{p} }^{\bs{\mu}_{p}}f(\bs{X}_p) \,  \Dp{ \bs{Y}_{q} }^{\bs{\rho}_{q}} g (\bs{Y}_q) \;.
\enq
where $\Ups \,=\,  \big\{ \bs{Z}_{p+q} \in (\R^{1,1})^{p+q } \, : \, \norm{\bs{Z}_{p+q}} \, \geq  \, \alpha |\la|  \big\}$.
Passing on to spherical coordinates, using the upper bounds
\eqref{ecriture borne sup fct Psi shiftee par lambda}-\eqref{ecriture borne sup produit fonctions f et g} and  the compactness of  the angular integration
one gets for some constant $\wt{C}>0$
\bem
\big| \de \mf{M}^{(\la \bs{v})}[f,g] \big| \, \leq \, \wt{C} \, \big( 1 + |\la| \big)^M \hspace{-1mm} \Int{  \a \la }{ +\infty }  \hspace{-1mm}  \dd R  R^{2(p+q)-1} \f{ (1+R)^M}{ (1+R)^Q }
\,  \leq  \, \wt{C} \,  \big( 1 + |\la| \big)^M \hspace{-2mm} \Int{  1+ \a \la }{ +\infty }  \hspace{-3mm}  \dd R \, R^{2(p+q)+M-Q-1}   \\
\, \leq  \,  \wt{C}^{\prime}   \f{ \big( 1 + |\la| \big)^M }{ Q- 2(p+q)-M  } (\a |\la| )^{2(p+q)+M-Q} \underset{\la \tend  \infty}{\rightarrow} 0 \;,
\label{ecriture convergence vers zero des delta M}
\end{multline}
provided that $Q$ is taken large enough.

\vspace{2mm}

We shall now establish that this preliminary result implies that $\mf{M}^{(\la \bs{v})}[f,g] \tend 0$ as $\la \tend \infty$, which is the sought result.
Recall that, by eq. \eqref{reecriture distribution M et Msg via dist N et N sg} of Lemma \ref{Lemme support distributions}, one has
\beq
\mf{M}^{(\la \bs{v})}[f,g] \; = \;  \mf{N}^{(\la\bs{v})}\big[ \mc{L}[ G_{p,q}^{ (\bs{0})} ] \big] \, = \,
\msc{F}\big[\mf{N}^{(\la\bs{v})}\big] \Big[ \, \msc{F}^{-1}\big[ \mc{L}[ G_{p,q}^{ (\bs{0})} ] \big] \,  \Big] \;.
\enq

Let $\vth_{\pm}$ be smooth partitions of unity on $\R$, $\vth_{+}+\vth_{-}=1$, such that
\beq
\vth_+=1 \quad \e{on} \quad  \intff{m}{+\infty} \qquad  \e{and} \qquad  \vth_-=1 \quad \e{on}  \quad \intff{-\infty}{-m} \, ,
\enq
$m$ being the mass scale appearing in \eqref{definition support masse} and \eqref{definition support masse 2}. Let
\beq
\chi_{\pm}(\bs{q}_1, \dots, \bs{q}_{p+q-1} ) \,  = \, \vth_{\pm}(q_{p;0}) \,  \msc{F}^{-1}\big[ \mc{L}[ G_{p,q}^{ (\bs{0})} ] \big] (\bs{q}_1, \dots, \bs{q}_{p+q-1} ) \;.
\enq
One has thus $\mf{M}^{(\la \bs{v})}[f,g] \; = \; \mf{N}^{(\la\bs{v})}\big[ \msc{F}[ \chi_+  +  \chi_-] \big] $. However,
\beq
\e{supp}[ \chi_-] \cap\mc{E}^{+,+}_{p+q-1,p}=\emptyset  \quad \e{so}\, \e{that} \quad  \mf{N}^{(\la\bs{v})}\big[ \msc{F}[  \chi_-] \big] =0 \, .
\enq
Observing further that
\beq
\e{supp}[ \chi_+] \cap\mc{E}^{+,-}_{p+q-1,p}=\emptyset\, ,\quad \e{so}\, \e{that} \quad  \mf{N}^{(\la\bs{v})}_{\sg}\big[ \msc{F}[  \chi_+] \big] =0 \, ,
\enq
one gets that
\beq
\mf{M}^{(\la \bs{v})}[f,g] \; = \; \mf{N}^{(\la\bs{v})}\big[ \msc{F}[ \chi_+ ] \big] \, - \,  \mf{N}^{(\la\bs{v})}_{\sg}\big[ \msc{F}[ \chi_+ ] \big] \;.
\label{ecriture decomposition M la v de f et g sur les N}
\enq
We now compute $\msc{F}[ \chi_+ ]\big( \bs{Y}_{p+q-1}  \big)$ explicitly with $\bs{Y}_{p+q-1} = \big( \bs{y}_1,\dots,\bs{y}_{p+q-1} \big)$
\beqa
\msc{F}[ \chi_+ ]\big( \bs{Y}_{p+q-1}  \big)  & = &\hspace{-4mm} \Int{ (\R^{1,1})^{p+q-1} }{} \hspace{-5mm} \dd^{p+q-1} \bs{Q} \,
\msc{F}^{-1}\big[ \mc{L}[ G_{p,q}^{ (\bs{0})} ] \big] (\bs{Q}_{p+q-1} )   \vth_+(q_{p;0}) \pl{a=1}{p+q-1} \Big\{ \ex{\i \bs{q}_a * \bs{y}_a }  \Big\} \\
 & =& \lim_{\eps\tend 0^+} \hspace{-4mm} \Int{ (\R^{1,1})^{p+q-1} }{} \hspace{-5mm} \dd^{p+q-1}\bs{Q}  \,
\msc{F}^{-1}\big[ \mc{L}[ G_{p,q}^{ (\bs{0})} ] \big] (\bs{Q}_{p+q-1} ) \,  \ex{-\eps q_{p;0}^2 }  \, \vth_+(q_{p;0}) \pl{a=1}{p+q-1} \Big\{ \ex{\i \bs{q}_a * \bs{y}_a }  \Big\} \\
& = & \lim_{\eps\tend 0^+} \Int{ \R }{} \f{ \dd u }{ 2\pi } \,
 \mc{L}[ G_{p,q}^{ (\bs{0})} ] ( \bs{Y}_{p+q-1}- u \bs{e}_{p;0}  ) \,  \msc{F}\big[ r_{\eps} \vth_+ \big] (u) \; .
\eeqa
We have passed on to the second line using dominated convergence since $\msc{F}^{-1}\big[ \mc{L}[ G_{p,q}^{ (\bs{0})} ] \big]  $ is in the Schwartz class and
$\vth_+$ is smooth and bounded. Further, the third line was obtained by taking explicitly the Fourier transform.  There, we have denoted
the canonical basis of $(\R^{1,1})^{p+q-1}$ by $\bs{e}_{s;a}$, $s\in \intn{1}{p+q-1}$ and $a\in \intn{0}{1}$, and have set $r_{\eps}(s) \, = \,  \ex{- \eps s^2 }$.
Now, going back to the explicit expression for  $\mc{L}[ G_{p,q}^{ (\bs{0})} ]$, we get
\bem
\msc{F}[ \chi_+ ]\big( \bs{Y}_{p+q-1}  \big)  \, = \,  \lim_{\eps\tend 0^+} \Int{ \R }{} \hspace{-1mm} \f{ \dd u }{ 2\pi }  \Int{ \R^{1,1} }{} \hspace{-1mm} \dd \bs{y}_{p+q} \, \msc{F}\big[ r_{\eps} \vth_+ \big] (u)
\, f\bigg( \sul{s=1}{p+q} \bs{y}_s \, - \, (u,0), \dots, \sul{s=p}{p+q} \bs{y}_s \, - \, (u,0) \bigg) \\
\times g\bigg(  \sul{s=p+1}{p+q} \bs{y}_s, \dots,  \bs{y}_{p+q}  \bigg) \;.
\label{ecritue representation integrale TF chi+}
\end{multline}
The convolution term appearing above can be recast as
\beqa
f_{\vth_+; \eps}(\bs{X}_p) & = &\Int{ \R }{} \hspace{-1mm} \f{ \dd u }{ 2\pi } \msc{F}\big[ r_{\eps} \vth_+ \big] (u)
f\big( \bs{x}_1 \, - \, (u,0), \dots, \bs{x}_p \, - \, (u,0) \big)  \\
& = &   \hspace{-2mm}  \Int{ (\R^{1,1})^p}{}  \hspace{-2mm} \dd^p \bs{Q} \, \msc{F}[f](\bs{Q}_p) \;  \vth_+\bigg( \sul{a=1}{p} Q_{a;0} \bigg) \;  \exp\bigg\{ -\eps  \Big( \sul{a=1}{p} Q_{a;0} \Big)^2 \bigg\} \;.
\eeqa
For any $\eps \geq 0$, $f_{\vth_+;\eps}$ is thus given by the Fourier transform of a function in the Schwartz class, so that $f_{\vth_+;\eps} \in \mc{S}\big( (\R^{1,1})^p \big)$.
It is easy to see by integration by parts that, uniformly in $\eps \geq 0$ and small enough,
\beq
\big| f_{\vth_+; \eps}(\bs{X}_p)  \big| \, \leq \, \f{ C_M }{  1+ \norm{X_p}^M }
\label{ecriture borne sup sur F vth+ eps}
\enq
for any $M>0$ and an $M$-dependent constant $C_M>0$.  Finally, by using that $ \msc{F}[f](\bs{Q}_p)$ is in the Schwartz class, one readily gets by dominated convergence that $f_{\vth_+;\eps}(\bs{X}_p) $
converges pointwise to $f_{\vth_+;0}(\bs{X}_p) $. Then, the pointwise convergence along with the domination estimates \eqref{ecriture borne sup sur F vth+ eps} allow one to apply dominated convergence on the level of
\eqref{ecritue representation integrale TF chi+}, which yields
\beq
\msc{F}[ \chi_+ ]\big( \bs{Y}_{p+q-1}  \big)  \, = \, \Int{ \R^{1,1} }{} \hspace{-1mm} \dd \bs{y}_{p+q} \,
\, f_{ \vth_+ ; 0 } \bigg( \sul{s=1}{p+q} \bs{y}_s , \dots, \sul{s=p}{p+q} \bs{y}_s \ \bigg)
\, g\bigg(  \sul{s=p+1}{p+q} \bs{y}_s, \dots,  \bs{y}_{p+q}  \bigg) \, = \, \mc{L}\big[ \wt{G}_{p,q}^{(\bs{0})}\big]\big( \bs{Y}_{p+q-1}  \big) \;,
\enq
where $\wt{G}_{p,q}^{(\la\bs{v})} \big( \bs{X}_p, \bs{Y}_q\big) \,= \, f_{ \vth_+ ; 0 }\big( \bs{X}_p \big) g\big( \bs{Y}_q\big)$.

Thus, going back to the representation \eqref{ecriture decomposition M la v de f et g sur les N}  and making use of the relations
\eqref{reecriture distribution M et Msg via dist N et N sg}, we get that
\beq
\mf{M}^{(\la\bs{v})}[f,g]  \, = \, \mf{M}^{(\la\bs{v})}[f_{ \vth_+ ; 0 },g] \, - \,  \mf{M}^{(\la\bs{v})}_{\sg} [f_{ \vth_+ ; 0 },g]
\limit{\la}{+\infty} \, 0 \;,
\enq
by \eqref{ecriture convergence vers zero des delta M} since $f_{ \vth_+ ; 0 } \in \mc{S}\big( (R^{1,1})^p \big)$.

\section*{Conclusion}

This work demonstrated that, under the convergence conjecture \ref{Conjecture convergence} of the defining series  representation, the multi-point correlation functions from \cite{KozSimonMultiPointCorrFctsSinhGordon}
and obtained by solving the integrable bootstrap equations
for the Sinh-Gordon quantum field theory satisfy the  Wightman axioms. The method we developed is universal to integrable quantum field theories in that we only used the structural
properties of the correlation functions stemming from the bootstrap equations and did not rely on some specificities of the model.
We are thus fairly confident that, upon minor modifications, it should be applicable to other, more complex quantum field
theories, the Sine-Gordon integrable quantum field theory in particular.

\section*{Acknowledgment}

The work of KKK and AS is supported by the ERC Project LDRAM : ERC-2019-ADG Project 884584. KKK acknowledges support from CNRS, from the joint AND-DFG TSF24 project ANR-24-CE92-0033
and the CNRS-IEA “Corrélateurs dynamiques de la chaîne XXZ en et hors équilibre thermal” grant.

\appendix

\section{Support of tempered distributions}

In this section we prove a result on the support of functions representing tempered distributions. This seems to be a very classical property. However, since we did not manage to find a reference
in the literature, we include a proof here for completeness.

\begin{prop}
\label{Proposition support fct dans thm structurel distributions temperees}

Let $N\in \mathbb{N}^*$ and $\op{D}$ be a tempered distribution on $\mc{S}(\R^N)$ such that $\e{supp}[\op{D}] \subset \Om^{\e{c}}$, with $\Om$ an open subset of $\R^N$. Then,
there exists $\bs{\mu}_N \,= \, (\mu_1, \dots, \mu_N)\in \mathbb{N}^N$ and $h \in \mc{C}^{0}(\R^N)$ of at most polynomial growth
\beq
\big| h(\bs{X}_p) \big| \, \leq \, C \cdot \big( 1+ \norm{\bs{X}_N}^M \big) \quad \e{for}\; \e{some} \quad C, M>0 \quad \e{with} \quad \norm{\bs{X}_N}^2 \,=\, \sul{s=1}{N} x_s^2\;,
\enq
satisfying  $\e{supp}[h] \subset \big(\Om^{\e{c}}\big)_{\eps}$ with $\eps>0$ and as small as need be, $A_{\eps}= \big\{ \bs{X}_N\in \R^{N} \, : \, \e{d}(A, \bs{X}_N ) \leq \eps \big\}$, and such that
\beq
\op{D}[f] \; =  \Int{ \R^N }{} \dd^N \bs{X}   \, h(\bs{X}_N) \,  \Dp{\bs{X}_N}^{\bs{\mu}_N} f(\bs{X}_N)
\label{ecriture represnetation distribution temperee avec restriction de support}
\enq
 for any $f \in \mc{S}(\R^N)$. There $\bs{X}_N=(x_1,\dots, x_N)\in \R^N$ and $\Dp{\bs{X}_N}^{\bs{\mu}_N} = \pl{a=1}{N} \Dp{x_a}^{\mu_a}$ .

\end{prop}

\Proof

The fact that there exists  $h \in \mc{C}^{0}(\R^N)$ and $\bs{\mu}_N \in \mathbb{N}^N$ such that \eqref{ecriture represnetation distribution temperee avec restriction de support}
holds is ensured by Theorem 8.3.1 in \cite{FriedlanderIntroDistributions}.
We now show that we may reduce to a setting where $\e{supp}[h] \subset (\Om^{\e{c}})_{\eps}$ with $\eps>0$ and as small as need be.

Let $\eps>0$  and pick $\Ups$ open in $\R^N$ such that $\ov{\Ups} \subset \Om$ and $\Ups^{\e{c}} \subset (\Om^{\e{c}})_{\eps}$. Let $f \in \mc{S}(\R^N)$ be such that $\e{supp}[f] \subset \Ups$.
Then, pick $\chi_{\sg}  \in \mc{C}^{\infty}_{\e{c}}(\R^N)$ such that $\e{supp}[\chi_{\sg}] \subset \intff{-\sg}{\sg}^N$ for $\sg$ small enough, $\chi_{\sg} \geq 0$
and $\chi_{\sg}$ integrates to $1$. Let
\beq
f_{\sg}(\bs{X}_N) \, = \,\Int{\R^N}{} \dd^N \bs{Y} \chi_{\sg}(\bs{X}_N- \bs{Y}_N) f(\bs{Y}_N) \;,
\enq
so that $\e{supp}[f_{\sg}] \, \subset \, (\e{supp}[f])_{\sg}$. In particular, for $\sg$ small enough, independent of $f$, $\e{supp}[f_{\sg}] \subset \Om$.
Thus, one has
\beqa
0 &=& \op{D}[f_{\sg}] \, = \, \Int{ \R^N }{} \dd^N \bs{X}  \Int{\R^N}{} \dd^N \bs{Y}  \, h(\bs{X}_N) \,     (\Dp{\bs{X}_N}^{\bs{\mu}_N}\chi_{\sg})(\bs{X}_N- \bs{Y}_N) f(\bs{Y}_N)  \\
&= & (-1)^{|\bs{\mu}_N|} \Int{\R^N}{} \dd^N \bs{Y}  f(\bs{Y}_N)  \, \,
 \Dp{\bs{Y}_N}^{\bs{\mu}_N} \underbrace{ \Int{ \R^N }{} \dd^N \bs{X}   \chi_{\sg}(\bs{X}_N- \bs{Y}_N)  h(\bs{X}_N) }_{ = h_{\sg}(\bs{Y}_N) } \;.
\eeqa
$h_{\sg}$ is smooth and since the relation holds for any Schwartz function $f$ with support inside of $\Ups$, one has
\beq
\Dp{ \bs{X}_N }^{ \bs{\mu}_N } h_{\sg} \big( \bs{X}_N \big) \, = \, 0 \qquad \e{with} \qquad \bs{X}_N \in \Ups  \;.
\enq
Thus, there exists a multivariate polynomial $P_{\sg}\in \Cx[X_1,\dots, X_N]$ with upper bounded partial degree in respect to $X_s$
$\e{d}_{s}[P_{\sg}] \, <\, \mu_s$ uniformly in $\sg$ small enough, and such that $h_{\sg}(\bs{X}_N) = P_{\sg}(\bs{X}_N)$ for any $\bs{X}_N \in \Ups $.
One has that $h_{\sg} \tend h $ pointwise on $\Ups$, so that $P_{\sg}(\bs{X}_N) \tend P_{0}(\bs{X}_N)$. Obviously, $P_0$ is a polynomial and
$\e{d}_{s}[P_{0}] \, <\, \mu_s$. Therefore, one has $h=P_0$ on $\Ups$. Now, obviously, for any $f \in \mc{S}(\R^N)$, it holds
\beq
\Int{\R^N}{} \dd^N \bs{Y}  P_0(\bs{Y}_N)  \, \Dp{\bs{Y}_N}^{\bs{\mu}_N}   f(\bs{Y}_N) \, = \,  0 \;.
\enq
One then sets $\wt{h}=h-P_0$ and $\e{supp}[\wt{h}] \subset \Ups^{\e{c}}  \subset (\Om^{\e{c}})_{\eps}$, and it obviously holds
\beq
\op{D}[f_{\eps}] \, = \, \Int{\R^N}{} \dd^N \bs{Y} \wt{h}(\bs{Y}_N)  \, \Dp{\bs{Y}_N}^{\bs{\mu}_N}   f(\bs{Y}_N) \;.
\enq
The polynomial growth of $\wt{h}$ follows from the one of $h$. \qed


\begin{thebibliography}{10}

\bibitem{BabujianFringKarowskiZapletalExactFFSineGordonBootsstrapI}
H.~Babujian, A.~Fring, M.~Karowski, and A.~Zapletal, \emph{{"Exact form factors
  in integrable quantum field theories: the sine-Gordon model."}}, Nucl. Phys.
  B \textbf{\bf{538}} (1999), 535--586.

\bibitem{BabujianKarowskiBreatherFFSineGordon}
H.~Babujian and M.~Karowski, \emph{{"Sine-Gordon breather form factors and
  quantum field equations."}}, J.Phys.A \textbf{\bf{35}} (2002), 9081--9104.

\bibitem{BogoliubiovLogunovTodorovIntroAxiomaticQFT}
N.~N. Bogolubov, A.~A. Logunov, and I.~T. Todorov, \emph{{"Introduction to
  Axiomatic Quantum Field Theory."}}, Mathematical physics monographs series,
  W. A. Benjamin, Inc., Advanced Book Program, Reading, Massachusetts, 1975.

\bibitem{BrazhnikovLukyanovFreeFieldRepMassiveFFIntegrable}
V.~Brazhnikov and S.~Lukyanov, \emph{{"Angular quantization and form factors in
  massive integrable models."}}, Nucl. Phys. B \textbf{\bf{512}} (1998),
  616--636.

\bibitem{FeiginLashkevichFreeFieldApproachToDescendents}
B.~Feigin and M.~Lashkevich, \emph{{"Form factors of descendant operators: free
  field construction and reflection relations."}}, J. Phys. A: Math. Theor.
  \textbf{42} (2009), 304014.

\bibitem{FriedlanderIntroDistributions}
F.G. Friedlander, \emph{"introduction to the theory of distributions"},
  Cambridge University Press, 1982.

\bibitem{GlimmJaffeSpencerP(phi)2ParRenormalisationConstructive}
J.~Glimm, A.~Jaffe, and T.~Spencer, \emph{{"The Wightman axioms and particle
  structure in the $P(\phi)_2$ quantum field model."}}, Ann. Math. \textbf{\bf
  100} (1974), 585--632.

\bibitem{GryanikVergelesSMatrixAndOtherStuffForSinhGordon}
V.M. Gryanik and S.N. Vergeles, \emph{{"Two-dimensional quantum field theories
  having exact solutions."}}, J. Nucl. Phys. \textbf{\bf 23} (1976),
  1324--1334.

\bibitem{GubinelliHofmanovaPhi4inD=2And3EucildianStochasticQuantisation}
M.~Gubinelli and M.~Hofmanov\'{a}, \emph{{" Global Solutions to Elliptic and
  Parabolic $\Phi^4$ Models in Euclidean Space."}}, Comm Math. Phys.
  \textbf{368} (2019), 1201–1266.

\bibitem{GuillarmouGunaratnamVargasGMCConstructionofSinhGMeasure}
C.~Guillarmou, T.S. Gunaratnam, and V.~Vargas, \emph{{"2d Sinh-Gordon model on
  the infinite cylinder."}}, math-pr:2405.04076.

\bibitem{GuillarmouKupiainenRhodesVargasProofOfDOZZByGaussianMultChaos}
C.~Guillarmou, A.~Kupiainen, R.~Rhodes, and V.~Vargas, \emph{{"Conformal
  bootstrap in Liouville theory."}}, Acta. Math. \textbf{233} (2024), 33--194.

\bibitem{HeisenbergSomeAspectsofSMatrixIdeas}
W.~Heisenberg, \emph{{"Der mathematische Rahmen der Quantentheorie der
  Wellenfelder."}}, Zeit. f\"{u}r Naturforschung \textbf{\bf{1}} (1946),
  608--622.

\bibitem{Jona-LasinioMitterStochasticQuantisationPhi4In2D}
G.~Jona-Lasinio and P.K. Mitter, \emph{{"On the stochastic quantization of
  field theory."}}, Comm. Math. Phys. \textbf{\bf{101}} (1985), 409--436.

\bibitem{KarowskiWeiszFormFactorsFromSymetryAndSMatrices}
M.~Karowski and P.~Weisz, \emph{{"Exact form factors in (1+1)-dimensional field
  theoretic models with soliton behaviour."}}, Nucl. Phys. B \textbf{\bf 139}
  (1978), 455--476.

\bibitem{KhamitovGLMeqnsForSinhGordonAndProofOfLocalCommutativity}
I.~M. Khamitov, \emph{{"A constructive approach to the quantum (cosh$\phi$)2
  model. I. The method of the Gel'fand-Levitan-Marchenko equations."}}, J. Sov.
  Math. \textbf{40} (1988), 115--148.

\bibitem{KirillovCombinatorialIdenititesForLocalCommutativityinSinhGordonQFTByKhamitov}
A.N. Kirillov, \emph{{"T-invariance, CPT-invariance, and local commutativity
  for the quantum (cosh $\phi$)2-model."}}, J. Sov. Math. \textbf{\bf 40}
  (1988), 6--13.

\bibitem{KirillovSmirnovFirstCompleteSetbootstrapAxiomsForQIFT}
A.N. Kirillov and F.A. Smirnov, \emph{{"A representation of the current algebra
  connected with the SU (2)-invariant Thirring model."}}, Phys. Rev. B
  \textbf{\bf 198} (1987), 506--510.

\bibitem{KirillovSmirnovUseOfbootstrapAxiomsForQIFTToGetMassiveThirringFF}
\bysame, \emph{{"Form-factors in the SU(2)-invariant Thirring model."}}, J.
  Soviet. Math. \textbf{\bf 47} (1989), 2423--2450.

\bibitem{KozConvergenceFFSeriesSinhGordon2ptFcts}
K.K. Kozlowski, \emph{{"On convergence of form factor expansions in the
  infinite volume quantum Sinh-Gordon model in 1+1 dimensions."}}, Inventiones
  mathematicae \textbf{233} (2023), 725--827.

\bibitem{KozSimonMultiPointCorrFctsSinhGordon}
K.K. Kozlowski and A.~Simon, \emph{{"Multi-point correlation functions in the
  infinite volume quantum Sinh-Gordon model in 1+1 dimensions."}}, math-ph:
  2502.03894.

\bibitem{KupiainenRhodesVargasProofOfDOZZByGaussianMultChaos}
A.~Kupiainen, R.~Rhodes, and V.~Vargas, \emph{{"Integrability of Liouville
  theory: proof of the DOZZ Formula."}}, Ann. of Math. \textbf{191} (2020),
  81--166.

\bibitem{LukyanovFirstIntroFreeField}
S.~Lukyanov, \emph{{"Free field representation for massive integrable
  models."}}, Comm. Math. Phys. \textbf{\bf{167}} (1995), 183--226.

\bibitem{HairerRegularityStructureOverviewAndPhi4in3D}
M.Hairer, \emph{{"Regularity structures and the dynamical $\Phi^4_3$ model."}},
  Current Developments in Mathematics, \textbf{\bf} (2014), 49pp.

\bibitem{OsterwalderSchraderAxiomsEQFTIWithError}
K.~Osterwalder and R.~Schrader, \emph{{"Axioms for Euclidean Green’s
  functions."}}, Comm. Math. Phys. \textbf{\bf 31} (1973), 83--112.

\bibitem{OsterwalderSchraderAxiomsEQFTIIAdditionalAxiomsAndProof}
\bysame, \emph{{"On the equivalence of the Euclidean and Wightman formulation
  of field theory."}}, Comm. Math. Phys. \textbf{\bf 37} (1974), 257--272.

\bibitem{ParisiWuStochasticQuantisationIdea}
G.~Parisi and Y.~Wu, \emph{{"Perturbation theory without gauge fixing."}},
  Scientia Sinica \textbf{\bf 24} (1981), 483--496.

\bibitem{SmirnovFormFactors}
F.A. Smirnov, \emph{{"Form factors in completely integrable models of quantum
  field theory."}}, Advanced Series in Mathematical Physics, vol.~14, World
  Scientific, 1992.

\bibitem{StreaterWightmanPCTStatisticsAllThat}
R.~Streater and A.~Wightman, \emph{{"PCT, Spin and statistics, and all
  that."}}, W.A. Benjamin inc., New York New York, vol.~1, The mathematical
  physics monograph series, 1964.

\bibitem{SummersStateOfTheArtConstructiveQFTAsOf2012}
S.J. Summers, \emph{{"A Perspective on Constructive Quantum Field Theory."}},
  math-ph: 1203.3991 (2012), 60pp.


\bibitem{WheelerFirstIntroConceptSmatrix}
J.A. Wheeler, \emph{{"On the mathematical description of light nuclei by the
  method of resonating group structure."}}, Phys. Rev. B \textbf{\bf 52}
  (1937), 1107--1106.

\end{thebibliography}
\end{document}